\documentclass[hyper]{JHEP3} 
\usepackage{epsfig,amsmath}

\newcommand{\newc}{\newcommand}
\newc\eg{{\it {e.g.}}}	\newc\vs{{\it {vs.}}}	\newc\etal{{\it {et al.}}}
\newc\etc{{\it {etc.}}}	\newc\ie{{\it {i.e.}}}	

\newcommand\vd{v_d}     \newcommand\vu{v_u} \newcommand\vdu{v_{d,u}}     

\newcommand\unitmatrix{ {\mathchoice {\rm 1\mskip-4mu l} {\rm 1\mskip-4mu l}
{\rm 1\mskip-4.5mu l} {\rm 1\mskip-5mu l}} }

\newcommand{\charn}{\chi^{-}} 

\newc{\mhalf}{m_{1/2}}      \newc{\mzero}{m_0}

\newc{\pleft}{P_L}      \newc{\pright}{P_R}

\newc{\sckm}{\mbox{\rm sCKM}}  \newc{\sckmzero}{\mbox{\rm sCKM}^{(0)}}

\newcommand\gluino{\widetilde{g}} \newcommand\mgluino{m_{\gluino}}

\newcommand{\stopone}{\wt{t}_1} \newcommand{\stoptwo}{\wt{t}_2} 
\newcommand{\mstopone}{m_{\stopone}} \newcommand{\mstoptwo}{m_{\stoptwo}}
\newcommand{\sbottom}{\wt{b}} 

\newcommand{\msquark}{m_{\wt{q}}}

\newc\bsgamma{b\rightarrow s\gamma} \newc\bxsgamma{B\rightarrow X_{s}\gamma}
\newc\brbsgamma{BR(B\rightarrow X_s\gamma)}

\newc{\tanb}{\tan\beta}
\newc{\azero}{A_0}
\newc{\at}{A_t} \newc{\abot}{A_b} \newc{\atau}{A_\tau}
\newc{\bmu}{B\mu}           \newc{\sgn}{{\rm sgn}}
\newc{\mone}{M_1}           \newc{\mtwo}{M_2}

\newc{\bino}{\widetilde B}              \newc{\wino}{\widetilde W_3}
\newc{\higgsinob}{{\widetilde H}^0_b}   \newc{\higgsinot}{{\widetilde H}^0_t}

\newc{\mtop}{m_{\rm t}}
\newc{\mbottom}{m_{\rm b}}

\newc{\mw}{m_{\rm W}}
\newc\msusy{M_{\rm SUSY}}
\newc{\mplanck}{M_{\rm P}}

\newc{\mub}{{\mu}_{\rm b}}	  \newc{\muw}{{\mu}_{\rm W}}
\newc{\mususy}{{\mu}_{\rm SUSY}}  \newc{\muzero}{\mu_{\rm 0}}
\newc{\deltaw}{\delta^{W}}	  \newc{\deltah}{\delta^{H}}
\newc{\deltas}{\delta^{S}}	  \newc{\deltacharn}{\delta^{\charn}}
\newc{\deltax}{\delta^{X}}	  
\newc{\deltaneut}{\delta^{\chi^0}}  \newc{\deltagluino}{\delta^{\gluino}}

\newc{\Ci}{C_i}	\newc{\Cip}{C_i^{\prime}}

\newc{\deltadll}{\delta^d_{LL}}	\newc{\deltadlr}{\delta^d_{LR}}
\newc{\deltadrl}{\delta^d_{RL}}	\newc{\deltadrr}{\delta^d_{RR}}

\newc{\abund}{\Omega h^2}
\newc{\abundchi}{\Omega_\chi h^2}
\newc{\rhocrit}{\rho_{crit}}
\newc{\rhochi}{\rho_{\chi}}

\newc{\xf}{x_f}
\newc{\jxf}{J({\xf})}
\newc{\VEV}[1]{\langle #1 \rangle}

\newcommand\tev{\,\mbox{TeV}}
\newcommand\gev{\,\mbox{GeV}}

\newc{\ra}{\rightarrow}
\newc{\beq}{\begin{equation}}
\newc{\eeq}{\end{equation}}
\newc{\bea}{\begin{eqnarray}}
\newc{\eea}{\end{eqnarray}}

\newcommand\lsim{\mathrel{\rlap{\lower4pt\hbox{\hskip1pt$\sim$}}
    \raise1pt\hbox{$<$}}}
\newcommand\gsim{\mathrel{\rlap{\lower4pt\hbox{\hskip1pt$\sim$}}
    \raise1pt\hbox{$>$}}}

\newcommand{\mdbare}{m_d^{(0)}}
\newcommand{\mubare}{m_u^{(0)}}

\newcommand{\Rn}[1]{{\uppercase\expandafter{\romannumeral#1}}}
\newcommand{\ovl}[1]{\overline{#1}}
\newcommand{\wt}[1]{\widetilde{#1}}

\newcommand{\msd}{m_{\wt{d}}}

\newcommand{\mg}{m_{\wt{g}}}

\newcommand{\mdq}{m_d}
\newcommand{\muq}{m_u}
\newcommand{\mdqw}{\ovl{m}_d}
\newcommand{\muqw}{\ovl{m}_u}

\newcommand{\mca}{ m_{\chi^{-}_{a}} }
\newcommand{\mnr}{ m_{\chi^{0}_{r}} }

\newcommand{\xsuica}{x^{{\widetilde u}_{I}}_{ \chi^{-}_{a}}}
\newcommand{\xsdinr}{x^{{\widetilde d}_{I}}_{ \chi^{0}_{r}}}
\newcommand{\xsdig}{x^{{\widetilde d}_{I}}_{ \widetilde{g}}}

\newcommand{\epsfile}[1]{\relax}
\title{Large Beyond--Leading--Order Effects in
 {\boldmath 
$b\rightarrow s\gamma$
}\\ in Supersymmetry with General Flavor Mixing }

\author{Ken-ichi Okumura\\
        Department of Physics, Lancaster University,
        Lancaster LA1 4YB, England\\
        E-mail: \email{K.Okumura@lancaster.ac.uk}}

\author{Leszek Roszkowski\\
        Department of Physics, Lancaster University,
        Lancaster LA1 4YB, England\\
        E-mail: \email{L.Roszkowski@lancaster.ac.uk}}

\abstract{We examine squark--gluino loop effects on the process
  $\bsgamma$ in minimal supersymmetry with general flavor mixing in
  the squark sector. In the regime of heavy squarks and gluino, we
  derive analytic expressions for the beyond--LO corrections to the
  Wilson coefficients and find them to be often large, especially at
  large $\tanb$ and $\mu>0$. The ensuing ranges of values of the
  Wilson coefficients are typically smaller than in the LO
  approximation, and sometimes even change sign. This has the effect
  of often reducing, relative to the LO, the magnitude of
  supersymmetric contributions to $\brbsgamma$. This ``focusing
  effect'' is caused by contributions from: {\it (i)} an RG evolution
  of the Wilson coefficients; {\it (ii)} a correction to the LO
  chargino contribution to the Wilson coefficients, which can
  considerably reduce the LO gluino contribution.  This partial
  cancellation of the two contributions takes place only in the case
  of general flavor mixing.  As a result, stringent lower bounds on
  the mass scale of superpartners, which apply in the case of minimal
  flavor violation, can be substantially reduced for even small
  departures from the scenario. The often disfavored case of $\mu<0$
  can also become allowed for $\msusy$ as small as $\sim200\gev$,
  compared to $\gsim500\gev$ at LO and over $2\tev$ in the case of
  minimal flavor violation. Limits on the allowed amount of flavor
  mixing among the 2nd and 3rd generation down--type squarks are also
  typically considerably weakened. The input CKM matrix element
  $K^{(0)}_{cb}$ can be larger than the experimental value by a factor
  of ten, or can be as small as zero.  }


\begin{document}


\section{ Introduction}\label{sec:intro}

It has long been recognized that the inclusive process $B\rightarrow
X_{s}\gamma$ plays a prominent role in testing ``new physics'' beyond
the Standard Model (SM)~\cite{hurth02}.  In the SM the leading
contribution comes from a virtual $W$--top (charm, up) quark loop,
which is accompanied by a real photon with GIM suppression.
New physics effects, which are also loop induced, can be of the same
order if a mass scale associated with the new physics is not much
larger than $\mw$. This in particular is the case with softly broken low--energy
supersymmetry (SUSY) where the scale of SUSY breaking $\msusy$ is
expected to be $\lsim {\cal O}(1\tev)$ on the grounds of naturalness.

Over the last decade, experimental precision has improved
considerably, and the world average of the measured branching ratio
has now been
determined with an uncertainty of some $10\%$,
\beq
{\brbsgamma}_{expt} = ( 3.34 \pm 0.38 )\times 10^{-4},
\label{bsgexptvalue:ref}
\eeq 
where results from four experiments~\cite{exp}, including
a recent one from BaBar, have been taken into account~\cite{expaverage}.

Theoretical calculations in the SM have been performed in several
steps~\cite{bm02,ggh03}, including the full next--to--leading order (NLO) QCD
correction which has recently been completed in~\cite{bcmu02}, and
have reached as similar precision\footnote{Somewhat different ranges
are quoted in the literature but they agree within their $1\,\sigma$
errors.}
\beq 
{\brbsgamma}_{SM} = ( 3.70 \pm 0.30 ) \times 10^{-4}.
\label{bsgsmvalue:ref}
\eeq 
Given the approximate agreement of the SM
prediction~(\ref{bsgsmvalue:ref}) with
experiment~(\ref{bsgexptvalue:ref}), any possible ``new physics''
effects must now be confined to the remaining, relatively narrow,
window of uncertainty.  This is normally expected to impose severe
constraint on extensions of the SM.  For example, in the two Higgs
doublet model (2HDM) extension of the SM, where full NLO QCD
corrections have also been completed~\cite{2hdnlo,cdgg97,gm01}, this
allows one to derive rather stringent lower limit on the mass of the
charged Higgs. It is worth noting, however, that the limit becomes
sizeably weaker when NLO QCD corrections are included, as compared
with the leading order (LO) approximation~\cite{cdgg97}.  This
demonstrates the importance of NLO QCD
corrections~\cite{2hdnlo,cdgg97} and NLO mass renormalization
effects~\cite{gm01}.

Supersymmetric contributions to $\brbsgamma$ have not yet been
calculated with a similar level of accuracy, although, since the first
detailed LO analysis~\cite{bbmr90}, much important work has been done
towards reaching this goal~\cite{hurth02}. On the other hand, one
might argue that, given little room left for ``new physics''
contributions, it may actually not be essential to include NLO QCD and
SUSY QCD corrections since all the relevant superpartners may anyway
have to be heavy enough in order for the SUSY contributions to be
suppressed. However,
in this regime, the magnitude of SUSY contributions is often
comparable with LO and NLO QCD corrections.  Furthermore, in some
cases separate SUSY effects may actually be quite large, but may
approximately cancel each other, thus allowing for lower $\msusy$.

More importantly, new effects are known to exist beyond LO, and these
may significantly modify SUSY contributions which in the LO
approximation appear to be large. One such important new effect is
induced by squark--gluino loop corrections to quark masses and
couplings. For example, the mass correction to the bottom quark can be
of order $\sim40\%$ in the regime of large $\tanb$~\cite{hrs}. The
same mechanism has also a strong impact on effective quark couplings
to gauge and Higgs bosons and to
superpartners~\cite{br99,cgnw99,dgg00,cgnw00}.  Its effect on
$\bsgamma$ has been explored~\cite{dgg00,cgnw00} in the framework of
minimal flavor violation (MFV)~\cite{mfv:ref} (where one assumes that
the CKM matrix is a unique source of flavor mixing among both quarks
and squarks), and shown to be sizeable at large $\tan\beta$ and for
large ratio of $\msusy/\mw$.

The MFV scenario has some appealing theoretical features and
is also partially motivated by stringent constraints from the
$K^0-\bar{K}^0$ system \cite{ggms96,ciuchinietal98,cfms03} 
on the allowed mixings among the squarks of the first two
generations. On the other hand, it is hard to believe that MFV is
strictly obeyed at the electroweak scale. It is usually assumed to
hold at some high scale $\Lambda$, like the grand unification scale,
in an attempt to relate soft SUSY breaking terms to the structure of
the Yukawa sector. However, the assumptions of MFV are not
RG--invariant\footnote{For a comprehensive discussion and
model--independent approach see~\cite{agis02}.} and sizeable
flavor--violating terms of order
$\ln\left(\Lambda^2/\mw^2\right)/\left(4\pi\right)^2$ are generated by
the running from the scale $\Lambda$ to the electroweak scale. Further
violations are also induced by threshold effects at the SUSY breaking
scale $\msusy\sim1\tev$. It is therefore sensible to assume that MFV
can be regarded as, at best, only an approximate scenario at the
electroweak scale.

A broader framework is that of general flavor mixing (GFM) among
squarks and/or among sleptons. Flavor mixings of this type are not
protected by any known symmetry, and can in principle be large. As a
result, without any additional ansatz, otherwise allowed
contributions to, \eg, the processes such as $\mu\to e\gamma$ or
$K^0-\bar{K}^0$ mixing can exceed experimental bounds by as much as
several orders of magnitude~\cite{en81}.
In this case, experimental
limits become an efficient tool in severely restricting possible
mixings among scalar superpartners, especially between the 1st and 2nd
generation, while still leaving considerable room for departures from the
MFV scheme.

In the case of $\bsgamma$, going beyond the MFV framework has particularly important
implications. While in the MFV case only the chargino and up--squark
loop exchange contributes to the process, in going beyond MFV
flavor--violating gluino (and, to a lesser extent, neutralino) and
down--squark loop exchange diagrams start playing a substantial
role. The importance of these additional diagrams has long been
recognized~\cite{bbm86}, in part because of the gluino exchange
enhancement by the strong coupling constant. 
Following the earlier work of~\cite{ht92,ggms96}, it has more recently
been carefully studied in the LO approximation~\cite{bghw99}, and also
in some special cases beyond LO, \eg~in~\cite{ekrww02}. (For a clear
and systematic discussion of SUSY contributions in the case of GFM see
ref.~\cite{bghw99}.)

Given the fact that beyond--LO effects in $\bsgamma$ have been shown
to be sizeable in MFV and that by going beyond MFV new important loop
contributions appear, it is worthwhile to consider the role of
beyond--LO effects in the framework of GFM among squarks, and to
examine their impact on bounds on supersymmetric masses and other
parameters. In particular, one may ask whether small departures from
MFV assumptions lead to only small perturbations on the various bounds
on SUSY derived in the case of MFV. We will show that this is not the
case.

In the previous paper~\cite{or1} we have presented our first results
showing a strong impact of the squark--gluino loop corrections on
$\bsgamma$ in the framework of GFM. We have pointed out that such
beyond--LO effects have an important effect of generally reducing the
magnitude of SUSY contribution to $\brbsgamma$, especially at large
$\tan\beta$ and for $\mu>0$. This effect of ``focusing'' on the SM
value is already present at some level in the case of MFV but becomes
strongly enhanced when GFM is allowed. This is because in this case
large beyond--LO contributions from the chargino and the gluino can
partially cancel each other. Furthermore, gluino--squark corrections
typically reduce SUSY contributions to Wilson coefficients $C_{7,8}$
relative to the LO approximation. This effect comes in addition to the
known effect of reducing the magnitude of the Wilson coefficients due
to a resummation of QCD logarithms in RG--running between $\msusy$ and
$\mw$ \cite{dgg00}.  In addition, in the framework of GFM, large
contributions to the chirality--conjugate Wilson coefficients
$C_{7,8}^{\prime}$ due to chargino exchange are induced which can
partially cancel the contribution from gluino exchange.

In this study, we present analytic expressions for dominant beyond--LO
corrections to the Wilson coefficients for the process $\bsgamma$ in
the Minimal Supersymmetric Standard Model (MSSM) with GFM in the
squark sector. In addition to including the NLO contributions from the
SM and the 2HDM and the full NLO QCD corrections from gluon exchange, we
compute dominant NLO--level SUSY QCD effects to relevant SUSY contributions,
which are enhanced by large--$\tanb$ factors.  While the full
two--loop calculation still remains to be performed (with the first
step already undertaken in~\cite{bhy03}), our goal is to calculate the
terms which are likely to be dominant beyond LO due to being enhanced at
large $\tan\beta$ and at large $\msusy/\mw$.  We will work in the
regime where all the colored superpartners are heavier than all the
other states.  We generalize the analyses
of~\cite{cdgg98,dgg00,cgnw00} to the case of GFM. In addition, we
include in our analysis an NLO anomalous dimension, an NLO QCD matching
condition and a resummation of $\tanb$--enhanced radiative corrections~\cite{cgnw00,dgg00}, 
which all further improve the accuracy of our analysis.
We use the squark mass eigenstate formalism which is more appropriate when
squark off--diagonal terms are not assumed to be small. We thus do not use
the mass insertion approximation, similarly as in ref.~\cite{bghw99}. 

Our main results can be summarized as follows. In the framework of
GFM, dominant $\tanb$--enhanced beyond--LO effects due to
squark--gluino loop corrections are quite large and have a strong
impact on the Wilson coefficients. Their main effect is that of: {\it
(i)} reducing, relative to LO, the size of the Wilson coefficients
$C_{7,8}$, and {\it (ii)} of generating additional contributions to
the Wilson coefficients $C_{7,8}^{\prime}$ which can be of similar
size.  As a result, the total values of $C_{7,8}^{(\prime)}$ can
significantly change. For $\mu>0$ these are usually have a smaller
value, but not necessarily magnitude, relative to the LO. In fact,
they can even change sign. (For $\mu<0$ the pattern is less clear but
the effect often remains sizeable.)  As a result, the supersymmetric
contribution to $\brbsgamma$ can often change dramatically, even for
small mixings among down--type squarks.

Our analysis demonstrates that various bounds on SUSY parameters
obtained in the context of the MFV framework at the same level of
accuracy can be highly unstable against even even small departures
from it.  An additional effect is that of strongly relaxing upper
limits on the allowed amount of mixing in the 2nd--3rd generation
squark sector relative to the LO case. Another important implication
of our analysis is that the elements of the input CKM matrix can be
larger than the experimentally measured values by a factor as large as
ten (which is significantly more than some $\sim40\%$ obtained
in~\cite{br95} with a specific boundary condition at the GUT scale),
or can be even zero, in which case the measured values of the CKM
matrix would have a purely radiative origin~\cite{banks87}.

The impact of large gluino--squark loops on Wilson coefficients,
considered here for the process of $\bsgamma$, appears to be quite
generic and is likely to play an important role in other rare
processes involving the bottom quark~\cite{for03}.

The paper is organized as follows. In sec.~\ref{sec:gfm} and in
appendices~\ref{sec:muw:effective-vertices}--\ref{sec:Yukawa-correction}
we provide a detailed discussion of the effect of the squark--gluino
loop corrections on quark masses and couplings and on the squark
sector. In sec.~\ref{sec:procedure} we outline our procedure of
computing the Wilson coefficients, and in sec.~\ref{sec:wilson} we
present their analytic expressions. Numerical effects are then presented in
sec.~\ref{ref:results} and our conclusions are summarized in
sec.~\ref{ref:conclusions}.  Several useful formulae are collected in
appendices~\ref{sec:PV-functions}--\ref{sec:nlo-qcd-corrections} for
completeness.

\section{Effective Quark Mass in SUSY with General Flavor Mixing}\label{sec:gfm}

We will first define an effective, supersymmetric, softly--broken
model with general flavor mixing in the squark sector as the framework
for our analysis. In particular, we will concentrate on the effect of the
squark--gluino loop corrections to down--type quark mass matrix on other
properties of the model, in particular on the squark sector, effective
couplings and the CKM matrix. Unless otherwise stated, we will work in
the $\overline{MS}$--scheme.

The quark mass terms in an effective Lagrangian in the super--CKM basis
read
\beq
-{\cal L}_{q}^{\rm mass}= 
\mbox{\boldmath ${\bar d}_{R}^{\dag} m^{(0)}_{d} d_{L}$} + 
\mbox{\boldmath ${\bar d}_{R}^{\dag} \delta m_{d} d_{L}$} + 
\mbox{\boldmath ${\bar u}_{R}^{\dag} m^{(0)}_{u} u_{L}$} + 
\mbox{\boldmath ${\bar u}_{R}^{\dag} \delta m_{u} u_{L}$} + h.c.,
\label{eq:quarklagr}
\eeq
where $\mbox{\boldmath $d_{L}$}=\left( {d_{L}}_i\right)$,
$\mbox{\boldmath $d_{R}$}=\left({d_{R}}_i\right)$, are the down--quark
fields, and $\mbox{\boldmath $u_{L}$}=\left({u_{L}}_i\right)$ and
$\mbox{\boldmath $u_{R}$}=\left({u_{R}}_i\right)$ are up--type quark
fields, with $\mbox{\boldmath $m^{(0)}_{ {d,u} }$}=
{\left(m^{(0)}_{d,u}\right)}_{ij}= \left( {m^{(0)}_{d,u}}_{ij}
\right)$ ($i,j=1,2,3$) denoting 
their respective uncorrected, or ``bare'', $3\times3$ mass matrices. These
are related to the respective input, or ``bare'', Yukawa coupling
matrices $\mbox{\boldmath $Y^{(0)}_{d,u}$}$ in the usual way $\mbox{\boldmath $
m^{(0)}_{d,u}$} = \vdu \mbox{\boldmath $Y^{(0)}_{d,u}$}$, where
$\vdu\equiv \langle H_{d,u}^0\rangle$.

The mass corrections $\mbox{\boldmath $\delta m_{ d,u }$}$ arise from
squark--gluino one--loop contributions.\footnote{For simplicity we
neglect all weak and Yukawa contributions as
sub--dominant~\cite{babukolda99}. They can be found in~\cite{pbmz97}.}
Of particular interest is the mass correction {\boldmath
$\delta m_{d}$} since it is enhanced at large $\tanb$~\cite{hrs}
and/or in the case when left--right soft terms are large,
\beq
\mbox{\boldmath $\delta m_{ d}$} = 
\vd\tanb\, \mbox{\boldmath $\delta Y^{(0)}_{d}$} +
\mbox{\boldmath $\delta m_{ d}^{\rm soft}$}.
\label{eq:deltad}
\eeq
Note that the latter are in general independent of the Yukawa coupling
and can also be sizeable.\footnote{In this paper, we adopt a purely
phenomenological approach at the electroweak scale and do not assume
any thoeretical constraints comming from. \eg, renormalization
group evolution above the electroweak scale and/or vacuum stability
conditions~\cite{ccb}.}  An explicit expression for $\mbox{\boldmath
$\delta m_{ d }$}$ and a self--consistent procedure for computing it
will be given below. Analogous corrections in the up--quark sector are
not enhanced by large $\tanb$--factors, although in general there
could be a substantial effect from left--right soft terms.

The effective (physical) quark mass matrices  $\mbox{\boldmath $m_{d,u}$}$ 
in the super--CKM basis are given by

\begin{eqnarray}
\label{eq:dquarkmassmatrix}
\mbox{\boldmath $m_{d}$} &=& \mbox{\boldmath $m^{(0)}_{d}$} +
\left(\frac{\alpha_s}{4\pi}\right) \mbox{\boldmath $\delta m_{d}$},\\
\nonumber 
&&\\
\label{eq:uquarkmassmatrix}
\mbox{\boldmath $m_{u}$} &=& \mbox{\boldmath $m^{(0)}_{u}$} +
\left(\frac{\alpha_s}{4\pi}\right) \mbox{\boldmath $\delta m_{u}$}.
\end{eqnarray}
Note that we have explicitely extracted the factor $\alpha_/4\pi$ in
the expressions above in order to stress the order of the leading SUSY
QCD (SQCD) one--loop correction. Subdominant effects, \eg, due to
chargino--stop loops considered in~\cite{dgg00}, can be added
linearly.

The effective (physical) quark mass matrices  $\mbox{\boldmath $m_{d,u}$}$ 
in the super--CKM basis are (by definition) diagonal:
 \begin{eqnarray}
\label{eq:dquarkmasseigen}
\mbox{\boldmath $m_{ d }$} &=&
\mbox{\boldmath $diag$} \left( m_{ d_{i} }\delta_{ij} \right)=
\mbox{\boldmath $diag$} \left( m_d, m_s, m_b \right)
,\\
\nonumber 
&&\\
\label{eq:uquarkmasseigen}
\mbox{\boldmath $m_{ u }$} &=& \mbox{\boldmath $diag$} \left( m_{ u_{i}
}\delta_{ij} \right)= \mbox{\boldmath $diag$} \left( m_u, m_c, m_t
\right),
\end{eqnarray}
where $m_{d,s,b}$ and $m_{u,c,t}$ denote the physical masses of the
quarks.  In contrast, in this basis the matrices {\boldmath
$m^{(0)}_{d,u}$} will in general be non--diagonal, and so will
{\boldmath $\delta m_{ d,u }$}.

The CKM matrix is defined as usual\footnote{We will generally follow
the conventions and notation of Gunion and Haber~\cite{gh84}, unless
otherwise stated.}
\beq
\mbox {\boldmath $K=V_{u_{L}} V_{d_{L}}^{\dag}$}.
\label{eq:ckmmatrix}
\eeq
The elements of $K$ are determined from experiment, except for a small
loop correction which can be found in
appendix~\ref{sec:muw:effective-vertices}.  The unitary matrices
{\boldmath $V_{{d,u}_{L,R} }$} rotate the interaction basis' quark
fields ${d^{o}_{L}}_i$, ${d^{o}_{R}}_i$, ${u^{o}_{L}}_i$,
${u^{o}_{R}}_i$ to the mass eigenstates ${d_{L}}_i$, ${d_{R}}_i$,
${u_{L}}_i$, ${u_{R}}_i$ ($i=1,2,3$) in the physical super--CKM
basis. For example, ${d_{L}}_i={ V_{d_L} }_{ij} {d^{o}_{L}}_j$,
${u_{R}}_i={ V_{u_R} }_{ij} {u^{o}_{R}}_j$ ($i,j=1,2,3$). Also
$\mbox{\boldmath $m^{(0)}_{d}$}= \mbox{\boldmath $ V_{d_R}\, m^{(0)
o}_{d}\, V_{d_L}^{\dag} $}$, where $\mbox{\boldmath $m^{(0) o}_{d}$}$
stands for the down--quark ``bare'' mass matrix in the interaction
basis, and analogously for the up--quarks.

For our purpose, it will be convenient to further introduce a ``bare''
super--CKM basis.\footnote{To be precise, one should use different
symbols to denote the ``bare'', or input, quantities, which may be
expressed in different bases, in particular in the interaction basis
and the super--CKM basis, and the various quantities, ``bare'' or
loop--corrected, appearing in the ``bare'' super--CKM basis. For the
sake of simplicity, we will use just one superscript ``$(0)$'' but
will define all the relevant terms as they appear.}  It is the basis
in which the ``bare'' mass matrices {\boldmath $m^{(0)}_{d,u}$} are
diagonal.  In the absence of the mass corrections {\boldmath $\delta
m^{(0)}_{d,u}$}, the ``bare'' and the physical mass matrices would of
course coincide and their eigenvalues would be equal to the physical
quark masses which we will treat as input. In other words, there would
be no need to distinguish between the ``bare'' and the physical
super--CKM basis. (In fact, this level of accuracy is sufficient to
compute supersymmetric contributions to rare decays in the LO
approximation.) Note however, that once the mass corrections are
switched on, it will be the ``bare'' Yukawa couplings, which initially
appear as free parameters in the theory, that will have to be adjusted
to compensate for the loop corrections.

The ``bare'' CKM matrix is defined as {\boldmath
$K^{(0)}=V^{(0)}_{u_{L}} V^{(0)\,\dag}_{d_{L}}$}, where
$\mbox{\boldmath $V^{(0)}_{{d,u}_{L,R}}$}$ denote the unitary matrices
that would transform the ``bare'' down-- and up--quark mass matrices
in the interaction basis into a diagonal form, in analogy with the
case of the corrected super--CKM basis.  As we will see later, in the
presence of the one--loop mass corrections, the elements of {\boldmath
$K^{(0)}_{cb}$} and {\boldmath $K_{cb}$} (the latter being
determined by experiment) can show large discrepancies, which can
reach a factor of ten, or can be reduced down to zero.

Turning next to the squark sector, in the super--CKM basis spanned by
the fields
$\mbox{\boldmath $\wt{d}_{L}$}=\left( { \wt{d}_{L_{i}} }
\right)$, $\mbox{\boldmath $\wt{d}_{R}$}=\left( { \wt{d}_{R_{i}} } \right)$
($i=1,2,3$), and $\mbox{\boldmath
$\wt{u}_{L}$}=\left( { \wt{u}_{L_{i}} } \right)$, $\mbox{\boldmath
$\wt{u}_{R}$}=\left( { \wt{u}_{R_{i}} } \right)$, the mass terms in the
Lagrangian read
\beq
-{\cal L}_{\widetilde{q},\, {\rm soft }}^{\rm mass}=
\left(\mbox{\boldmath$\wt{d}_{L}^{\dag}$},
\mbox{\boldmath$\wt{d}_{R}^{\dag}$}\right) \left( \mbox{\boldmath ${\cal
 M}_{\widetilde{d}}^{2}$} \right) 
\left(
\begin{array}{ll}
\mbox{\boldmath$\wt{d}_{L}$} \\
\mbox{\boldmath$\wt{d}_{R}$}
\end{array}
\right) 
+
\left(\mbox{\boldmath$\wt{u}_{L}^{\dag}$},
\mbox{\boldmath$\wt{u}_{R}^{\dag}$}\right) \left( \mbox{\boldmath ${\cal
 M}_{\widetilde{u}}^{2}$} \right) 
\left(
\begin{array}{ll}
\mbox{\boldmath$\wt{u}_{L}$} \\
\mbox{\boldmath$\wt{u}_{R}$}
\end{array}
\right).
\label{eq:squarklagr}
\eeq
The $6\times6$ down--squark mass matrix {\boldmath ${\cal
 M}_{\widetilde{d}}^{2}$} and the up--squark matrix {\boldmath ${\cal
 M}_{\widetilde{u}}^{2}$} are decomposed into $3\times3$ block
 sub--matrices as follows
\begin{eqnarray}
&&
\mbox{\boldmath ${\cal M}_{\widetilde{d}}^{2}$}
=
\left(
\begin{array}{ll}
\mbox{\boldmath $m^{2}_{d,LL}$} + 
\mbox{\boldmath $F_{d,LL}$} +
\mbox{\boldmath $D_{d,LL}$}~~&  
\mbox{\boldmath $m^{2}_{d,LR}$} + 
\mbox{\boldmath $F_{d,LR}$} 
\label{eq:msddef} \\
\left(
\mbox{\boldmath $m^{2}_{d,LR}$} 
+ 
\mbox{\boldmath $F_{d,LR}$} \right)^\dag &  
\mbox{\boldmath $m^{2}_{d,RR}$} + 
\mbox{\boldmath $F_{d,RR}$} +
\mbox{\boldmath $D_{d,RR}$}  
\end{array}
\right),\\
&& 
\nonumber \\
&&
\nonumber \\
&&
\mbox{\boldmath ${\cal M}_{\widetilde{u}}^{2}$}
=
\left(
\begin{array}{ll}
\mbox{\boldmath $m^{2}_{u,LL}$} + 
\mbox{\boldmath $F_{u,LL}$} +
\mbox{\boldmath $D_{u,LL}$}~~&  
\mbox{\boldmath $m^{2}_{u,LR}$} + 
\mbox{\boldmath $F_{u,LR}$} 
\label{eq:msudef} \\
\left(
\mbox{\boldmath $m^{2}_{u,LR}$} 
+ 
\mbox{\boldmath $F_{u,LR}$} \right)^\dag &  
\mbox{\boldmath $m^{2}_{u,RR}$} + 
\mbox{\boldmath $F_{u,RR}$} +
\mbox{\boldmath $D_{u,RR}$}  
\end{array}
\right).
\label{eq:fudef} %
\end{eqnarray}
The  terms appearing in~(\ref{eq:msddef})--(\ref{eq:msudef}) are
related to their more familiar 
counterparts in the interaction basis by the same unitary
transformations {\boldmath $V_{{d,u}_{L,R} }$} that appear in the
quark sector:  squark fields transform in the same way as the respective
quark fields from the interaction basis 
$\mbox{\boldmath $\wt{d}_{L}^{o}$}=\left(\wt{d}_{L_i}^{o}\right)$, 
$\mbox{\boldmath $\wt{d}_{R}^{o}$}=\left(\wt{d}_{R_i}^{o}\right)$ ($i=1,2,3$), and
$\mbox{\boldmath $\wt{u}_{L}^{o}$}=\left(\wt{u}_{L_i}^{o}\right)$,
$\mbox{\boldmath $\wt{u}_{R}^{o}$}=\left(\wt{u}_{R_i}^{o}\right)$ to
the super--CKM basis which is spanned by $\mbox{\boldmath $\wt{d}_{L,R}$}$
and $\mbox{\boldmath $\wt{u}_{L,R}$}$.
For example, $\mbox{\boldmath $\wt{d}_{L}$} =
\mbox{\boldmath $V_{d_{L}} \wt{d}_{L}^{o}$}$ and $\mbox{\boldmath
$\wt{u}_{L}$} = \mbox{\boldmath $V_{u_{L}} \wt{u}_{L}^{o}$}$, \etc\
If one writes the Lagrangian for the soft SUSY breaking terms in the
interaction basis as
\bea
-{\cal L}_{\widetilde{q},\, {\rm soft }}^{o,\,\rm mass}= &&
\mbox{\boldmath $\wt{d}_{L}^{o\, \dag} m^2_{Q} \wt{d}_{L}^{o}$} + 
\mbox{\boldmath $\wt{d}_{R}^{o\, \dag} m^2_{D} \wt{d}_{R}^{o}$} + 
\left[ \mbox{\boldmath $\wt{d}_{L}^{o\,\dag} $} 
\left(\vd \mbox{\boldmath $A_d^{o\,\ast} $} \right) 
\mbox{\boldmath $\wt{d}_{R}^{o}$}+ h.c. \right]+
\nonumber
\\
&&
\mbox{\boldmath $\wt{u}_{L}^{o\, \dag} m^2_{Q} \wt{u}_{L}^{o}$} + 
\mbox{\boldmath $\wt{u}_{R}^{o\, \dag} m^2_{U} \wt{u}_{R}^{o}$} + 
\left[ \mbox{\boldmath $\wt{u}_{L}^{o\,\dag} $} 
\left(\vu \mbox{\boldmath $A_u^{o\,\ast} $}\right)
\mbox{\boldmath $\wt{u}_{R}^{o}$} + h.c.\right],
\label{eq:squarklagrint}
\eea
then the soft mass terms appearing in the mass matrix {\boldmath ${\cal
 M}_{\widetilde{d}}^{2}$}
are given as follows
\begin{eqnarray}
\label{eq:mdll}
&&\mbox{\boldmath $m^{2}_{d,LL}$}=\mbox {\boldmath $V_{{d}_{L} }
m^{2}_{Q} V_{{d}_{L}}^{\dag}$},\\
\label{eq:mdrr}
&&\mbox{\boldmath $m^{2}_{d,RR}$} = \mbox{\boldmath $V_{{d}_{R} }
m^{2}_{D} V_{{d}_{R}}^{\dag}$},\\
\label{eq:mdlr}
&&\mbox{\boldmath $m^{2}_{d,LR}$}= \mbox{\boldmath
$V_{{d}_{L} }$} \left( \vd \mbox{\boldmath $A_d^{ \ast}$} \right)
\mbox{\boldmath $V_{{d}_{R}}^{\dag}$},
\end{eqnarray}
and similarly for the up sector. (In particular, $\mbox{\boldmath
$m^{2}_{u,RR}$} = \mbox{\boldmath $V_{{u}_{R} } m^{2}_{U}
V_{{u}_{R}}^{\dag}$}$). The matrices $\mbox{\boldmath
$m^{2}_{d\,u,LL}$}$ and $\mbox{\boldmath $m^{2}_{d\,u,RR}$}$ are all
hermitian and are in general non--diagonal, as are their counterpart
soft mass parameters $\mbox {\boldmath $m^{2}_{Q}$}$, $\mbox
{\boldmath $m^{2}_{D}$}$ and $\mbox {\boldmath $m^{2}_{U}$}$
in~(\ref{eq:squarklagrint}).  Note that {\boldmath $m^{2}_{d,LL}$} and
{\boldmath $m^{2}_{u,LL}$} are related by $SU(2)_L$ invariance. Since
in the interaction basis they are both equal to $\mbox{\boldmath
$m^2_{Q}$}$, in the super--CKM basis one finds $\mbox{\boldmath
$m^{2}_{d,LL}$} = \mbox{\boldmath $K^{\dag} m^{2}_{u,LL} K$}$.

The trilinear $3\times3$ matrices {\boldmath $A_{d,u}$}
that appear in the $LR$ soft terms are in general non--hermitian. Note that
we do not assume {\boldmath $A_{d,u}$ } 
to be necessarily proportional to their respective Yukawa couplings.

The $F$--terms are given by 
\begin{eqnarray}
\label{eq:fdll}
&&\mbox{\boldmath$F_{d,LL}$}=\mbox{\boldmath ${\mdbare}^{\dag} \mdbare$},\\
\label{eq:fdrr}
&&\mbox{\boldmath$F_{d,RR}$}=\mbox{\boldmath $\mdbare {\mdbare}^{\dag}$},\\
\label{eq:fdlr}
&&\mbox{\boldmath $F_{d,LR}$}=-\mu\tanb\, \mbox{\boldmath
${\mdbare}^{\dag}$},
\end{eqnarray}
and similarly for the up squarks (except $\mbox{\boldmath
$F_{u,LR}$}=-\mu\cot\beta\, \mbox{\boldmath ${\mubare}^{\dag}$}$).
Note it is the ``bare'' mass matrices $\mbox{\boldmath
$m^{(0)}_{d,u}$}$ that appear in the $F$--terms. This is because
these are derived from the superpotential which contains ``bare''
Yukawa couplings and other ``bare'' parameters. For this reason, in
the super--CKM basis the $F$--terms are {\em not} diagonal, which will
lead to important effects. On the other hand, the $D$--terms remain
diagonal in flavor space. They read
\begin{eqnarray}
\label{eq:ddll}
&&\mbox{\boldmath$D_{d,LL}$}=m_Z \cos2\beta\left[T_{3
    d}-Q_d\sin^2\theta_W \right] \mbox{\boldmath $\unitmatrix$},\\
\label{eq:ddrr}
&&\mbox{\boldmath$D_{d,RR}$}=\left(m_Z \cos2\beta\, Q_d\sin^2\theta_W
\right) \mbox{\boldmath $\unitmatrix$},
\end{eqnarray}
where $T_{3 d}=-1/2$ and $Q_d=-1/3$ (and analogously for the up--sector,
with $T_{3 u}=1/2$ and $Q_d=2/3$).

The mass matrices $\mbox{\boldmath ${\cal M}^2_{{\widetilde d}} $}$
($\mbox{\boldmath ${\cal M}^2_{{\widetilde u}} $}$) in the super--CKM
basis can be diagonalized by unitary matrices which are represented by
two $6\times3$ sub--matrices $\mbox{\boldmath $\Gamma_{d\,L}$}$ and
$\mbox{\boldmath $\Gamma_{d\,R}$}$ ($\mbox{\boldmath $\Gamma_{u\,L}$}$
and $\mbox{\boldmath $\Gamma_{u\,R}$}$),
\begin{equation}
\left(
\mbox{\boldmath $\Gamma_{d\,L}$}, \mbox{\boldmath $\Gamma_{d\,R}$} 
\right)
\mbox{\boldmath ${\cal M}_{\widetilde{d}}^{2}$}
\left( 
\begin{array}{l} 
\mbox{\boldmath $\Gamma_{d\,L}^{\dag}$} \\
\mbox{\boldmath $\Gamma_{d\,R}^{\dag}$}
\end{array}
\right) 
= \mbox{\boldmath $diag$}
\left( m_{\widetilde{d}_{1}}^{2},\ldots, m_{\widetilde{d}_{6}}^{2}\right)
\label{eq:gammaddef}
\end{equation}
and
\begin{equation}
\left(
\mbox{\boldmath $\Gamma_{u\,L}$}, \mbox{\boldmath $\Gamma_{u\,R}$} 
\right)
\mbox{\boldmath ${\cal M}_{\widetilde{u}}^{2}$}
\left( 
\begin{array}{l} 
\mbox{\boldmath $\Gamma_{u\,L}^{\dag}$} \\
\mbox{\boldmath $\Gamma_{u\,R}^{\dag}$}
\end{array}
\right) 
= \mbox{\boldmath $diag$}
\left( m_{\widetilde{u}_{1}}^{2},\ldots, m_{\widetilde{u}_{6}}^{2}\right),
\label{eq:gammaudef}
\end{equation}
where $m_{\widetilde{d}_{I}}$ and $m_{\widetilde{u}_{I}}$
($I=1,\ldots,6$) are the physical masses of the down--,
$\mbox{\boldmath $\widetilde{d}$} = \left(\widetilde{d}_{I}\right)$, and up--type,
$\mbox{\boldmath $\widetilde{u}$} = \left(\widetilde{u}_{I}\right)$,  squark mass eigenstates,
respectively. These are related to the corresponding fields 
$\mbox{\boldmath $\widetilde{d}_{L,R},\widetilde{u}_{L,R}$}$  in the
super--CKM basis by
\begin{equation}
\mbox{\boldmath $\widetilde{d}$} = 
\mbox{\boldmath $\Gamma_{d\,L,R}$}\: \mbox{\boldmath $\widetilde{d}_{L,R}$}
,~~~~~~~~~~~~~~~~
\mbox{\boldmath $\widetilde{u}$} = 
\mbox{\boldmath $\Gamma_{u\,L}$}\: \mbox{\boldmath $\widetilde{u}_{L,R}$}.
\label{eq:squarkfieldtfm}
\end{equation}

We will also need squark mass eigenvalues and diagonalizing matrices 
in the limit of vanishing
mass corrections {\boldmath $\delta m_{ d,u }$}. Squark mass matrices
in this limit will be denoted by
\begin{eqnarray}
&&
\mbox{\boldmath ${\cal M}_{\widetilde{d}}^{(0)\,2}$}
= \mbox{\boldmath ${\cal M}_{\widetilde{d}}^{2}$} 
\left( \mbox{\boldmath $\delta m_{d}$} \ra \mbox{\boldmath $0$}\right),\\
\label{eq:msd0def} 
&&
\nonumber \\
&&
\mbox{\boldmath ${\cal M}_{\widetilde{u}}^{(0)\,2}$}
=
\mbox{\boldmath ${\cal M}_{\widetilde{u}}^{2}$} 
\left( \mbox{\boldmath $\delta m_{u}$} \ra \mbox{\boldmath
  $0$}\right).
\label{eq:msu0def}
\end{eqnarray}
Since in this limit the ``bare'' and
the physical super--CKM bases coincide, the $F$--terms are now
diagonal and are given by
$\mbox{\boldmath $F_{d,LL}^{(0)}$}=
\mbox{\boldmath $F_{d,RR}^{(0)}$} = \mbox{\boldmath $diag$}
\left(m_d^2, m_s^2,m_b^2\right) $ and $\mbox{\boldmath
$F_{d,LR}^{(0)}$} = -\mu\tanb\, \mbox{\boldmath $diag$}
\left(m_d,m_s,m_b\right)$ 
(and analogously for the up--squarks). This
is because the ``bare'' mass matrices {\boldmath $m^{(0)}_{d,u}$}, that
appear in the $F$--terms, are in this limit diagonal, with
the physical quark masses along the diagonal.

We will further introduce the masses of the
down-- and up--type squarks, respectively, in the limit of vanishing
mass corrections {\boldmath $\delta m_{ d,u }$}
\begin{equation}
m^{(0)}_{\widetilde{d}_{I}} = 
m_{\widetilde{d}_{I}} 
\left( \mbox{\boldmath $\delta m_{ d }$} \ra \mbox{\boldmath
  $0$}\right),~~~~~~~
m^{(0)}_{\widetilde{u}_{I}} = 
m_{\widetilde{u}_{I}} 
\left( \mbox{\boldmath $\delta m_{ u }$} \ra \mbox{\boldmath
  $0$}\right)~~~~~~~(I=1,\ldots,6),
\label{eq:msu0massdef}
\end{equation}
as well as the diagonalizing matrices 
$\mbox{\boldmath $\Gamma_{d\,L}^{(0)}$} = 
\mbox{\boldmath $\Gamma_{d\,L}$}
\left( \mbox{\boldmath $\delta m_{ d }$} \ra \mbox{\boldmath
  $0$}\right)$ and 
$\mbox{\boldmath $\Gamma_{d\,R}^{(0)}$} = 
\mbox{\boldmath $\Gamma_{d\,R}$}
\left( \mbox{\boldmath $\delta m_{ d }$} \ra \mbox{\boldmath
  $0$}\right)$,
and analogously for the up--squarks.

\begin{figure}[b]
\begin{center}
\psfig{figure=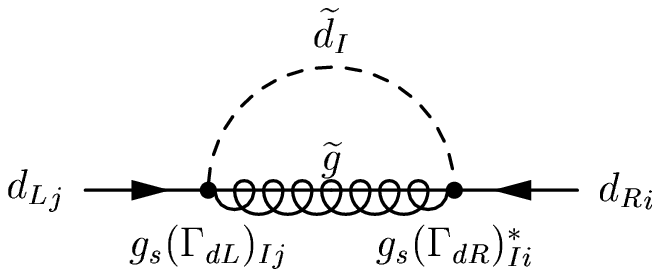, angle=0, width=6.0cm}
\caption{Gluino--squark correction to the down--type quark mass matrix.
Direction of the arrows indicates chirality of the quark field.
\label{fig:mass_correction}}
\end{center}
\end{figure}

We will now apply the above formalism to describe our procedure of
computing the one--loop mass correction {\boldmath $\delta
m_{d}$}~(\ref{eq:deltad}). The relevant diagram is given by
fig.~\ref{fig:mass_correction} for which one obtains~\cite{br95} 
\beq
\left(\delta m_{ d }\right)_{ij}
= 2 C_2(3) 
\sum_{I=1}^6 
\left(\Gamma_{d\,R}^{\ast}\right)_{Ii} \left(\Gamma_{d\,L}\right)_{Ij}
m_{\wt{g}} B_0( {m^2_{\wt{g}}}, m_{\wt{d}_I}^2), 
\label{eq:deltamq}
\eeq
where the quartic Casimir operator for $SU(3)$ is 
$C_2(3) = \frac{4}{3}$, $m_{\wt{g}}$ denotes the mass of the gluino and 
$B_0$ is one of the Passarino--Veltman functions which are
collected in appendix~\ref{sec:PV-functions}. (An analogous expression
for the mass correction $\left(\delta m_{ u }\right)_{ij}$ can be
obtained from~(\ref{eq:deltamq}) by simply replacing $d\ra u$ and
$\wt{d}\ra \wt{u}$.)

In our analysis we take the physical masses $m_{d_{1,2,3}}$ as
input. On the other hand, in computing the diagram of
fig.~\ref{fig:mass_correction}, one uses the elements of the ``bare''
mass matrix {\boldmath $m^{(0)}_{d}$} and of the squark mass matrix
{\boldmath ${\cal M}_{\widetilde{d}}^{2}$}, which in turn are
functions of {\boldmath $m^{(0)}_{d}$} and other ``bare''
parameters. The final result must be equal to the physical quark
masses.  This means that we need to employ an iterative procedure to
simultaneously determine $\mbox{\boldmath $m^{(0)}_{d}$}$,
$\mbox{\boldmath $\delta m^{(0)}_{d}$} (\mbox{\boldmath
$m^{(0)}_{d}$})$ and $\mbox{\boldmath ${\cal M}_{\widetilde{d}}^{2}$}
(\mbox{\boldmath $m^{(0)}_{d}$})$ to a desired level of accuracy.

In the first iteration one neglects {\boldmath $\delta m_{d}$} and
thus identifies {\boldmath $m^{(0)}_{d}$} with {\boldmath
$m_{d}$}. One then computes {\boldmath $\delta m_{d}$}
by applying~(\ref{eq:deltamq}) where the elements of the squark sector are
determined by {\boldmath ${\cal M}_{\widetilde{d}}^{(0)\,2}$}, and
not by {\boldmath ${\cal M}_{\widetilde{d}}^{2}$}. In other
words, in evaluating~(\ref{eq:deltamq}) one initially makes the substitutions
$m_{\wt{d}_I}\rightarrow m^{(0)}_{\wt{d}_I}$ and {\boldmath
$\Gamma_{d\,RL}$} $\rightarrow$ {\boldmath $\Gamma^{(0)}_{d\,RL}$}.

In the second step one uses {\boldmath $\delta m_{d}$} from the first
step to compute $\mbox{\boldmath$m^{(0)}_{d}$}= \mbox{\boldmath
$m_{d}$} - \left(\alpha_s/4\pi\right)\,\mbox{\boldmath $\delta
m_{d}$}$. The new values of the elements of $\mbox{\boldmath
$m^{(0)}_{d}$}$ are then used in~(\ref{eq:msddef})
and~(\ref{eq:gammaddef}) to determine the new eigenvalues
$m_{\widetilde{d}_{I}}$ ($I=1,\ldots,6$) and the new elements of
{\boldmath $\Gamma_{d\,L}$} and {\boldmath $\Gamma_{d\,R}$}. These are
then used to compute {\boldmath $\delta m_{d}$} via~(\ref{eq:deltamq})
and {\boldmath $m^{(0)}_{d}$} in the third step, and so on. The
procedure converges quite rapidly, although not as well at very large
$\tanb$.

The above iterative procedure amounts to resumming $\tanb$ terms to
all orders in perturbation theory. The importance of resummation has
been stressed in~\cite{cgnw00,dgg00} for extending the validity of
perturbative calculation to the case of large $\tanb$. Below we will
provide a simple formula for $\tanb$--resummation effect on the bottom
mass in the limit of MFV.

Squark--gluino loop corrections also modify various couplings
involving gauge and Higgs bosons, as well as those of the chargino,
neutralino and gluino fields. The effective couplings are summarized
in the appendices~\ref{sec:muw:effective-vertices}
and~\ref{sec:Yukawa-correction}. They are computed in the same
self--consistent way as the quark and squark masses and mixings.

We will now illustrate the above procedure of computing quark mass
corrections in the limit of vanishing
inter--generational mixings among quarks. Concentrating on the
correction to the mass of the b--quark one obtains
\beq
\delta m_{b}= \left(\delta m_{d}\right)_{33} = C_2(3)
\sin 2\theta_{\sbottom}\mgluino \left[
B_0\left(\mgluino^2,m^2_{{\widetilde b}_1}\right) - 
B_0\left(\mgluino^2,m^2_{{\widetilde b}_2}\right) \right],
\eeq
where now the two sbottom mass states are ${\widetilde
b}_1={\widetilde d}_3$ and ${\widetilde b}_2={\widetilde d}_6$, and
they result from the mixing of only the interaction states ${\widetilde
b}_L$ and ${\widetilde b}_R$, with $\theta_{\sbottom}$ denoting the
mixing angle,
\beq
\left( \begin{array}{cc} {\bf \Gamma_{d\,L}}, {\bf \Gamma_{d\,R}} \end{array}
\right) =
\left( \begin{array}{cc} 
 \cos\theta_{\sbottom} & \sin\theta_{\sbottom} \\
-\sin\theta_{\sbottom} & \cos\theta_{\sbottom} \end{array}
\right),
~~~
\sin 2\theta_{\sbottom} = \frac{2 \left({\cal M}^2_{{\widetilde d}}\right)_{36}
}
{ m^2_{{\widetilde b}_{1}} - m^2_{{\widetilde b}_{2}} },
\eeq
where $\left({\cal M}^2_{{\widetilde d}}\right)_{36} = \left(
m_{d,LR}^2 \right)_{33} + \left( F_{d,LR} \right)_{33} $.
After some simple steps one obtains
\beq
\delta m_{b}= 2 C_2(3) 
\frac{ \left({\cal M}^2_{{\widetilde d}}\right)_{36}} {\mgluino} 
H_2\left( 
\frac{m^2_{{\widetilde b}_{1}} }{\mgluino^2},
\frac{m^2_{{\widetilde b}_{2}} }{\mgluino^2} 
\right), 
\eeq
where the function $H_2$ is given by
\beq
H_2(x,y) = \frac{x\ln x}{(1-x)(x-y)}
          +\frac{y\ln y}{(1-y)(y-x)}.
\eeq
In the regime of moderate to large $\tanb$, and after neglecting $\left(
m_{d,LR}^2 \right)_{33}$, one recovers a familiar
expression~\cite{dgg00,cgnw00}

\beq
m_b=\left(\sqrt{2} m_W\frac{Y_b^{(0)}}{g}\right)\cos\beta\,
\left(1+\epsilon_b\tan\beta\right)=m_b^{(0)}\left(1+\epsilon_b\tan\beta\right), 
\eeq
where $m_b^{(0)}$ is the ``bare'' bottom quark mass and $\epsilon_b$ is
given by
\beq
\label{eq:deltambdgg00}
\epsilon_b = -\frac{2\alpha_s}{3\pi}\frac{\mu}{\mg} 
H_2\left( 
\frac{m^2_{{\widetilde b}_{1}} }{\mgluino^2},
\frac{m^2_{{\widetilde b}_{2}} }{\mgluino^2} 
\right).
\eeq

The resummation of large radiative corrections at large
$\tanb$~\cite{cgnw00,dgg00} to all orders in perturbation theory is in
our case achieved by employing the iterative procedure described
above. In the case of the b--quark mass considered here, after the first iteration one
obtains $m^{(0)}_b(1^{\rm st})=m_b\left( 1 -
\epsilon_b^{(1)}\tanb\right)$, where $\epsilon_b^{(1)}$ denotes the
correction~(\ref{eq:deltambdgg00}) obtained in the {\em first} step of
iteration, where one makes the replacement $m_{{\widetilde b}_{1,2}}
\rightarrow m^{(0)}_{{\widetilde b}_{1,2}}$.  If one denotes by
$\epsilon_b^{(k)}$ the correction~(\ref{eq:deltambdgg00}) obtained in
the $k$--th step of iteration then, after n steps one obtains
\beq
m^{(0)}_b(n^{\rm th})=m_b \left[
1 - \epsilon_b^{(n)}\tanb + 
\epsilon_b^{(n)} \epsilon_b^{(n-1)}\tan^2\beta 
+ \ldots 
+\left(-1\right)^{n}\epsilon_b^{(n)}\epsilon_b^{(n-1)}\ldots\epsilon_b^{(1)}
\tan^n\beta 
\right].
\label{eq:mbresum}
\eeq

We will now discuss the case of MFV as a limit of the GFM
scenario. While in the literature the notion of MFV is not unique, and
often depends on a model, the scenario is usually defined as the one
where, at some scale, the $LL$ and $RR$ soft mass matrices are
proportional to the unit matrix and the $LR$ ones are proportional to
the respective Yukawa matrices. A convenient basis here is the
super--CKM basis~\cite{dgh84} where the Yukawa matrices are
diagonal. In this case MFV corresponds to all off--diagonal entries in
the soft mass matrices being zero. Thus it is the off--diagonal
elements of $\mbox{\boldmath $m^{2}_{d\,u,LL}$}$, $\mbox{\boldmath
$m^{2}_{d\,u,RR}$}$ and $\mbox{\boldmath $m^{2}_{d\,u,LR}$}$ which in
this basis are responsible for flavor change~\cite{hkr85}.  This is
because, in the super--CKM basis, the couplings of the neutral
gauginos are flavor--diagonal while the mixing in the charged
gaugino--quark--squark couplings are all set by the SM CKM matrix
$\mbox{\boldmath $K$}$.

As mentioned above, in the absence of quark mass loop corrections,
there is no need to distinguish between the ``bare'' and the physical
super--CKM basis. However, once these are taken into account, an
ambiguity arises. In one approach, one can insist that the $LR$ soft
mass matrices remain proportional to the ``bare'' Yukawa couplings,
which now become modified with respect to their tree--level
values. This view can be motivated by an underlying assumption that,
at some high scale, the two sets of quantities are ultimately related.
Alternatively, in a more phenomenological approach like this one, one
can assume that the $LR$ soft mass matrices do not change and thus
remain proportional to the physical Yukawa couplings (which are the
same as the ``bare'' {\em tree--level} Yukawa couplings) which in the
super--CKM basis give the physical masses of the
quarks. Here we choose the latter option. Fortunately, the difference
between the two choices is numerically small.

We thus define the MFV scenario as the one in which, in the physical
super--CKM basis, one has

\beq
\label{eq:mfvdef}
\mbox{\boldmath $m^{2}_{d,LL}$} \propto \mbox{\boldmath $\unitmatrix$},\ \ \ \
\mbox{\boldmath $m^{2}_{d,RR}$} \propto \mbox{\boldmath $\unitmatrix$},\ \ \ \
\mbox{\boldmath $m^{2}_{d,LR}$} \propto \mbox{\boldmath $m_d$}/\vd,
\eeq
and analogously for the up--sector. Note that in the last relation of
eq.~(\ref{eq:mfvdef}) one also assumes that
$\mbox{\boldmath$A_{d,u}^{T}$}\propto\mbox{\boldmath$Y_{d,u}$}$
which in the case of GFM does not have to be the case.

In order to parametrize departures from the MFV scenario, it is
convenient to introduce the following dimensionless ($3\times3$
matrix) parameters $\mbox{\boldmath $\delta^d_{LL}$}=\left(
\delta^d_{LL}\right)_{ij}$, $\mbox{\boldmath $\delta^d_{LR}$}$, \etc,
where $i,j=1,2,3$,
\beq
{\left( \delta^d_{LL} \right)}_{ij}=~~~~~
\frac{ {\left( m^{2}_{d,LL} \right)}_{ij} }
{ \sqrt{ {\left( m^{2}_{d,LL} \right)}_{ii} 
         {\left( m^{2}_{d,LL}\right)}_{jj} }},~~~
{\left( \delta^d_{LR} \right)}_{ij} =
\frac{ {\left( m^{2}_{d,LR} \right)}_{ij} }
{ \sqrt{ {\left( m^{2}_{d,LL} \right)}_{ii} 
         {\left( m^{2}_{d,RR}\right)}_{jj} }},
\label{deltalldef:eq}
\eeq



as well as

\beq
{\left( \delta^d_{RR}\right)}_{ij}=~~~~~
\frac{ {\left( m^{2}_{d,RR} \right)}_{ij} }
{ \sqrt{ {\left( m^{2}_{d,RR} \right)}_{ii} 
         {\left( m^{2}_{d,RR} \right)}_{jj} }},~~~
{\left( \delta^d_{RL} \right)}_{ij} =
\frac{ {\left( m^{2}_{d,RL} \right)}_{ij} }
{ \sqrt{ {\left( m^{2}_{d,RR} \right)}_{ii} 
         {\left( m^{2}_{d,LL} \right)}_{jj} }},
\label{deltarrdef:eq}
\eeq
and analogously for the up--sector. The $\delta$'s can be evaluated at
any scale equal to or above the typical scale of the soft terms. We
will compute them at the scale $\msusy$ which we will assume to be the
characteristic mass scale for the soft mass terms.  Note that the
definitions~(\ref{deltalldef:eq})--(\ref{deltarrdef:eq}) remain the
same with and without loop quark mass corrections and are therefore
particularly convenient for comparing the effects of LO and dominant
beyond--LO corrections to $\brbsgamma$ considered
here.\footnote{Alternatively, we can define the dimensionless
parameters in ``bare'' super--CKM basis in view of the correlation
between the soft masses and the ``bare'' Yukawa coupling, as stated
above. The iterative procedure we have introduced works with this
definition equally well.}

\section{Outline of the Procedure}\label{sec:procedure}

Before plunging into technical details, we will first provide a general
outline of our procedure for treating beyond--LO corrections to
$\brbsgamma$ in the framework of GFM. One  starts with an
effective Lagrangian~\cite{cmm97} evaluated at some scale $\muzero$,
\beq {\cal L}
= \frac{4 G_F}{\sqrt{2}} K_{ts}^{\ast} K_{tb} \sum \left[ \Ci({\muzero}){\it
P}_{i} + \Cip({\muzero}){\it P}_{i}^{\prime} \right].
\label{smoperators:ref}
\eeq 
The Wilson coefficients $C_i(\muzero)$ and $C_{i}^{\prime}(\muzero)$
associated with the operators ${\it P}_{i}$ and their
chirality--conjugate partners ${\it P}_{i}^{\prime}$ play the
role of effective coupling constants.  The $\Cip$'s are obtained from
$\Ci$'s by a chirality exchange $L\leftrightarrow R$.

At $\muzero=\muw$, the Wilson coefficient of the four--fermion
operator $C_2$ has a tree level contribution. Dominant loop
contributions determine the coefficients $C_{7,8}$ of the magnetic and
chromomagnetic operators, respectively, which are also most sensitive
to ``new physics''.  These coefficients mix with each other when they
are next evolved by renormalization group equations (RGEs) from $\muw$ down
to $\mub$. Then $\brbsgamma$ is evaluated at $\mub$ as described in~\cite{gm01},
with updated numerical values for the ``magic numbers'' taken from~\cite{bcmu02}.

\subsection{SM and 2HDM Contributions}\label{sec:sm}

In the SM+2HDM, in addition to the $W$--loop, there is a charged
Higgs boson $H^{-}$ contribution to $C_{7,8}$ which always adds
constructively to the one from the SM. Both contribute to ${\it
C}_i^{\prime}$ as well, but this is suppressed by $m_s/m_b$ and can be
ignored. In contrast, this is not necessarily the case in SUSY
extensions of the SM.

In the SM+2HDM case the sum in eq.~(\ref{smoperators:ref}) extends to
eight, while in SUSY in general several additional operators
arise~\cite{bghw99}. In our analysis we will henceforth restrict ourselves to
the ``SM basis'' of operators but will consider below
to  what extent it is justified to neglect the additional operators.

Corrections from hard gluons to the vertices
involving the $W$--boson, $H^-$ and $G^-$ (where $G^-$
represents the unphysical scalar appearing in $R_{\xi}$ gauge)  are
evaluated at the scale $\muw=\mw$. 
In computing them we follow the
NLO QCD calculation of the matching condition at $\muw$~\cite{2hdnlo,cdgg97}.

\subsection{SUSY Contributions}\label{sec:susypart}

In SUSY, additional loop contributions arise from diagrams involving
the chargino, neutralino and gluino exchange, along with appropriate
squarks. Furthermore, as has been shown in~\cite{br99,cgnw00,dgg00},
NLO SUSY QCD corrections to one--loop diagrams involving the gluino
field can be as important as those involving hard gluons.

The diagrams involving a loop exchange of the wino and higgsino
components of the charginos (the superpartners of the $W^{-}$ and the
$H^{-}$) with the up--type squarks add to the SM part constructively
or destructively, depending on the sign of $\mu$.  The
chargino/up--type squark exchange is the only SUSY contribution if one
imposes MFV assumptions. In the case of GFM, the diagrams involving
the exchange of the gluino (and neutralino) and down--type squarks
with flavor violating squark mass mixings are also allowed and, as
discussed in sec.~\ref{sec:gfm}, in general play a substantial role.

In  the softly--broken low energy SUSY model, with either MFV or GFM, an
additional complication arises that is related to the fact that
several new mass scales appear which are associated with the masses of
the contributing SUSY partners. In this analysis we will assume that
all color--carrying superpartners (gluinos and squarks) are
considerably heavier (by at least a factor of a few) than other states
\beq 
\mususy\sim{\cal
O}(\mgluino,\msquark,\mstopone,\mstoptwo)\gg \muw\sim{\cal
O}(\mw,\mtop, m_{H^{\pm}})\gg \mub\sim{\cal
O}(\mbottom).\label{eq:hierarchy} 
\eeq 
Such a hierarchy is often realized in low--energy effective models
with a gravity--mediated SUSY breaking mechanism because of different
RG evolution of soft mass parameters from the unification scale
down. It thus appears justified to associate the heaviest
superpartners with the SUSY soft--breaking scale $\msusy$. In
particular, we assume both superpartners of the top quark to be heavy.
Below the scale $\mususy=\msusy$ we will then deal with an effective
model with the gluino and squark fields decoupled. Their effect will
nevertheless be present in modifying the couplings of the states below
$\mususy$.

As stated above, for the states at $\muw$ ($W$, $G^-$ and $H^-$), the
contributions to the Wilson coefficients are evaluated at $\muw$. The
beyond--LO SUSY QCD corrections to these vertices, which are of order
${\cal O}(\alpha_s)$, are absorbed into the effective vertices
involving Yukawa couplings by integrating out the gluino field using
the effective coupling method which has been introduced in~\cite{cdgg98}. In our case, the
effective couplings are computed at $\mususy$ and then evolved down to
$\muw$ using RGEs for the SM QCD $\beta$--functions with six quark
flavors. For simplicity, in the effective couplings we neglect the
running of the Yukawa and gauge couplings other than $\alpha_s$. We
have extended to the case of GFM in the squark sector the calculation
of the effective vertices and of the complete NLO matching condition
at $\muw$ which in~\cite{cdgg98} has been done in the framework of MFV.

The contributions to the Wilson coefficients from the
chargino, gluino and neutralino vertices, including NLO QCD and SUSY
QCD corrections, are first evaluated at $\mususy$ and then evolved down to
$\muw$ with the NLO anomalous dimension corresponding to six quark
flavors~\cite{misiakmunz95}, and with NLO QCD matching
condition~\cite{bmu00} at $\mususy$ within the SM operator basis.  The
leading SUSY QCD corrections beyond LO to the vertices of the above superpartner
have been computed in refs.~\cite{cdgg98,dgg00,cgnw00} in the
case of MFV. Here we extend the calculation to the case of GFM among
squarks.  

In light of the assumed mass hierarchy~(\ref{eq:hierarchy}), the
chargino, gluino and neutralino vertices all involve some massive
($\sim\msusy$) fields. In this case expansion by external momentum is
not justified and the effective coupling method, which has been
applied above to the states at $\muw$, cannot be used
anymore~\cite{cdgg98}.  In order to treat them properly, one would
have to calculate full 2--loop diagrams involving heavy fields, which
is beyond the scope of this study. This has not yet even been done in
the less complicated case of MFV. Instead, in~\cite{dgg00,cgnw00}
finite threshold corrections to the bottom quark Yukawa coupling,
which are enhanced at large $\tanb$, have been considered in the case
of MFV. Here we extend this method to the GFM scenario and compute
$\tanb$--enhanced (and also A--term enhanced) finite threshold
corrections to the matrix of the Yukawa couplings of the down--type
quarks.  (Analogous corrections to the up--type quarks are not
enhanced at large $\tanb$, although can possibly be by the soft terms.)
These beyond--LO corrections are likely to be dominant due to being
enhanced by $\tanb$ factors relative to other corrections.  As
mentioned above, we introduce NLO resummation of QCD logarithms to
$C_{7,8}$ and $C_{7,8}^{\prime}$ assuming the
hierarchy~(\ref{eq:hierarchy}).  In computing the chargino, gluino and
neutralino loop contributions we follow~\cite{bmu00} in implementing
NLO QCD matching conditions.

The full set of operators in the sum in~(\ref{smoperators:ref}) that
arises in SUSY has been systematically analyzed in ref.~\cite{bghw99}.  Box
diagrams involving the gluino field generate scalar and tensor type
four--quark operators, which are absent in the SM, at the matching
scale $\mususy$.  As a result of RG evolution from $\mususy$ to
$\mub$, in contrast to the vector type operators in the SM, at 1--loop
level these new operators mix with the magnetic and chromo--magnetic
operators and the contribution due to the mixing from these new
operators could be comparable to the initial loop contribution to the
(chromo--) magnetic operators.

This mixing effect appears, however, to be subdominant at least at
LO.\footnote{We thank T.~Hurth for providing us with this argument.}
In the presence of the new operators, the (chromo--)magnetic operators
are classified as dimension 5 or 6 depending on whether the chirality
flip of the operator originates from the mass of the gluino in the
loop or from that of the external quark~\cite{bghw99}.\footnote{In our
numerical calculations, we do not distinguish these two contributions
and use the same operators and corresponding anomalous dimensions as
for the SM contribution.  This is justified when the new operators are
neglected~\cite{bghw99}.} The new operators mix only with the
dimension 6 operators and their contribution to the $\bsgamma$
amplitude is suppressed by $m_b/\mgluino$ relative to that of the
dimension 5 ones.  Because the new operators always appear with the
dimension 5 operators, which are generated by the same set of flavor
mixing among squarks, their effect is subdominant.  This suppression
has been numerically confirmed in~\cite{bghw99}.

In the absence of the NLO anomalous dimensions, matrix elements,
bremsstrahlung corrections and matching conditions for the complete
set of operators, we are unable to evaluate their contribution at NLO at
the same level of accuracy as for the SM set.  However, the above
suppression mechanism is based on the chirality structure and we can
see no obvious reason for any sizeable enhancement from the new
operators at NLO which would overcome an additional $\alpha_s$
suppression with respect to the already subdominant LO contribution.
Based on the above argument, in this work we restrict ourselves to the
SM set of operators.\footnote{At large $\tanb$, some of the box
diagrams are enhanced by $\tan^2\beta$, instead of $\tanb$, as in the
penguin diagrams. However, they always come with a suppression factor
of $(m_b/\mgluino)^2$ and we assume that these contributions remain to be
subdominant even at large $\tanb$ if the gluino is heavy. We also
neglect a neutral Higgs mediated contribution to the
new operators, which could be important at large
$\tan\beta$~\cite{agis02}.}


\vspace{0.15cm}
\noindent
\section{Wilson coefficients}\label{sec:wilson}

In this section, we present expressions for the supersymmetric
contributions to $C_{7,}^{(\prime)}(\mu_W)$ and $C_8(\mu_W)$ obtained
using the procedure outlined above. 

In the MSSM, the Wilson coefficients for magnetic and chromo--magnetic
operators $C_7$ and $C_8$, in addition to the $W^-$--$t$ 
and $H^-$--$t$ loops, receive contributions from the
$\chi^{-}_{1,2}$--$\wt{u}_{1,\ldots,6}$,
$\chi^0_{1,\ldots,4}$--$\wt{d}_{1,\ldots,6}$  and
$\wt{g}$--$\wt{d}_{1,\ldots,6}$ loops, 
\beq
C_{7,8}(\muw) =  \deltaw C_{7,8}(\muw) + \deltah
C_{7,8}(\muw) +\deltas C_{7,8}(\muw),
\eeq
where $S=\chi^-,\chi^0,\wt{g}$,

\beq
\deltas C_{7,8}(\muw) = \deltacharn C_{7,8}(\muw) +
\deltaneut C_{7,8}(\muw) + \deltagluino C_{7,8}(\muw).
\eeq
Analogous expressions apply to $C^{\prime}_{7,8}$ except for the cases
explicitly mentioned below.

The coefficients $\deltax C_{7,8}$, where $X=W,H$, 
including order 
$\alpha_s$ NLO QCD and SUSY QCD corrections are evaluated at
$\muw$
\begin{equation}
\deltax C_{7,8}(\muw) = \deltax C^{(0)}_{7,8}(\muw)
 +\frac{\alpha_s(\muw)}{4\pi} \deltax C^{(1)}_{7,8}(\muw),
\end{equation}
where $\deltax C^{(0)}_{7,8}$ and $\deltax C^{(1)}_{7,8}$ denote the
LO and NLO contributions, respectively.\footnote{Our notation for the
Wilson coefficients follows that of ref.~\cite{cdgg98} but is extended
to the case of GFM. Note that in our case we do not assume them to be
necessarily proportional to CKM matrix elements.  }

The $H^-$ and $W^-$ contributions to the Wilson coefficients 
at $\muw$ are matched with an effective theory with five quark flavors.
Their LO contributions are given by~\cite{inamilim80,cdgg98},
\begin{eqnarray}
\deltaw C^{(0)}_{7,8}(\muw) &=&
F^{(1)}_{7,8}\left(\frac{\ovl{m}_t^2(\muw)}{m_W^2}\right), \\ 
\deltah C^{(0)}_{7,8}(\muw) &=& 
\frac{1}{3\tan^2\beta}  F^{(1)}_{7,8}\left(\frac{\ovl{m}_t^2(\muw)}{m_H^2}\right)
                       +F^{(2)}_{7,8}\left(\frac{\ovl{m}_t^2(\muw)}{m_H^2}\right),
\end{eqnarray}
where $\ovl{m}_t(\muw)$ is the $\ovl{MS}$ running mass of top quark at
$\muw$ and $m_H$ denotes the mass of the charged Higgs. The mass
functions $F^{(1,2)}_{7,8}$ are given in
appendix~\ref{sec:mass-functions}.

The NLO matching condition of the Wilson coefficients $\delta^{W,H}
 C^{(1)}_{7,8}$ at $\muw$ reads \beq
\label{eq:nlomcmw}
\delta^{W,H} C^{(1)}_{7,8}(\muw) =
                                 \delta^{W,H}_{g} C_{7,8}^{(1)}(\muw)
                                +
                                 \delta^{W,H}_{\wt{g}} C_{7,8}^{(1)}(\muw), 
\eeq
where $\delta^{W,H}_{g} C_{7,8}^{(1)}$ represents an NLO QCD
correction from two--loop diagrams involving one gluon line and
analogously $\delta^{W,H}_{\wt{g}} C_{7,8}^{(1)}$ stand for
corrections from two--loop diagrams with a gluino line, which are
integrated out at $\mususy$.  Explicit expressions for
$\delta^{W,H}_{g} C^{(1)}_{7,8}$ are given in refs.~\cite{cdgg98,
bmu00} and collected in appendix~\ref{sec:nlo-qcd-corrections}.

As outlined in sec.~\ref{sec:procedure}, in computing
$\delta^{W,H}_{\wt{g}} C_{7,8}^{(1)}$ we use the effective coupling
method~\cite{cdgg98} and replace in the relevant LO diagrams the tree--level
couplings for the $W^-$ ($G^-$) and $H^-$ vertices by corresponding
effective couplings.  These are given in
appendix~\ref{sec:muw:effective-vertices} and are first computed in
the regime of eq.~(\ref{eq:hierarchy}) at $\mususy$ and next evolved
down to $\muw$ using RGEs with SM QCD $\beta$--functions. The effect
of the running of the Yukawa and gauge couplings other than $\alpha_s$ is
neglected as subdominant. One finally obtains
\begin{eqnarray}
\deltaw_{\wt{g}} C_{7,8}^{(1)}(\muw) &=&
                  \frac{2}{3 g_2^2  K^{\ast}_{ts} K_{tb}}
                  \frac{{\mw}^2}{\ovl{m}_t^2(\muw)}\times \nonumber\\
              &&  \left[ \left(C^{Gu_3 d_2 (0)\ast}_{L} C^{Gu_3 d_3 (1)}_{L}
                    +C^{Gu_3 d_2 (1)\ast}_{L}C^{Gu_3 d_3 (0)}_{L}\right)
                   F_{7,8}^{(1)}
                             \left(\frac{\ovl{m}_t^2(\muw)}{{\mw}^2}\right) \right.
\\
              && 
\left.
                    +\frac{\ovl{m}_t}{\ovl{m}_b(\muw)}
                    \left(C^{Gu_3 d_2 (0)\ast}_{L} C^{Gu_3 d_3(1)}_{R} 
                    + C^{Gu_3 d_2 (1)\ast}_{L} C^{Gu_3d_3(0)}_{R}\right)
                   F_{7,8}^{(2)}
                             \left(\frac{\ovl{m}_t^2(\muw)}{{\mw}^2}\right)
                  \right],
\nonumber
\\
\delta^H_{\wt{g}} C_{7,8}^{(1)}(\muw) &=&
                  \frac{2}{3 g_2^2 K^{\ast}_{ts} K_{tb}}
                  \frac{{\mw}^2}{\ovl{m}_t^2(\muw)}\times
\nonumber\\
              &&  \left[ (C^{Hu_3 d_2 (0)\ast}_{L} C^{Hu_3 d_3 (1)}_{L}
                    +C^{Hu_3 d_2 (1)\ast}_{L} C^{Hu_3 d_3 (0)}_{L})
                    F_{7,8}^{(1)}
                             \left(\frac{\ovl{m}_t^2(\muw)}{m_H^2}\right)
\right.
\\
              && 
\left.
                        +\frac{\ovl{m}_t(\muw)}{\ovl{m}_b(\muw)}
                      (C^{Hu_3 d_2 (0)\ast}_{L} C^{Hu_3 d_3 (1)}_{R}
                    +C^{Hu_3 d_2 (1)\ast}_{L} C^{Hu_3 d_3 (0)}_{R})
                    F_{7,8}^{(2)}
                             \left(\frac{\ovl{m}_t^2(\muw)}{m_H^2}\right)
                  \right],
\nonumber
\end{eqnarray}
where $\ovl{m}_t(\muw)$ and $\ovl{m}_b(\muw)$ are $\ovl{MS}$ running
top and bottom masses
at $\muw$. 
Explicit expressions for the effective couplings
$C^{Gu_i d_j (0,1)}_{L,R}$ and $C^{Gu_i d_j (0,1)}_{L,R}$ at $\muw$
and other effective couplings 
are given in appendix~\ref{sec:muw:effective-vertices}.  Analogous expressions
for $\delta^{W,H} C^{\prime}_{7,8}$ can be obtained by simply interchanging L
and R in the formulae above.

In light of eq.~(\ref{eq:hierarchy}), the LO and NLO
 supersymmetric contributions to the Wilson coefficients
 $\deltas C_{7,8}$ are first computed at
$\mu_{SUSY}$,
\begin{eqnarray}
\deltas C_{7,8}(\mususy) &=& \deltas C^{(0)}_{7,8}(\mususy)
 +\frac{\alpha_s(\mususy)}{4\pi} \deltas C^{(1)}_{7,8}(\mususy).
\end{eqnarray}
Next, they are evolved down to $\muw$ with the QCD renormalization
group equation. 
 For this purpose we use the NLO anomalous dimension
obtained in~\cite{misiakmunz95}, assuming six quark flavors, and
employ the NLO expression given in~\cite{buras92} for matrix evolution
\beq
\label{eq:susymc}
\left( 
\begin{array}{c}
\delta^S C_7(\muw) \\
\delta^S C_8(\muw) 
\end{array}
\right)
= \left( 
\mbox{\boldmath $\unitmatrix$} + \frac{\alpha_s(\muw)}{4\pi} 
\mbox{\boldmath $J$} \right)
\mbox{\boldmath $U^{(0)}$} 
\left(
\mbox{\boldmath $\unitmatrix$} - \frac{\alpha_s(\mususy)}{4\pi}
\mbox{\boldmath $J$} \right)
\left(
\begin{array}{c}
\delta^S C_7(\mususy) \\
\delta^S C_8(\mususy) 
\end{array}
\right), 
\eeq
\beq
\mbox{\boldmath $U^{(0)}$} =
\left(
\begin{array}{rr}
\eta^{\frac{16}{21}} & 
-\frac{8}{3}\left( \eta^{\frac{16}{21}} - \eta^{\frac{2}{3}} \right) \\
0 &
\eta^{\frac{2}{3}}
\end{array} 
\right)
,~~~~~
\mbox{\boldmath $J$} = 
\left(
\begin{array}{rr}
-\frac{440}{49} & \frac{28136}{30429} \\
0 & -\frac{73}{9}  
\end{array}
\right),
\label{eq:evolution-matrix}
\eeq
where $\eta = \alpha_s(\mususy)/\alpha_s(\muw)$. After extracting the
LO and NLO parts of Wilson coefficients at both scales,
eq.~(\ref{eq:susymc}) reduces to a more familiar form in the
case of six quark flavors,\footnote{
In our numerical analysis, we actually use eq.~(\ref{eq:susymc}) and
and thus keep ${\cal O}(\alpha_s^2)$ and higher order
corrections in $\delta^S C_{7,8}(\muw)$ that are induced by the
evolution.}

\begin{eqnarray}
\deltas C^{(0)}_7(\muw) &=& 
\eta^{\frac{16}{21}} \deltas C^{(0)}_7(\mu_{SUSY})
+\frac{8}{3}\left( \eta^{\frac{2}{3}}-\eta^{\frac{16}{21}} \right)
 \deltas C^{(0)}_8(\mu_{SUSY}), \\
\deltas C^{(0)}_8 (\muw) &=&
\eta^{\frac{2}{3}} \deltas C^{(0)}_8 (\mu_{SUSY}),\\ \nonumber
\deltas C^{(1)}_7(\muw) &=& 
+\frac{440}{49}\left( \eta^{\frac{37}{21}}-\eta^{\frac{16}{21}} \right)
 \deltas C^{(0)}_7(\mu_{SUSY}) \\ \nonumber
&&
+\left( 
\frac{584}{27} \eta^{\frac{5}{3}}
-\frac{76256}{3381}\eta^{\frac{37}{21}}
-\frac{14296}{621}\eta^{\frac{2}{3}} 
+\frac{3520}{147}\eta^{\frac{16}{21}}
\right) \deltas C^{(0)}_8(\mu_{SUSY}) \\
&&
+\eta^{\frac{37}{21}} \deltas C^{(1)}_7(\mu_{SUSY})
+\frac{8}{3}\left( \eta^{\frac{5}{3}}-\eta^{\frac{37}{21}} \right)
 \deltas C^{(1)}_8(\mu_{SUSY}), \\ 
 \deltas C^{(1)}_8 (\muw) &=&
\eta^{\frac{5}{3}} \deltas C^{(1)}_8 (\mu_{SUSY}) 
+\frac{73}{9}\left( \eta^{\frac{5}{3}} - \eta^{\frac{2}{3}} \right)
\deltas  C^{(0)}_8 (\mu_{SUSY}).
\end{eqnarray}

While we work within the SM basis of operators, SUSY loop induced
${\it O}(\alpha_s)$ contributions to $C_{i}$ ($i=1,\ldots,6$) at
$\mususy$ also in principle mix with $C_{7,8}$ during the course of
evolution.  However, the corresponding $P_{i}$ ($i=1,\ldots,6$)
conserve chirality and are not enhanced by $\tan\beta$. We have
numerically verified that their typical values are ${\cal
O}(10^{-3})$--${\cal O}(10^{-4})$ if squarks/gluino masses are not
very close to $\muw$.  Their contributions to $C_{7,8}$ are further
suppressed by mixing factors.  Because the purpose of this paper is to
present large $\tanb$--enhanced beyond--leading--order effects in
$\brbsgamma$, for simplicity we neglect these small contributions
compared to the ${\cal O}(10^{-1})$ SM ones.

The NLO supersymmetric contributions to the Wilson coefficients
at $\mususy$ reads
\begin{eqnarray}
\label{eq:nlomcmsusy}
\deltas C^{(1)}_{7,8}(\mususy) &=&
\deltas_{g} C_{7,8}^{(1)}(\mususy) + 
\deltas_{\wt{g}} C_{7,8}^{(1)}(\mususy)~~~~~(S=\chi^-,\chi^0,\wt{g}),
\label{eq:SUSY-Wilson}
\end{eqnarray}
where $\deltas_g C_{7,8}^{(1)}$ and $\deltas_{\wt{g}} C_i^{(1)}$
denote respective gluon and gluino NLO corrections, analogously to the
$W$ and $H$ contributions at $\muw$.  Explicit expressions for
$\delta^{S}_{g} C^{(1)}_{7,8}$ can be found in ref.~\cite{bmu00} and
appendix~\ref{sec:nlo-qcd-corrections}.


\begin{figure}[t!]
\begin{center}
\begin{minipage}{12.0cm}
\centerline
{\hspace*{-.2cm}\psfig{figure=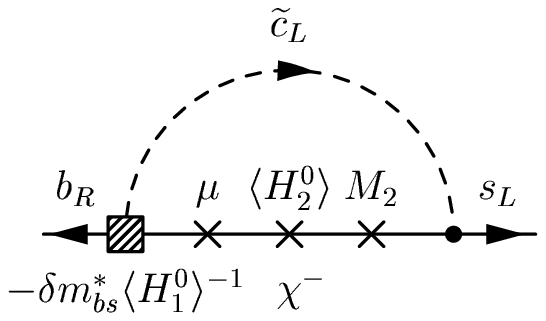, angle=0,width=6.0cm}
\hspace*{-.2cm}\psfig{figure=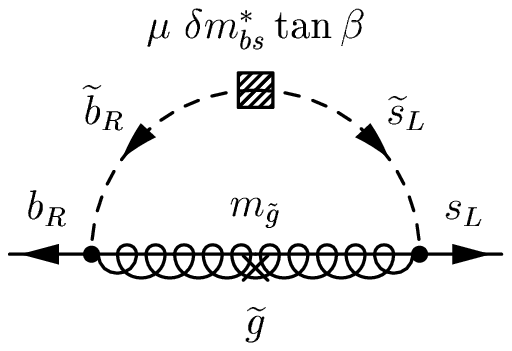, angle=0,width=6.0cm}
}
\end{minipage}
\caption{Leading diagrams that contribute to
$\delta^{\chi^-}_{\wt{g}} C_{7,8}^{(1)}$ (left) and
$\delta^{\wt{g}}_{\wt{g}} C_{7,8}^{(1)}$ (right).
The photon (or gluon) line is attached in every possible manner.
\label{fig:feynmancl7}}
\end{center}
\end{figure}

\begin{figure}[t!]
\begin{center}
\begin{minipage}{12.0cm}
\centerline
{\hspace*{-.2cm}\psfig{figure=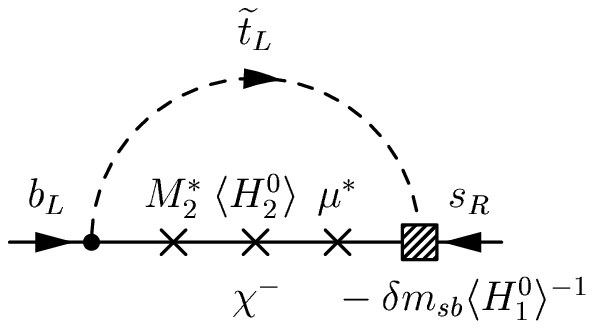, angle=0,width=6.0cm}
\hspace*{-.2cm}\psfig{figure=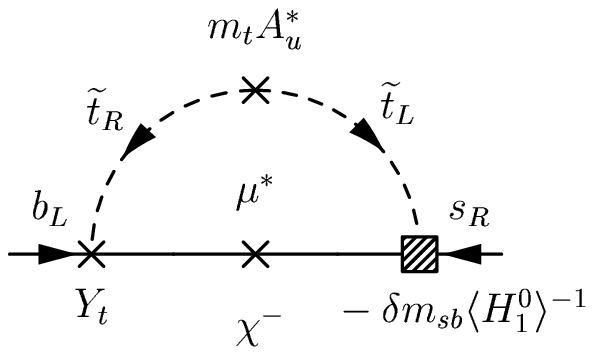, angle=0,width=6.0cm}
}
\vspace{1em}
\end{minipage}
\begin{minipage}{12.0cm}
\centerline
{\hspace*{-.2cm}\psfig{figure=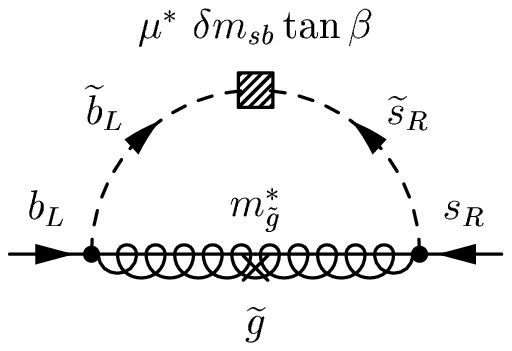, angle=0,width=6.0cm}
}
\end{minipage}
\caption{Leading diagrams which contribute to
$\delta^{\chi^-}_{\wt{g}} {C^{\prime\,(1)}_{7,8}}$ (upper) and
$\delta^{\wt{g}}_{\wt{g}} {C^{\prime\,(1)}_{7,8}}$ (lower).
\label{fig:feynmancr7}}
\end{center}
\end{figure}

In computing gluino corrections to supersymmetric contributions
$\deltas_{\wt{g}} C_{7,8}^{(1)}$, the effective vertex method does not
work, as explained in sec.~\ref{sec:procedure}. Instead of calculating
the full two--loop diagrams, we take into account $\tanb$--enhanced
corrections to the Yukawa couplings, as well as $LR$ mixings due to
terms involving the trilinear soft terms {\boldmath $A_{d,u}$} which
in principle can also be sizeable.  In the absence of full 2--loop
calculation of $C_{7,8}$, such effective Yukawa vertices contain only
the dominant corrections.  (In fig.~\ref{fig:feynmancl7} and
\ref{fig:feynmancr7}, we show the Feynman diagrams that contribute the
leading corrections to $C_{7,8}$.)  We indicate the approximation by
replacing $\deltas_{\wt{g}} C_{7,8}^{(1)}\rightarrow \deltas_{\wt{g}}
C_{7,8}^{(Y)}$. We obtain
\begin{eqnarray}
\label{eq:charlonlo}
\left( {\deltacharn} C_{7,8}^{(0)} + \frac{\alpha_s}{4\pi}
        {\delta^{\chi^-}_{\wt{g}}} C_{7,8}^{(Y)} \right) (\mususy) 
&=&\frac{1}{ g_2^2 K^{\ast}_{ts} K_{tb}}
                  \sum^2_{a=1}\sum^6_{I=1}
                  \frac{m_W^2}{ \mca^2 } \times \\ \nonumber
              &&  \left[ \left(C_{d\,R}\right)_{2aI} \left(C_{d\,R}\right)^{\ast}_{3aI}
                  H_1^{[7,8]}
                  \left(\xsuica \right)\right.\\  
\nonumber
	      &&  \left. +\frac{\mca}{\ovl{m}_b(\mususy)}
                      \left(C_{d\,R}\right)_{2aI} \left(C_{d\,L}\right)^{\ast}_{3aI}
                      \left( H_2^{[7,8]}
                      \left(\xsuica\right)
                      + \lambda^{[7,8]} \right)
                  \right],
\\
\label{eq:neutlonlo}
\left( \delta^{\chi^0} C_{7,8}^{(0)}+\frac{\alpha_s}{4\pi}
\delta^{\chi^0}_{\wt{g}} C_{7,8}^{(Y)} \right) (\mu_{SUSY}) &=&
                  \frac{1}{ g_2^2 K^{\ast}_{ts} K_{tb}}
                  \sum^4_{r=1}\sum^6_{I=1}
                  \frac{m_W^2}{\mnr^2}\times
\\
              &&  \left[ \left(N_{d\,R}\right)_{2rI}\left(N_{d\,R}\right)^{\ast}_{3rI}
                   H_3^{[7,8]}
                   \left(\xsdinr\right)
\right.
\nonumber\\
              && 
\left.
                     +\frac{\mnr}{\ovl{m}_b(\mususy)}
                      \left(N_{d\,R}\right)_{2rI}\left(N_{d\,L}\right)^{\ast}_{3rI}
                      H_4^{[7,8]}
                      \left(\xsdinr\right)
                  \right], 
\nonumber\\
\label{eq:gluilonlo}
\left( \delta^{\wt{g}} C_{7,8}^{(0)}+\frac{\alpha_s}{4\pi}
    \delta^{\wt{g}}_{\wt{g}} C_{7,8}^{(Y)} \right)(\mu_{SUSY}) &=&
                  \frac{4}{3 g_2^2 K^{\ast}_{ts} K_{tb}}
                  \sum^6_{I=1}
                  \frac{m_W^2}{\mg^2}\times
\\
              &&  \left[ \left(G_{d\,R}\right)_{2I}\left(G_{d\,R}\right)^{\ast}_{3I}
                  H_5^{[7,8]}
                  \left(\xsdig\right)
\right.
\nonumber\\
              && 
\left.
                        +\frac{\mg}{\ovl{m}_b(\mususy)}
                      \left(G_{d\,R}\right)_{2I}\left(G_{d\,L}\right)^{\ast}_{3I}
                      H_6^{[7,8]}
                             \left(\xsdig\right)
                  \right],
\nonumber
\end{eqnarray}
where 
\beq
x^X_Y = \frac{(m_X)^2}{(m_Y)^2}~~~~~(X=\wt{u}_I,\wt{d}_I,
    ~Y=\chi^-_a, \chi^0_r, \wt{g}), 
\eeq
and $H_{1,\ldots,6}^{[7,8]}$ and $\lambda^{[7,8]}$ are given in
appendix~\ref{sec:mass-functions}.
Here $\ovl{m}_b(\mususy)$ denotes 
the running bottom mass at $\mususy$. The couplings
$\left(C_{d\,R,L}\right)_{iaI}$, $\left(N_{d\,R,L}\right)_{irI}$ and
$\left(G_{d\,R,L}\right)_{iI}$ correspond to the chargino, neutralino
and gluino vertices including gluino corrections and are given in
appendix~\ref{sec:Yukawa-correction}.  Note that in the
mass functions we use physical squark masses $m_{\widetilde{d}_{I}}$
and $m_{\widetilde{u}_{I}}$, where $I=1,\ldots,6$.

The LO contributions at $\mususy$ $\delta^{S} C_{7,8}^{(0)}$
($S=\chi^-,\chi^0,\wt{g}$) can be obtained from the above expressions
by replacing $m_{\widetilde{d}_{I}}\ra m^{(0)}_{\widetilde{d}_{I}}$
and $m_{\widetilde{u}_{I}}\ra m^{(0)}_{\widetilde{u}_{I}}$ above and
by replacing the effective chargino, neutralino and gluino couplings
with their tree--level values, as discussed in
appendix~\ref{sec:Yukawa-correction}.  Expressions for beyond--LO
contributions $\delta^{S} C_{7,8}^{(Y)}$ can then be obtained by
subtracting $\delta^{S} C_{7,8}^{(0)}$ from the full expressions for
$\left(\delta^{S} C_{7,8}^{(0)}+\frac{\alpha_s}{4\pi}
\delta^{S}_{\wt{g}} C_{7,8}^{(Y)}\right) $ in
eqs.~(\ref{eq:charlonlo})--(\ref{eq:gluilonlo}).  Finally, expressions
for $\delta^{S} {C^{\prime}}^{(0)}_{7,8}$ and $\delta^{S}
{C^{\prime}}^{(Y)}_{7,8}$ can be obtained by interchanging the indices
L and R in the above formulae.

\begin{figure}[t!]
\begin{center}
\begin{minipage}{12.0cm}  
\centerline
{\hspace*{-.2cm}\psfig{figure=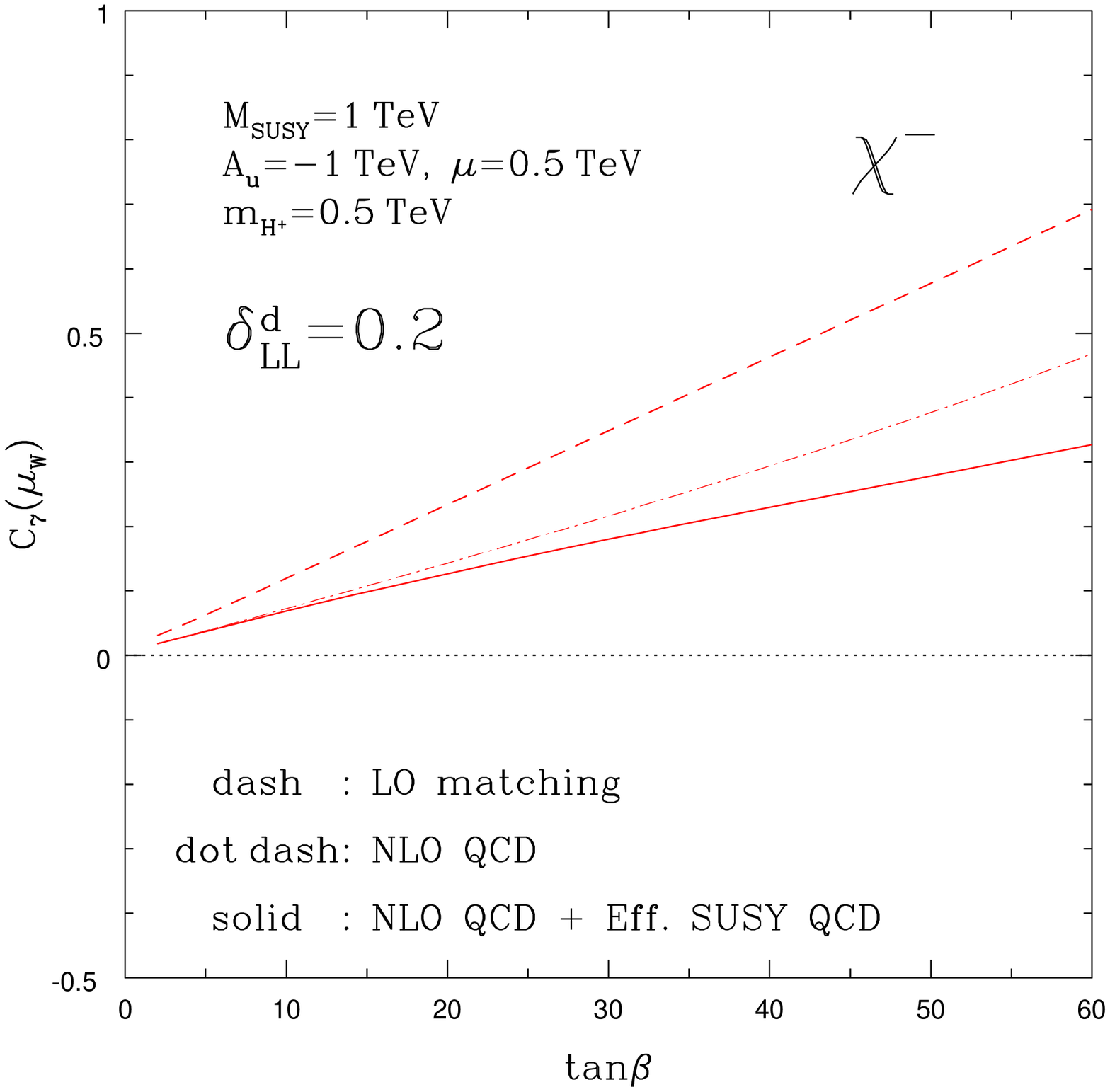, angle=0,width=6.0cm}
 \hspace*{-.2cm}\psfig{figure=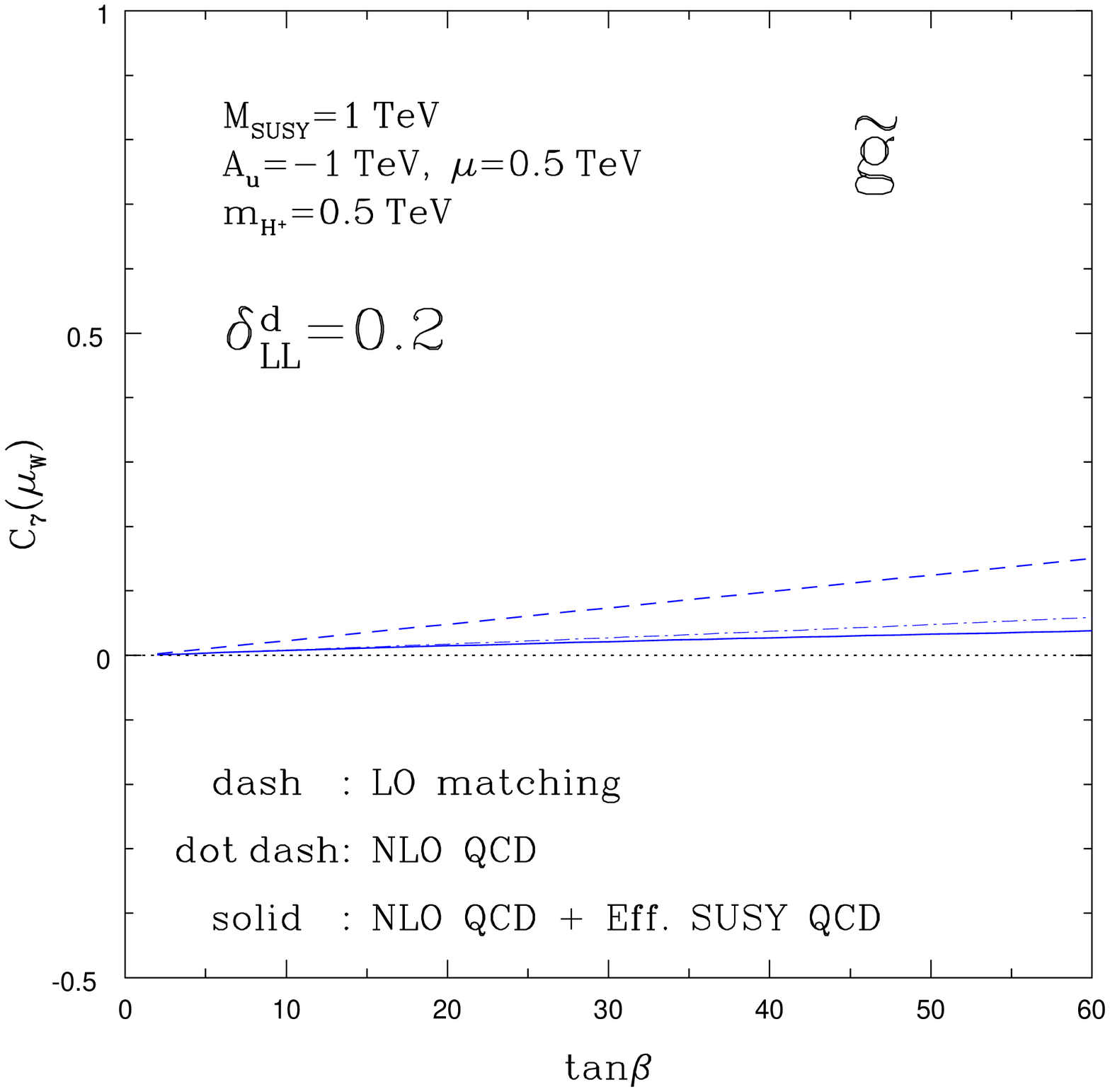, angle=0,width=6.0cm}
}
\end{minipage}
\psfig{figure=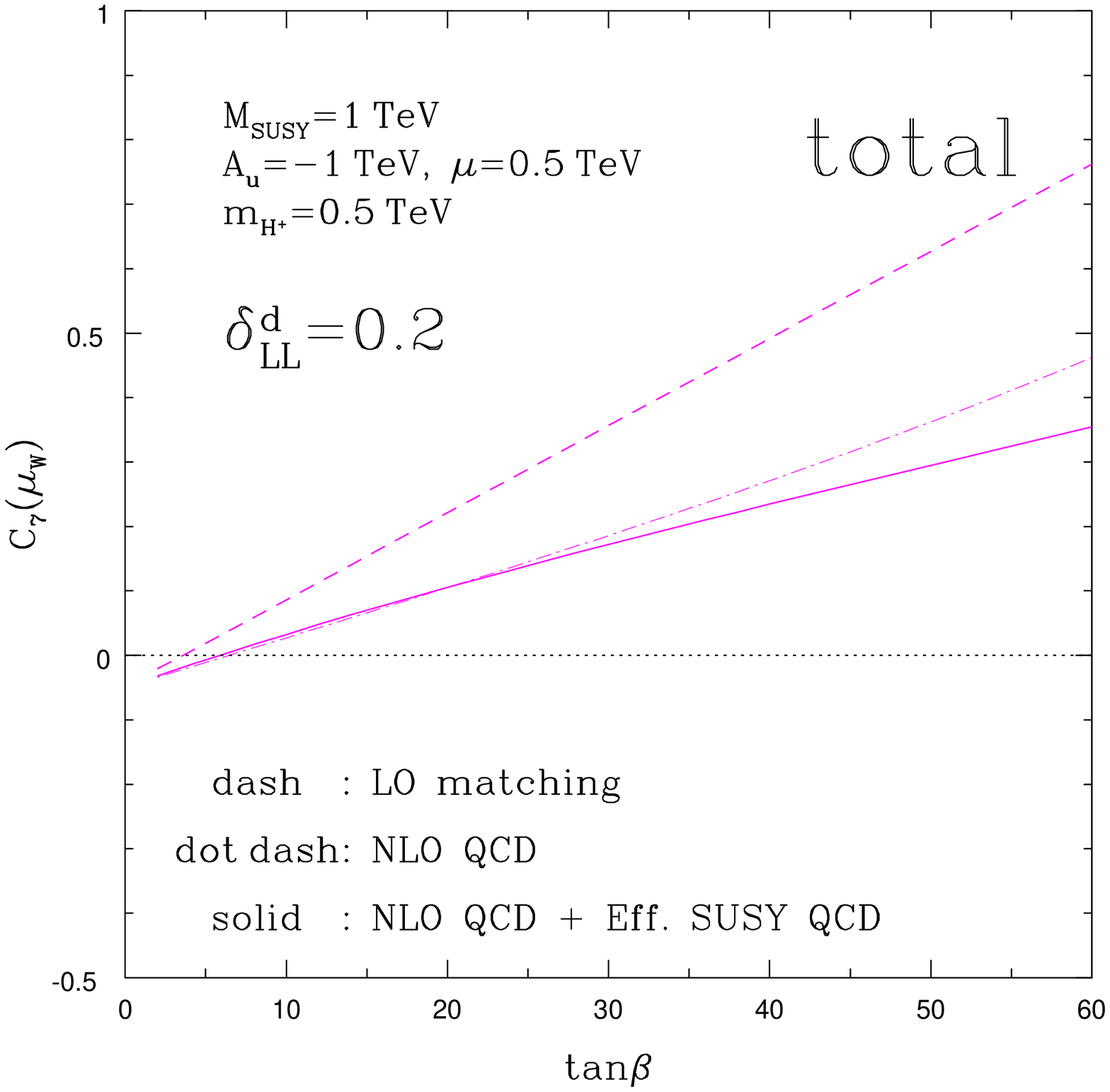, angle=0,width=8.0cm}
\caption{\label{fig:cl7gll02} {\small The Wilson coefficient
    $C_7(\muw)$ \vs\ $\tanb$ for $\msusy=1\tev$, 
    $\mgluino=\sqrt{2}\,m_{\widetilde q}=\msusy$, $\mu=0.5\tev$,
    $A_d=0$, $A_u=-1\tev$, $m_{H^+}=0.5\tev$ and
    $\delta^d_{LL}=0.2$. All the other $\delta^{d,u}_{.~.}$'s are set
    to zero.  In the upper left (right) window we show the chargino
    (gluino) contribution to $C_7(\muw)$, and its total value from new
    physics in the main (lower) window.  The dashed line shows the
    case of LO matching, the dot--dashed one includes the effect of
    NLO QCD, while the solid one includes also the effect of
    beyond--LO corrections from squark--gluino
    loops. 
} 
}
\end{center}
\end{figure}

\begin{figure}[t!]
\begin{center}
\begin{minipage}{12.0cm}  
\centerline
{\hspace*{-.2cm}\psfig{figure=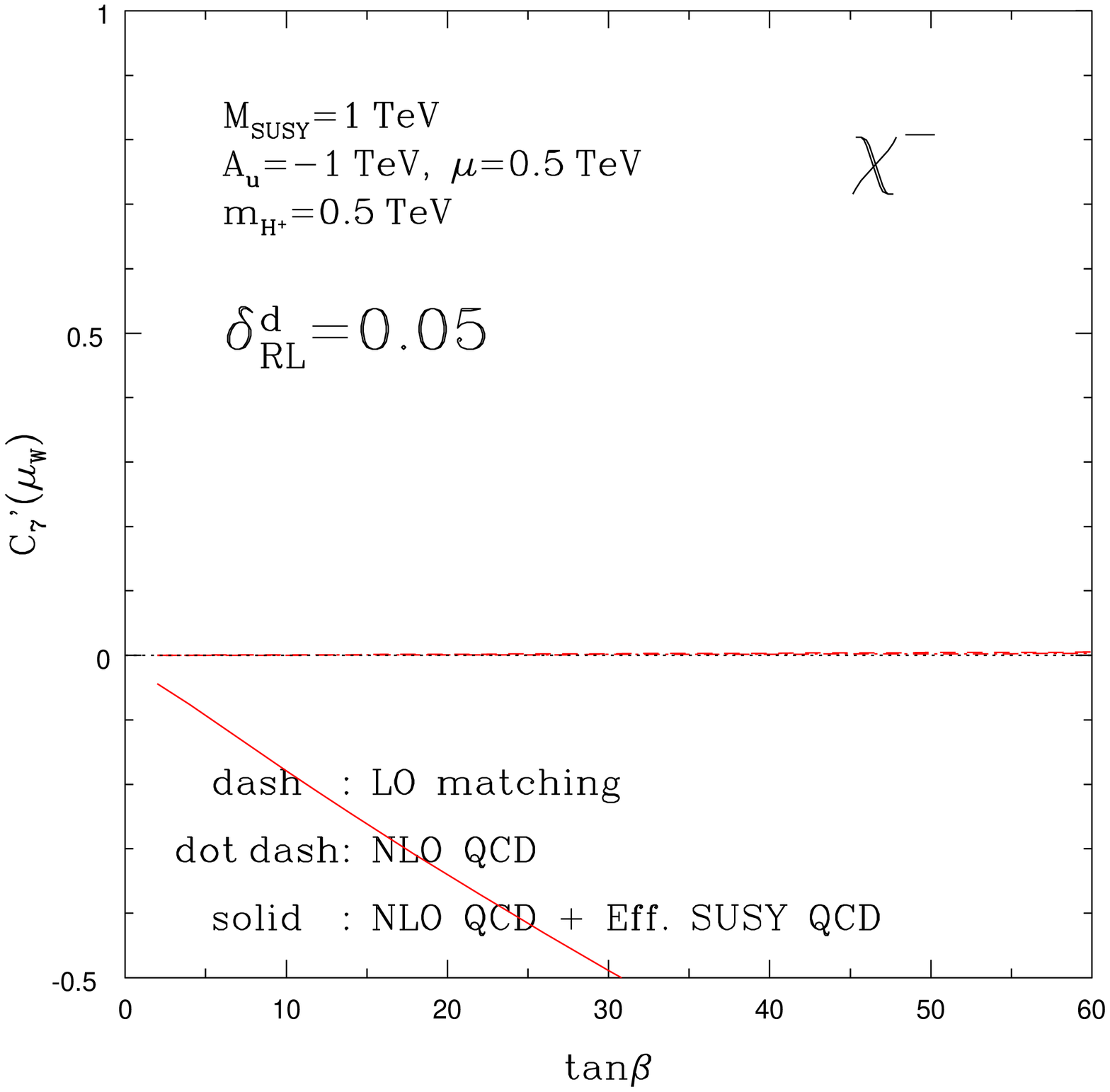, angle=0,width=6.0cm}
 \hspace*{-.2cm}\psfig{figure=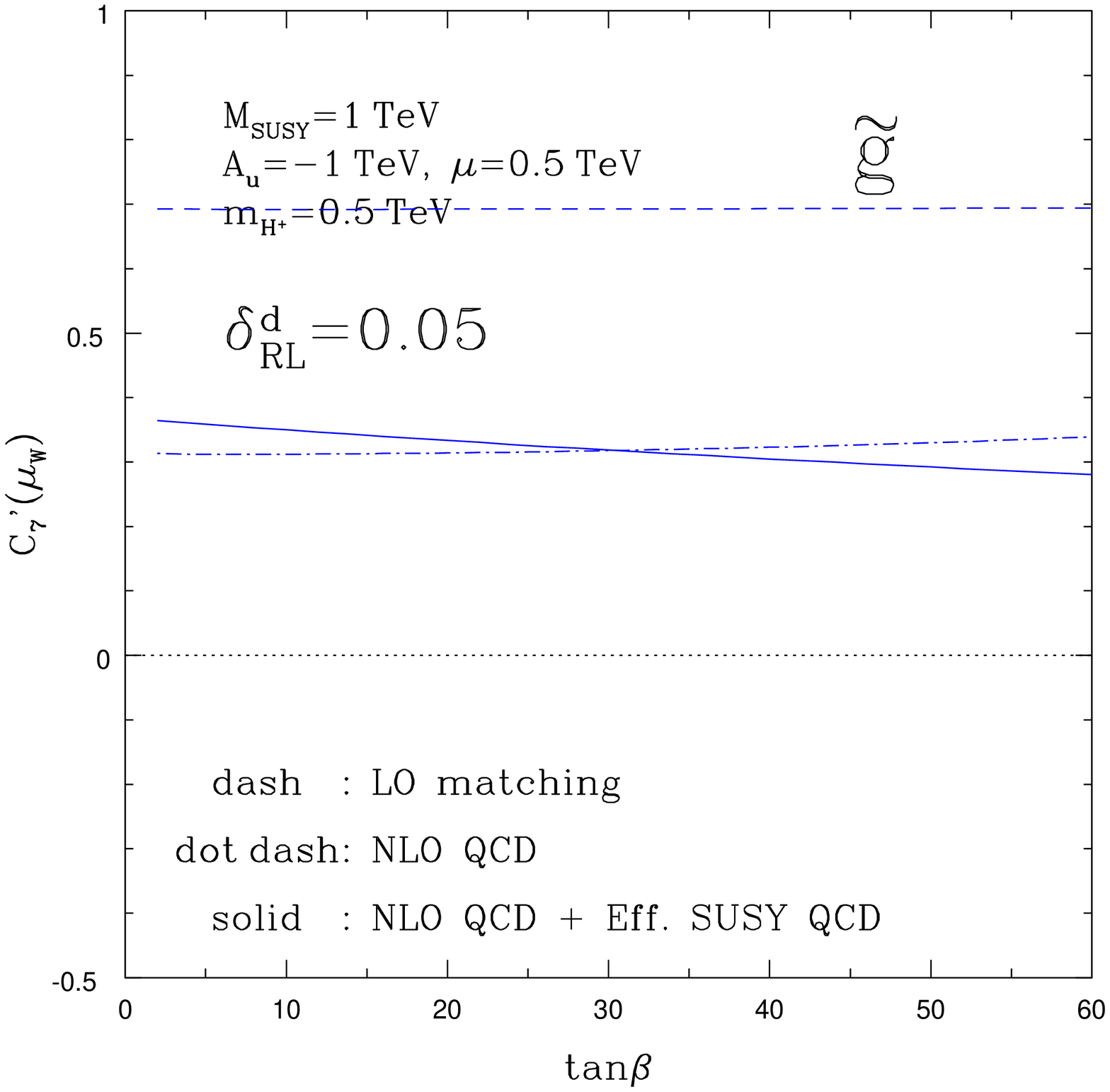, angle=0,width=6.0cm}
}
\end{minipage}
\psfig{figure=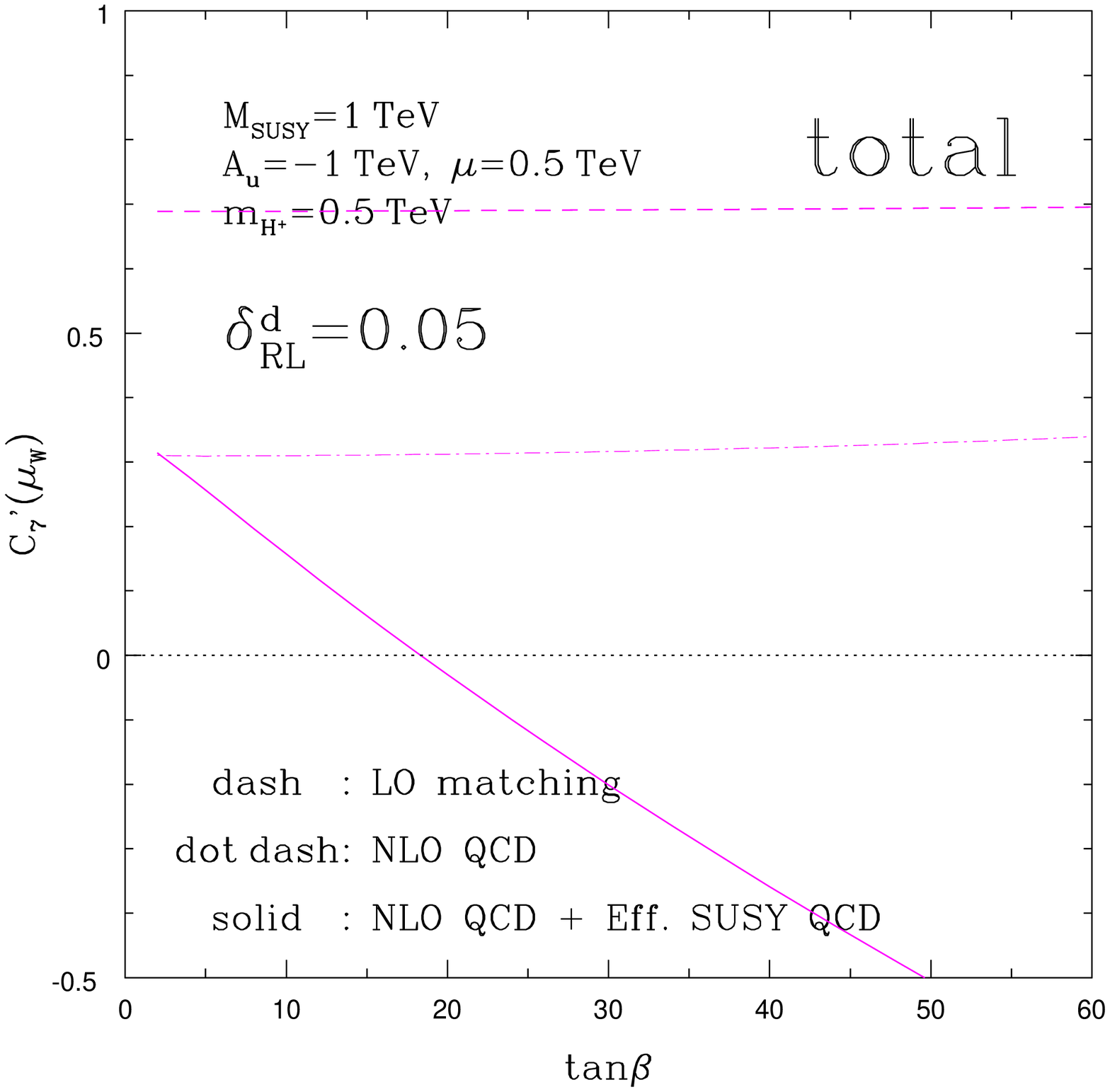, angle=0,width=8.0cm}
\caption{\label{fig:cr7grl005} {\small The Wilson coefficient
    $C_7^{\prime}(\muw)$ \vs\ $\tanb$ for $\delta^d_{RL}=0.05$ and for the
    other parameters as in fig.~\protect\ref{fig:cl7gll02}. } }
\end{center}
\end{figure}

\begin{figure}[t!]
\begin{center}
\begin{minipage}{12.0cm}  
{\hspace*{-.2cm}\psfig{figure=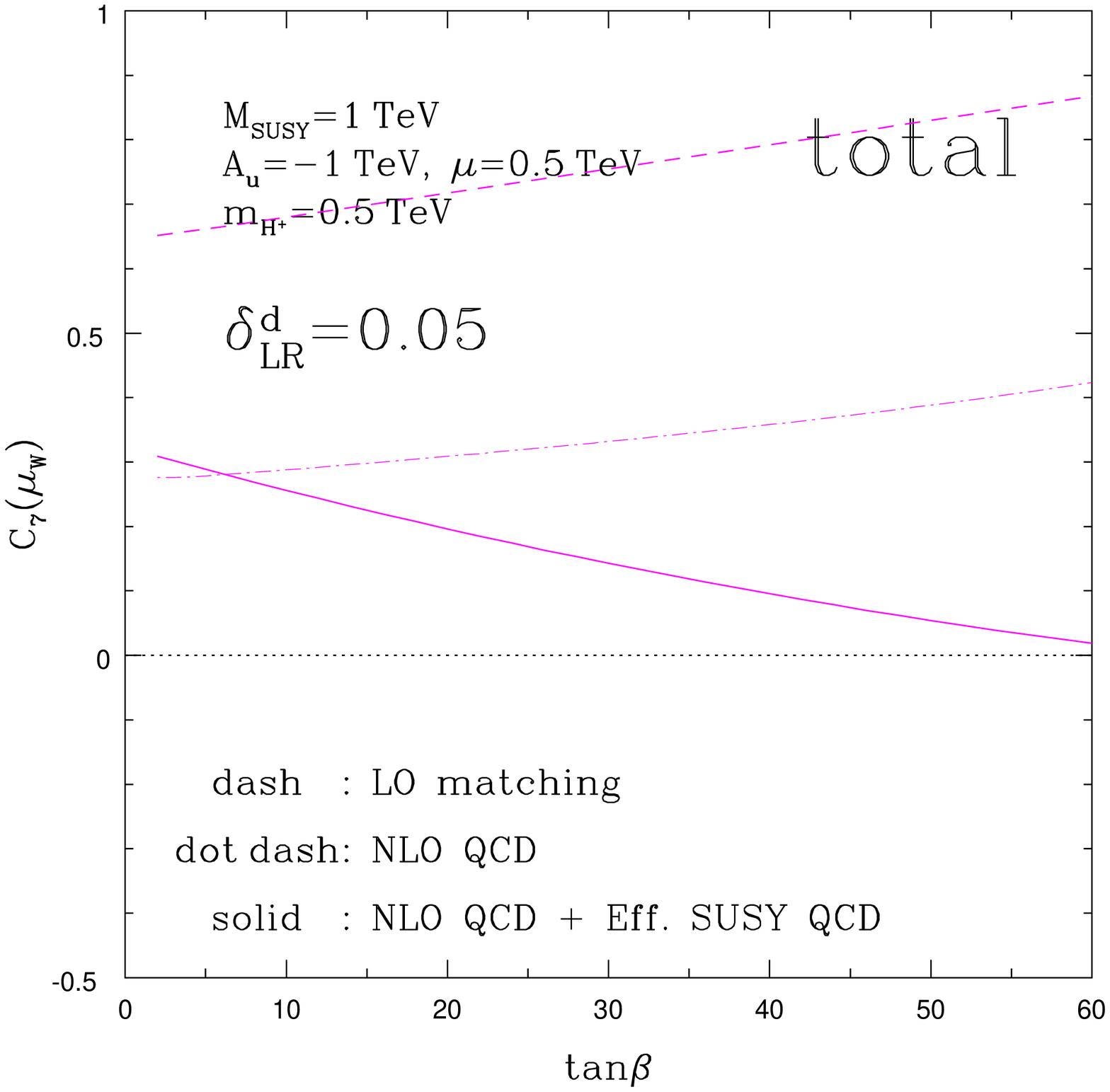, angle=0,width=6.0cm}
 \hspace*{-.2cm}\psfig{figure=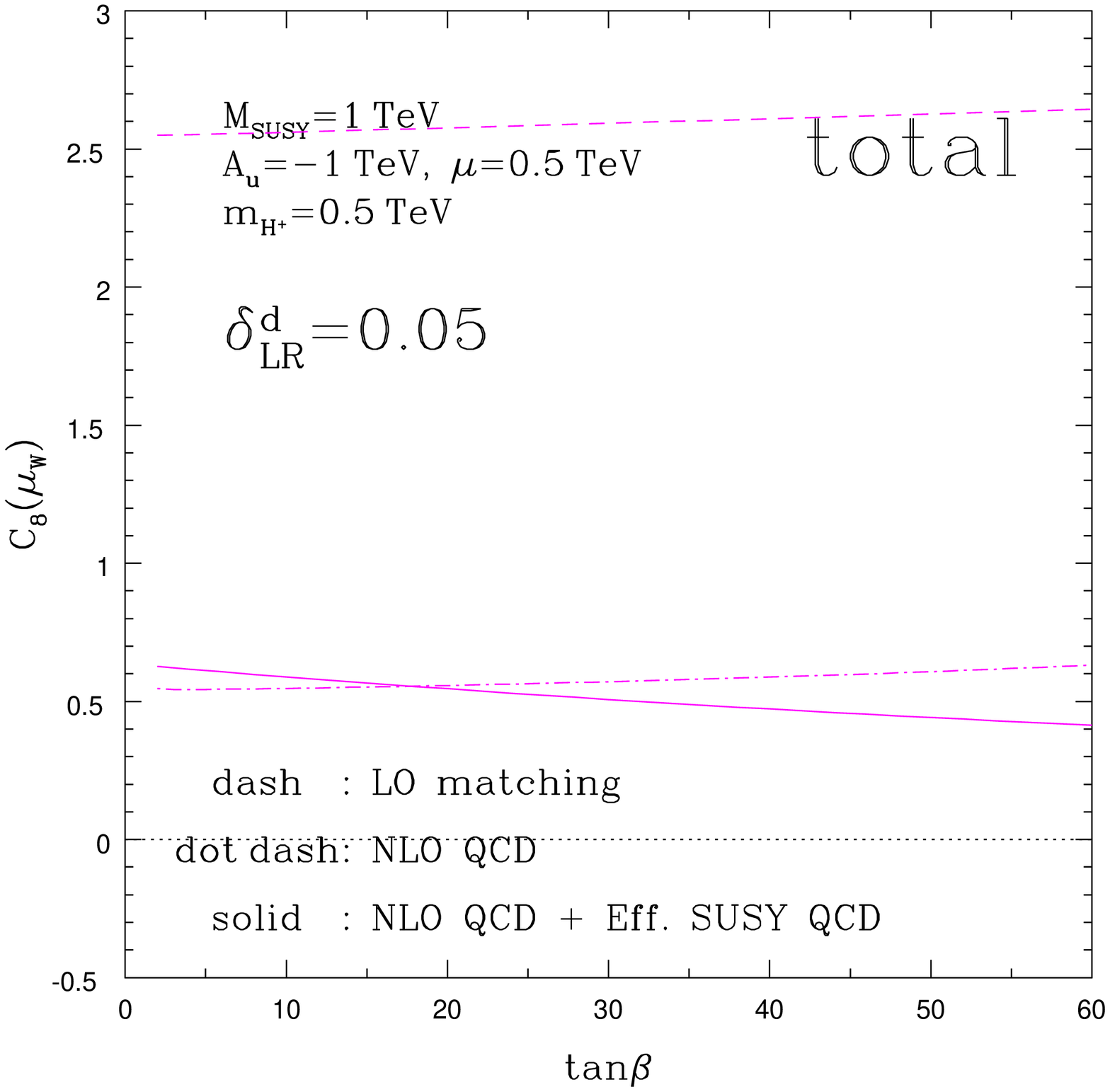, angle=0,width=6.0cm}
}
\end{minipage}
\end{center}
\vspace*{-.35in} 
\hspace*{-.70in}
\begin{center}
\begin{minipage}{12.0cm}  
{\hspace*{-.2cm}\psfig{figure=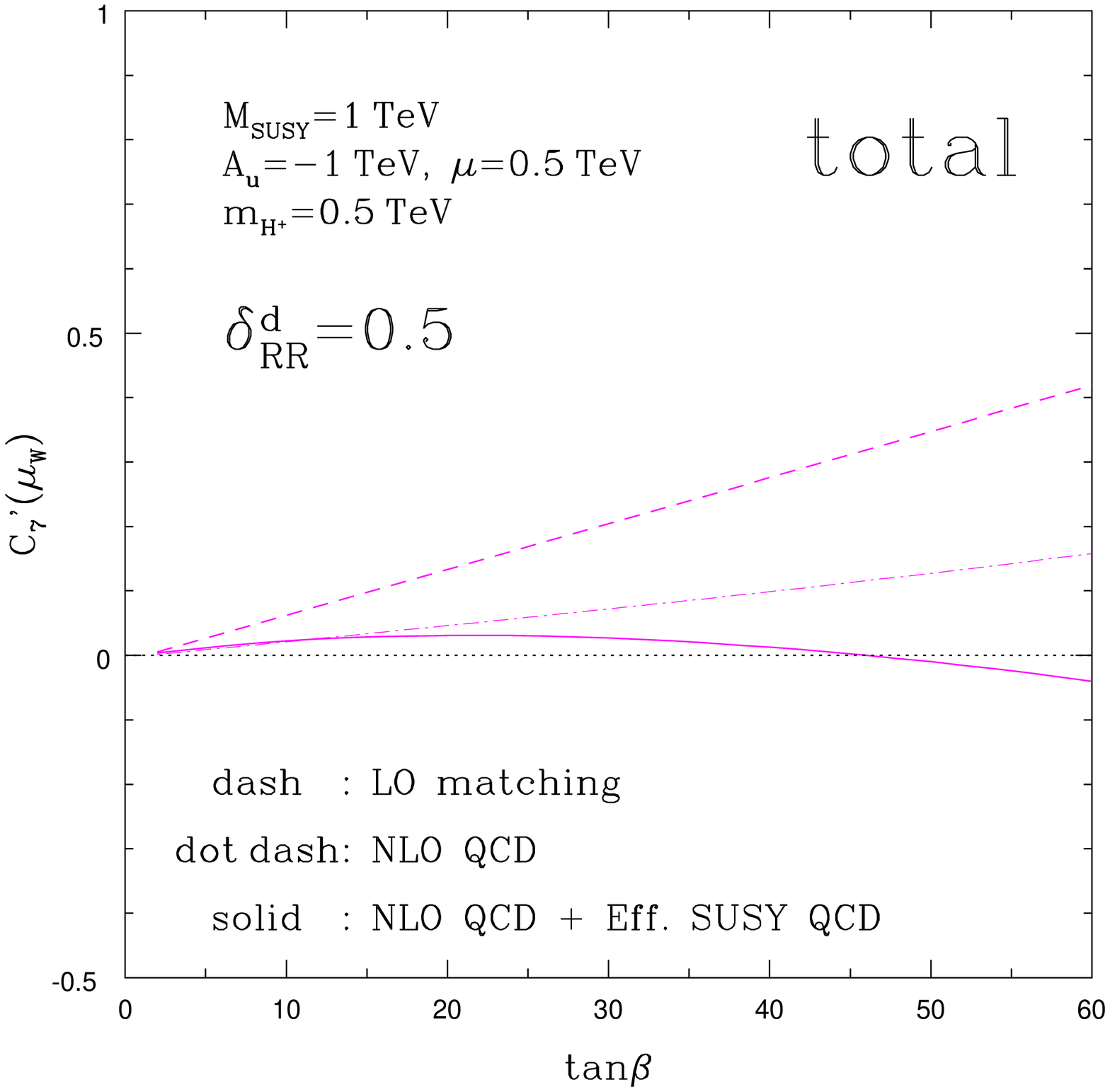, angle=0,width=6.0cm}
 \hspace*{-.2cm}\psfig{figure=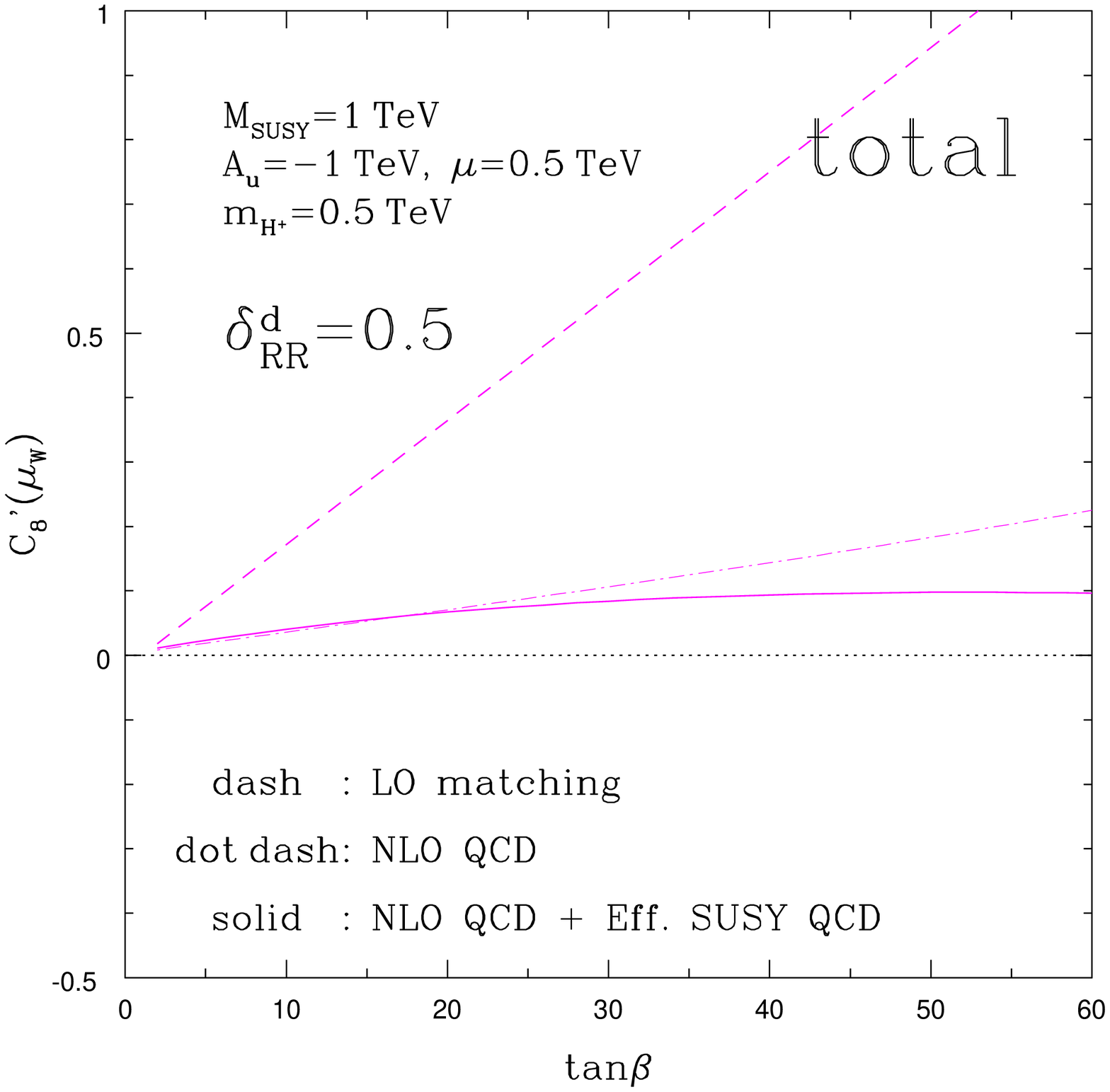, angle=0,width=6.0cm}
}
\end{minipage}
\caption{\label{fig:clr78mup} {\small The Wilson coefficients $C_7
    (\muw)$ (upper left) and $C_8(\muw)$ (upper right) \vs\ $\tanb$
    for $\delta^d_{RL}=0.05$; and $C_7^{\prime}(\muw)$ (lower left)
    and $C_8^{\prime}(\muw)$ (lower right) for
    $\delta^d_{RR}=0.5$. All other parameters as in
    fig.~\protect\ref{fig:cl7gll02}. } }
\end{center}
\end{figure}

\begin{figure}[t!]
\begin{center}
\begin{minipage}{12.0cm}  
{\hspace*{-.2cm}\psfig{figure=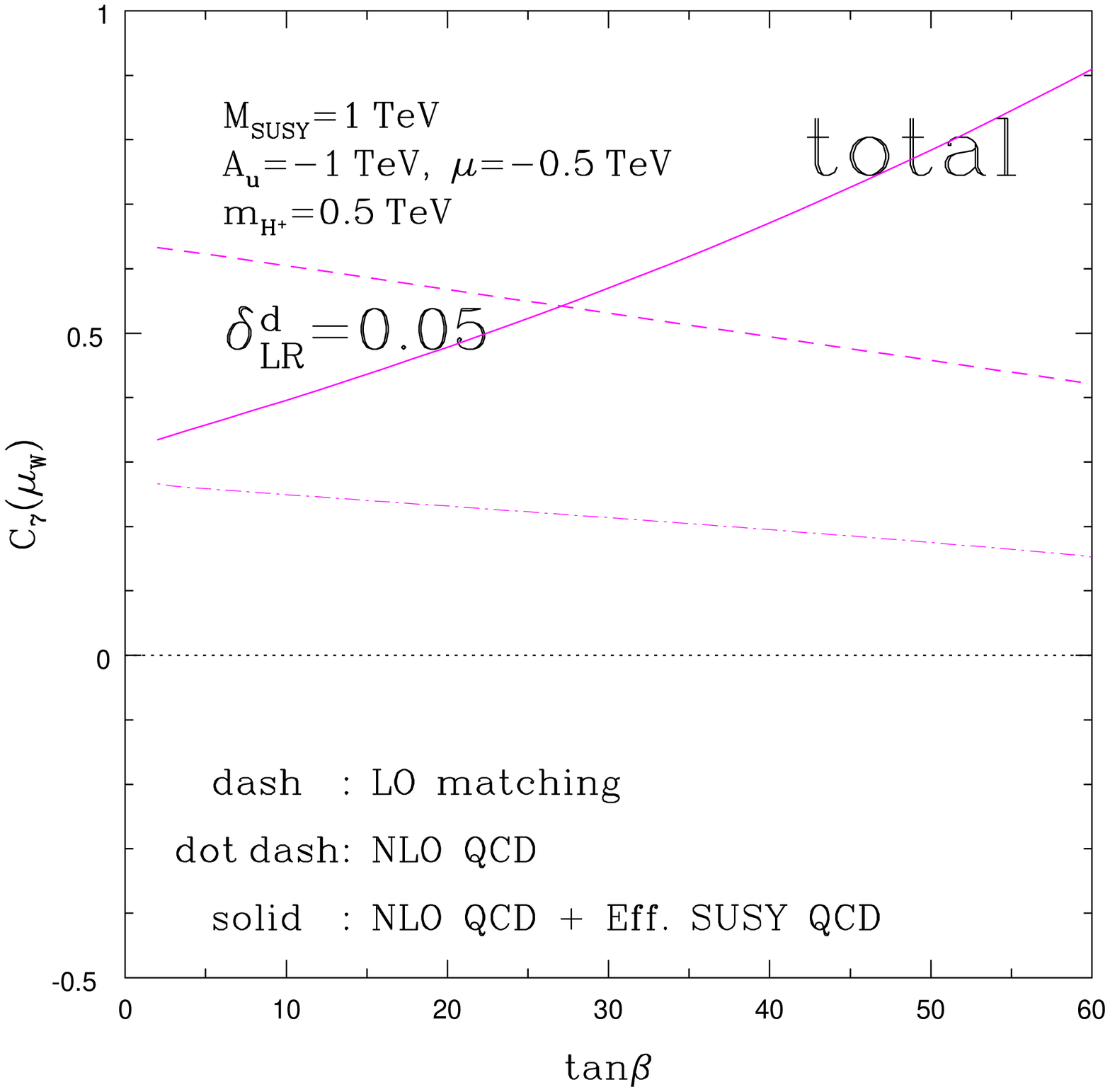, angle=0,width=6.0cm}
 \hspace*{-.2cm}\psfig{figure=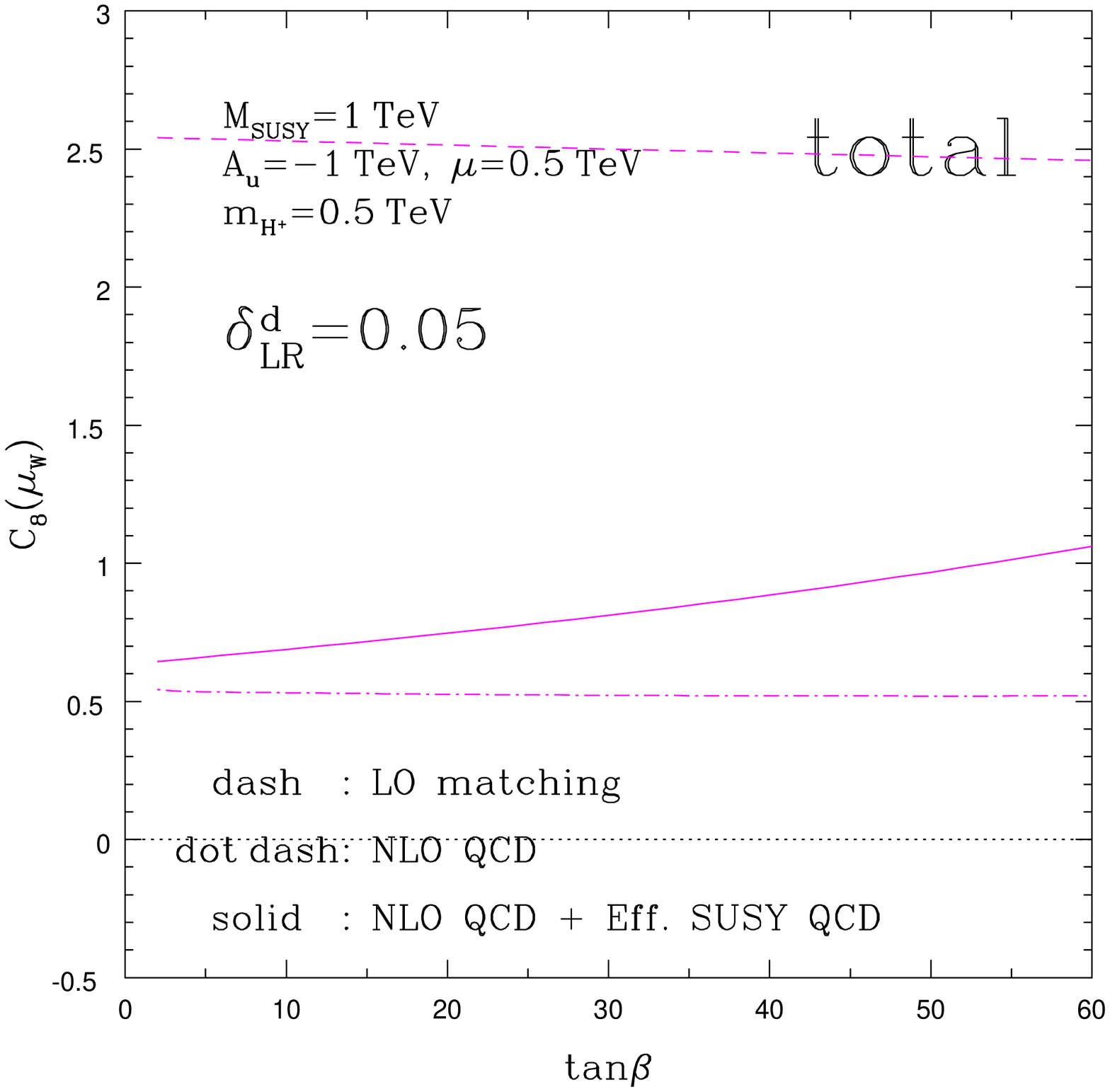, angle=0,width=6.0cm}
}
\end{minipage}
\end{center}
\vspace*{-.35in} 
\hspace*{-.70in}
\begin{center}
\begin{minipage}{12.0cm}  
{\hspace*{-.2cm}\psfig{figure=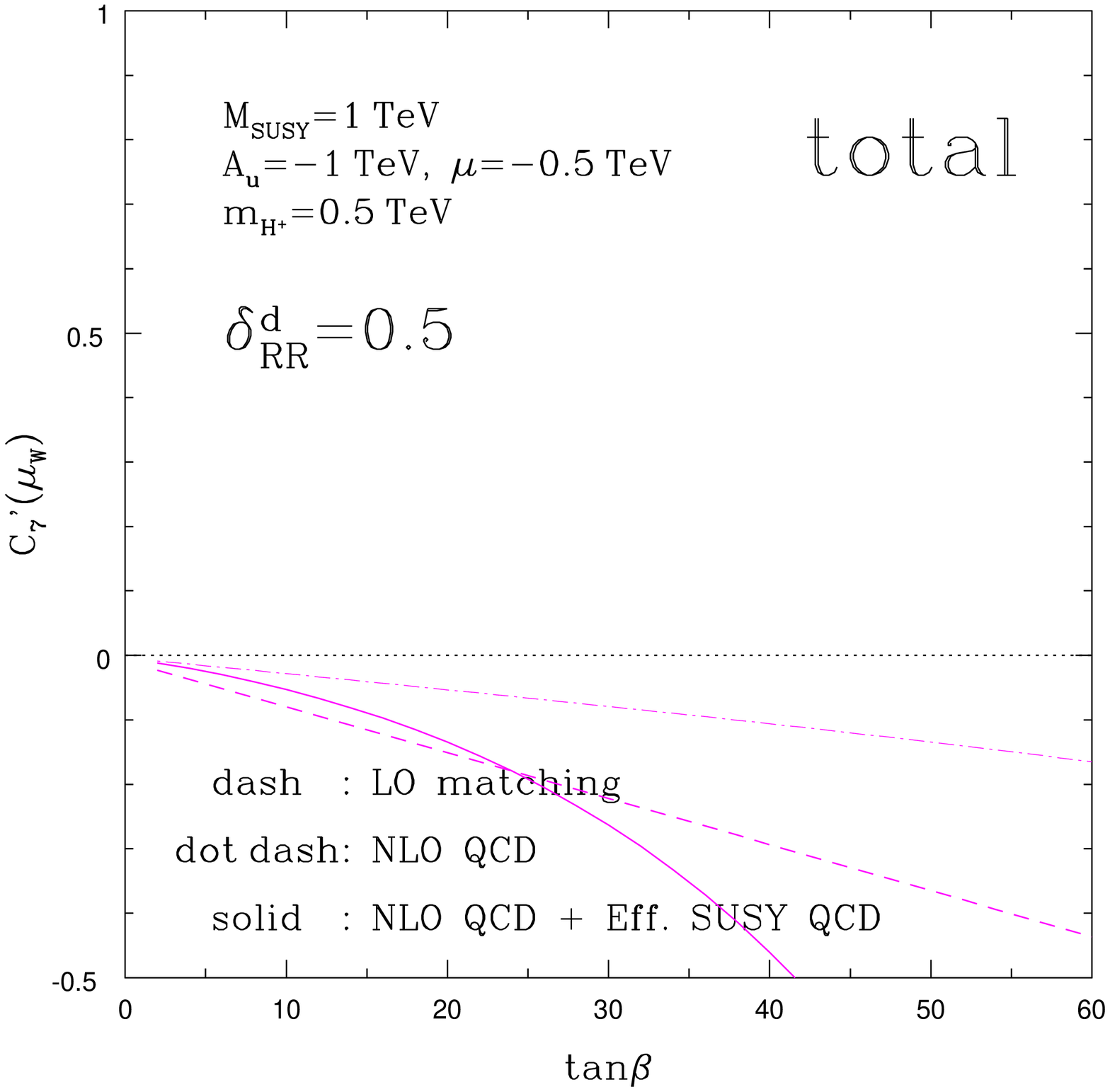, angle=0,width=6.0cm}
 \hspace*{-.2cm}\psfig{figure=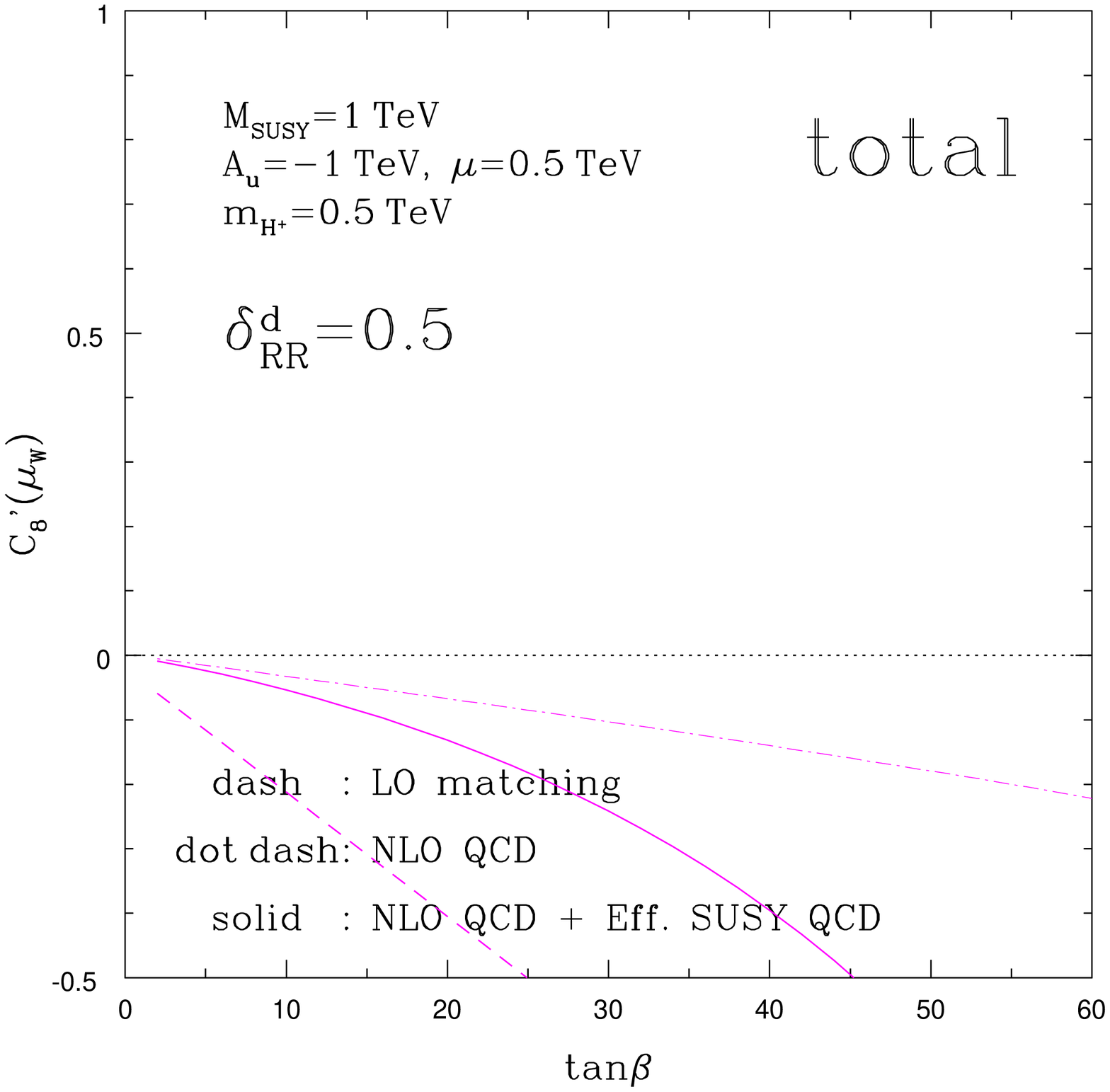, angle=0,width=6.0cm}
}
\end{minipage}
\caption{\label{fig:clr78mun} {\small The same as in 
fig.~\protect\ref{fig:cl7gll02} but for $\mu<0$.
 } }
\end{center}
\end{figure}

\begin{figure}[t!]
\begin{center}
\begin{minipage}{12.0cm}  
{\hspace*{-.2cm}\psfig{figure=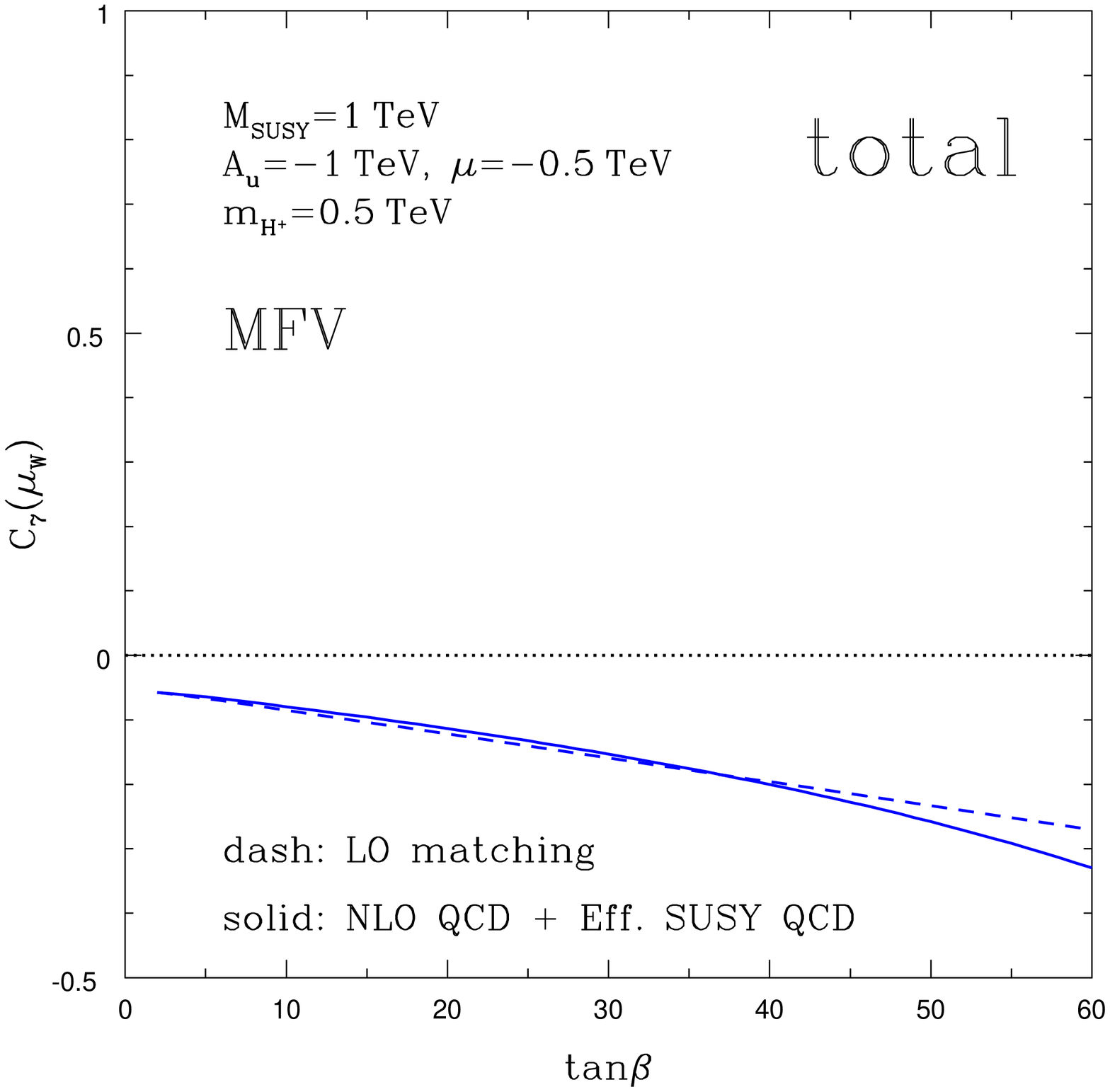, angle=0,width=6.0cm}
 \hspace*{-.2cm}\psfig{figure=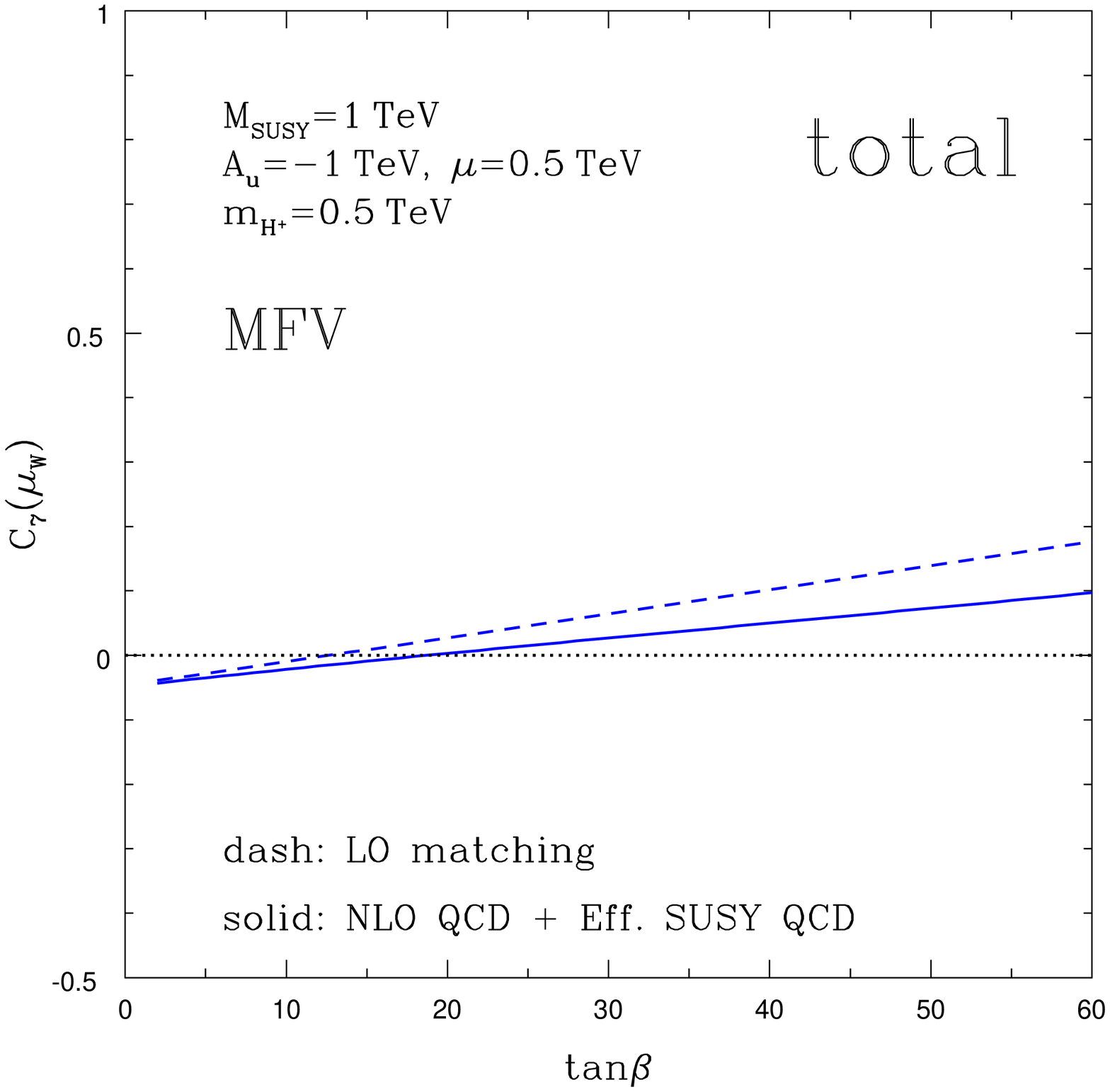, angle=0,width=6.0cm}
}
\end{minipage}
\end{center}
\begin{center}
\caption{\label{fig:cl7mfv} {\small The coefficient $C_7 (\muw)$ \vs\
    $\tanb$ in MFV for $\mu<0$ (left) and $\mu>0$ (right). All other
    parameters as in fig.~\protect\ref{fig:cl7gll02}. Relative to GFM,
    the difference between LO and beyond--LO effects is rather mild. 
} }
\end{center}
\end{figure}

\begin{figure}[t!]
\begin{center}
\begin{minipage}{12.0cm}  
{\hspace*{-.2cm}\psfig{figure=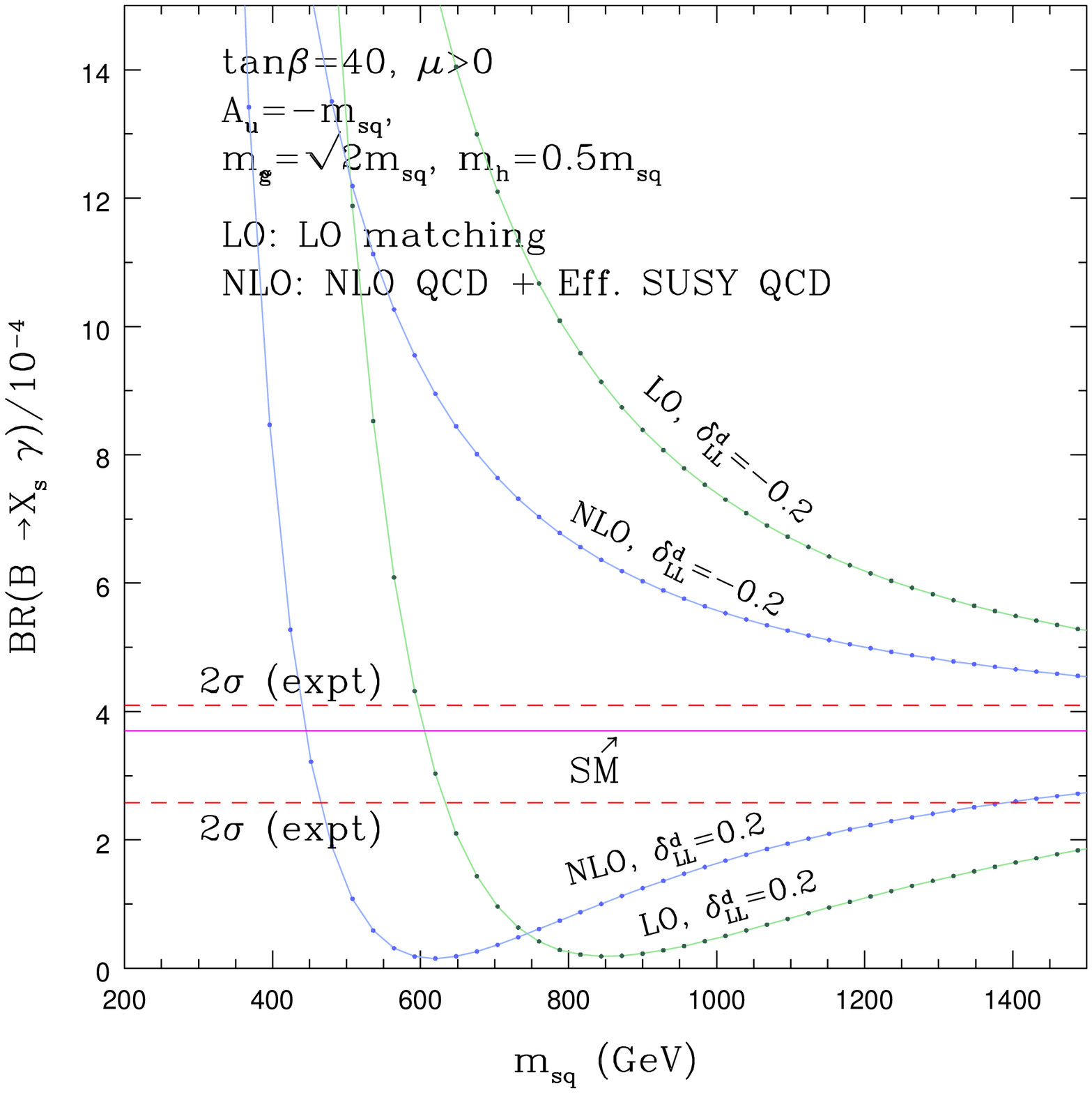, angle=0,width=6.0cm}
 \hspace*{-.2cm}\psfig{figure=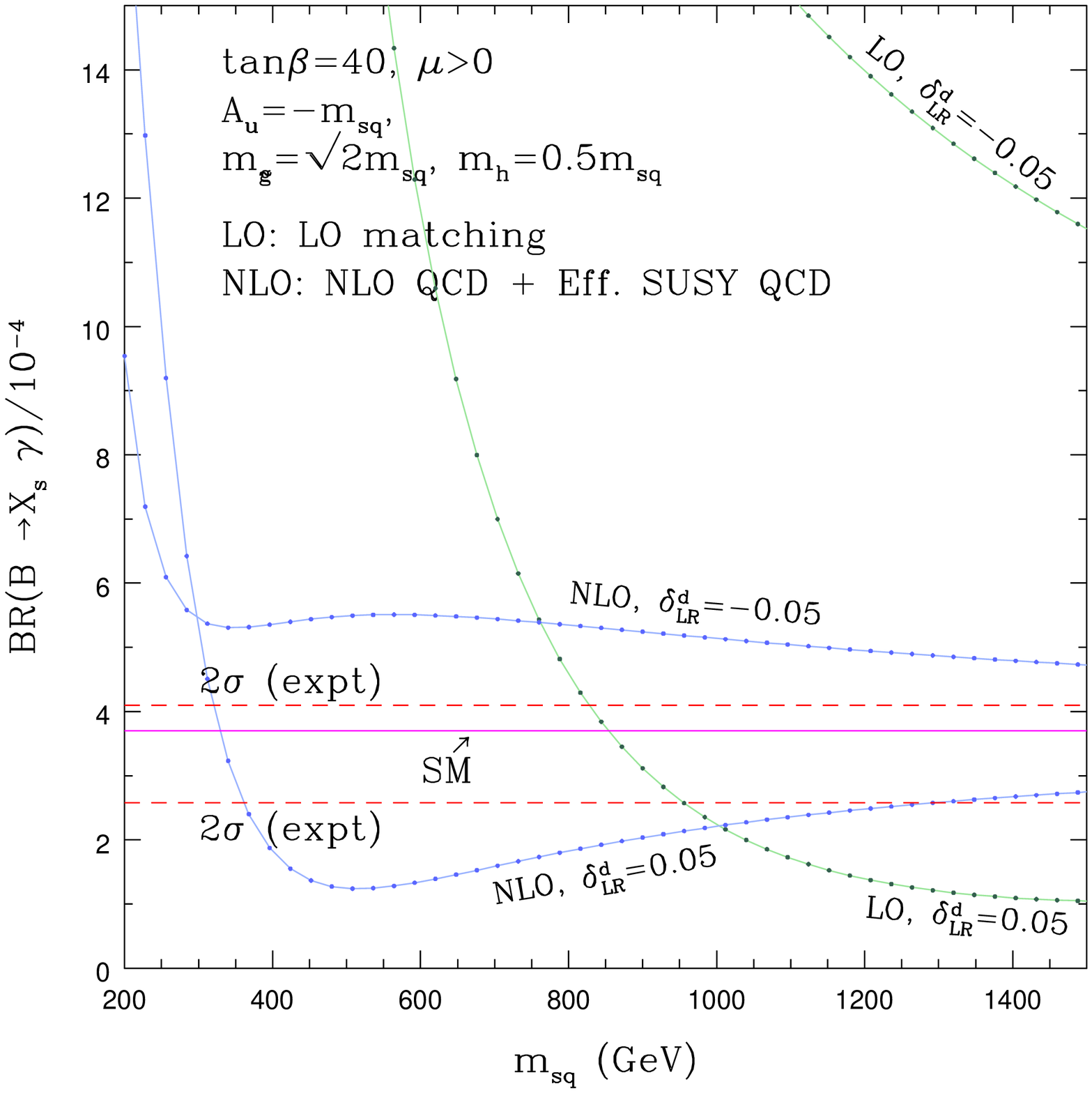, angle=0,width=6.0cm}
}
\end{minipage}
\end{center}
\vspace*{-.35in} 
\hspace*{-.70in}
\begin{center}
\begin{minipage}{12.0cm}  
{\hspace*{-.2cm}\psfig{figure=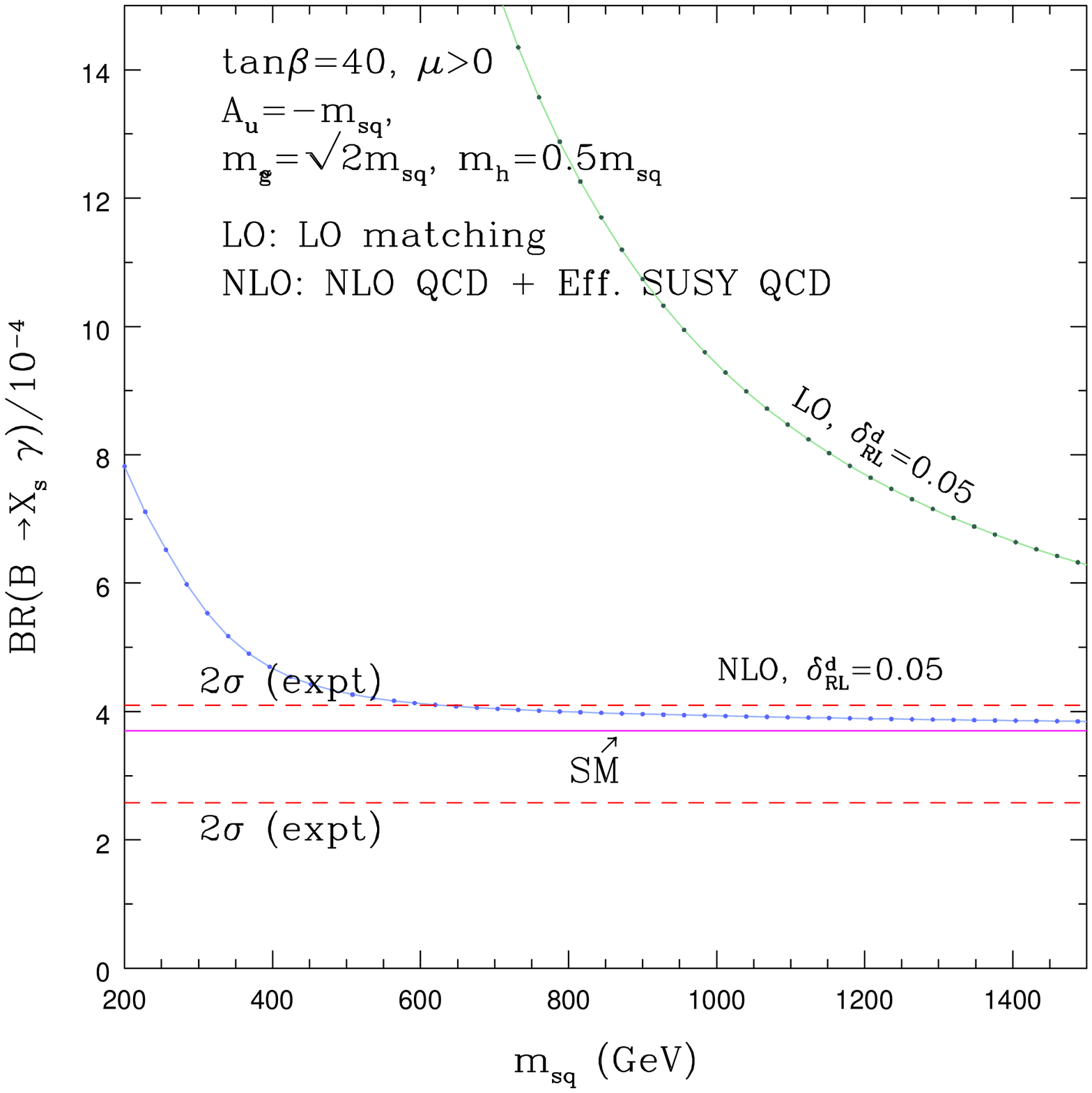, angle=0,width=6.0cm}
 \hspace*{-.2cm}\psfig{figure=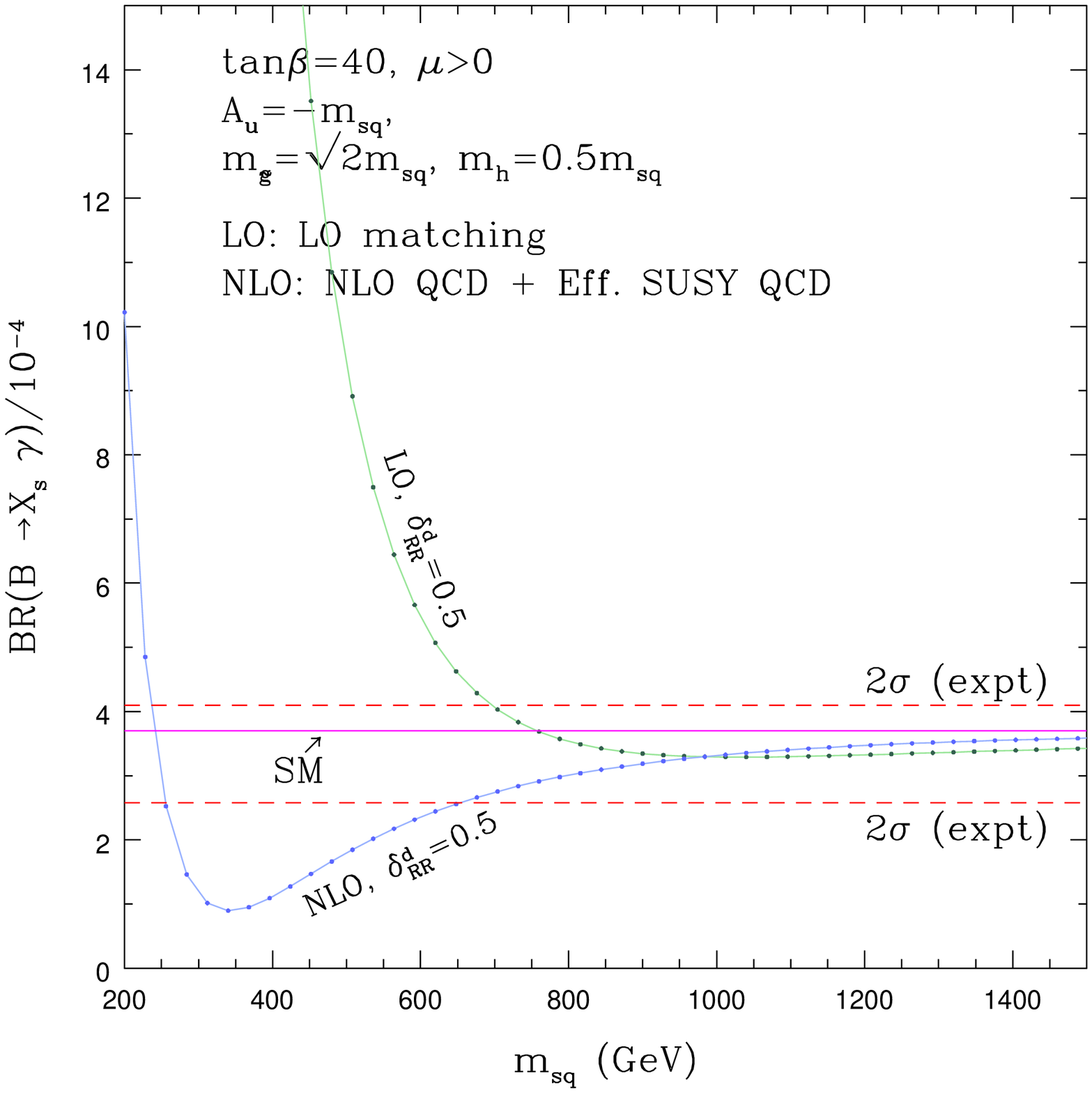, angle=0,width=6.0cm}
}
\end{minipage}
\caption{\label{fig:focus} {\small The branching ratio
    $\brbsgamma/10^{-4}$ \vs\ $m_{\widetilde q}=m_{\rm sq}=\msusy$ for
    in the LO and beyond--LO approximation for $\delta^d_{LL}=\pm0.2$
    (upper left), $\delta^d_{LR}=\pm0.05$ (upper right), $\delta^d_{RL}=0.05$
    (lower left) and $\delta^d_{RR}=0.5$
    (lower right). All other parameters as in
    fig.~\protect\ref{fig:cl7gll02}.
} }
\end{center}
\end{figure}

\begin{figure}[t!]
\begin{center}
\begin{minipage}{12.0cm}  
{
\hspace*{-.2cm}\psfig{figure=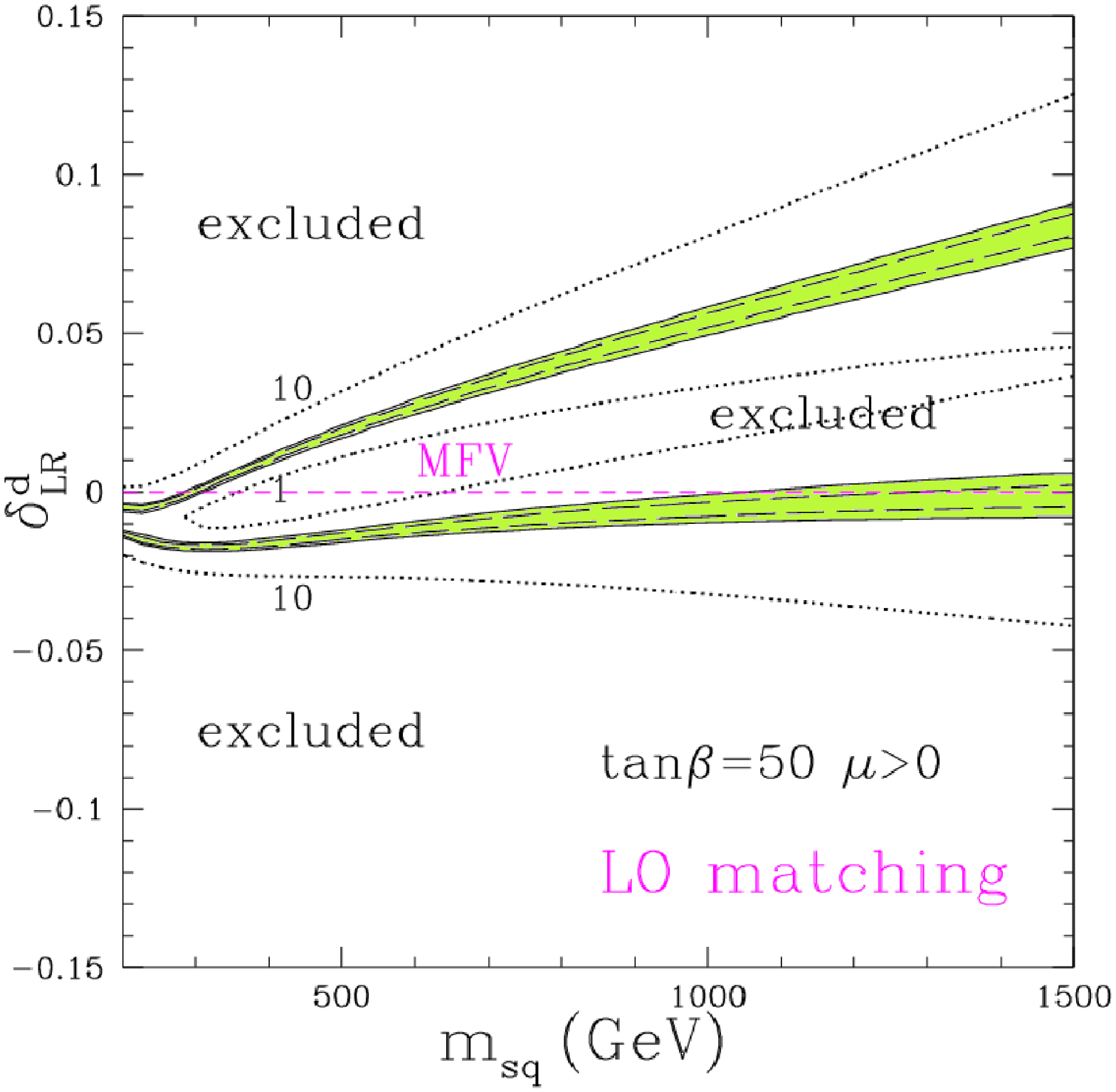, angle=0,width=6.0cm}
\hspace*{-.2cm}\psfig{figure=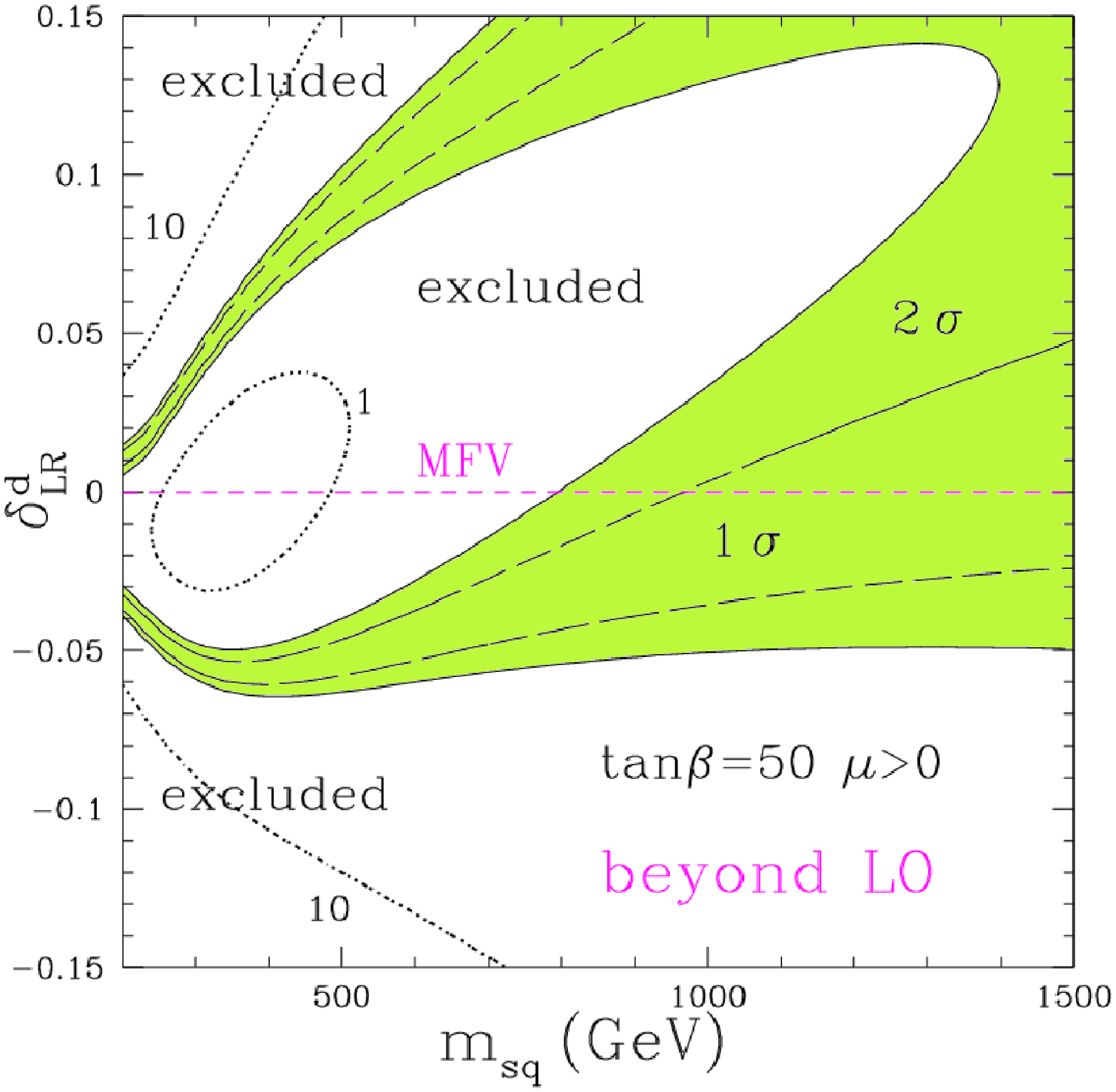, angle=0,width=6.0cm}
}
\end{minipage}
\caption{\label{fig:br-2-025-025-nlovslo} {\small Contours of the
    branching ratio $\brbsgamma/10^{-4}$ in the plane of
    $m_{\widetilde q}=m_{\rm sq}=\msusy$ and $\delta^d_{LR}$ in LO
    approximation (left) and with dominant beyond--LO corrections
    included (right). Regions of the plane of $m_{\widetilde q}$ and
    $\delta^d_{LR}$ where the branching ratio $\brbsgamma$ is
    consistent with experiment, eq.~(\protect\ref{bsgexptvalue:ref}),
    at $1\sigma$ ($2\sigma$) are delineated with a long--dashed
    (solid) lines. We take $\tanb=50$,
    $\mgluino=\sqrt{2}\,m_{\widetilde q}$, $A_d=0$,
    $A_u=-m_{\widetilde q}$, $\mu=0.5\, m_{\widetilde q}$, and
    $m_{H^+}=0.5\, m_{\widetilde q}$. The magenta short--dashed line
    marks the case of MFV. All the other $\delta^d$'s are set to
    zero. } }
\end{center}
\end{figure}

\begin{figure}[t!]
\begin{center}
\begin{minipage}{12.0cm}  
{
\hspace*{-.2cm}\psfig{figure=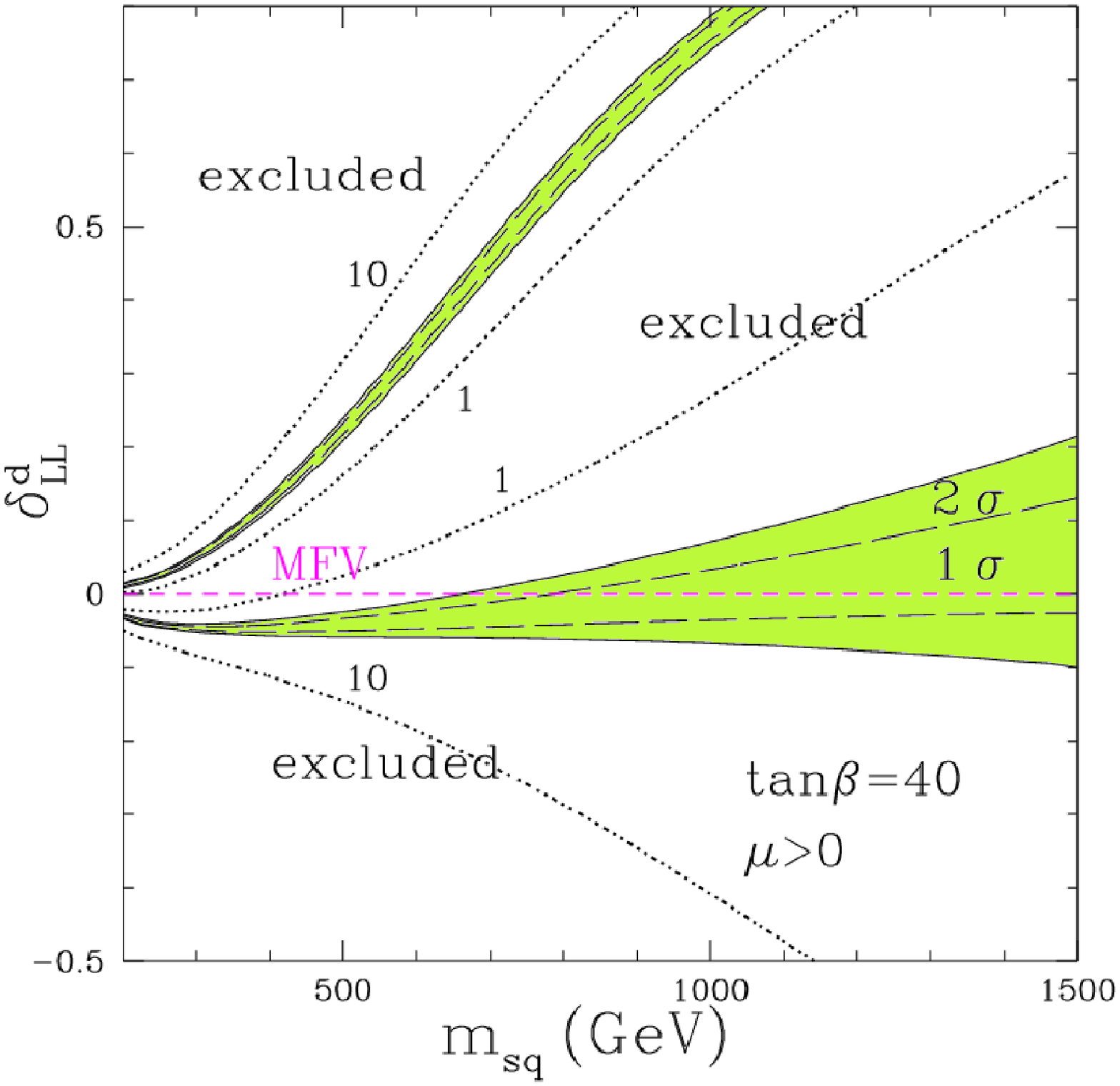, angle=0,width=6.0cm}
\hspace*{-.2cm}\psfig{figure=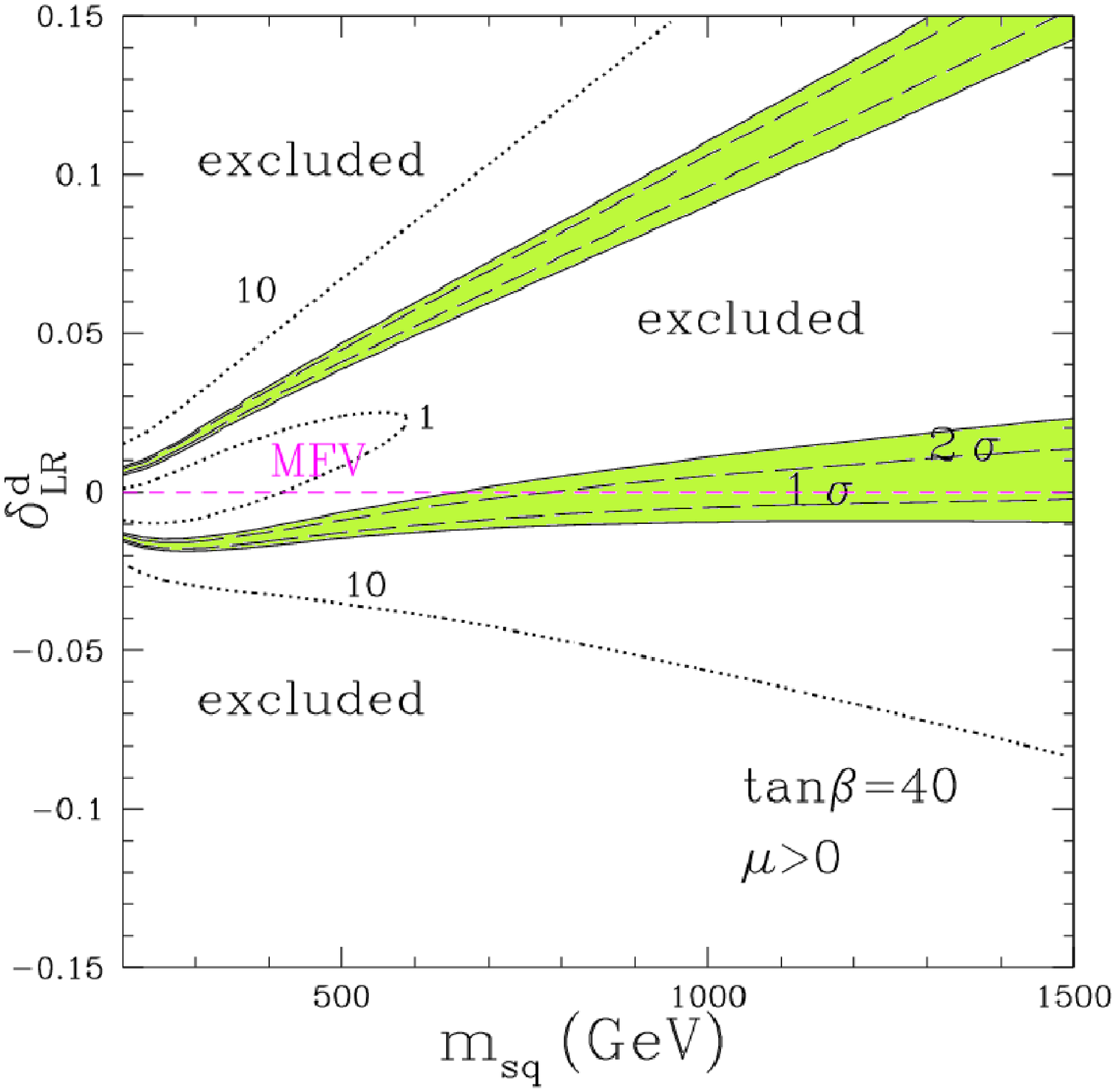, angle=0,width=6.0cm}
}
\end{minipage}
\end{center}
\vspace*{-.35in} 
\hspace*{-.70in}
\begin{center}
\begin{minipage}{12.0cm}  
{
\hspace*{-.2cm}\psfig{figure=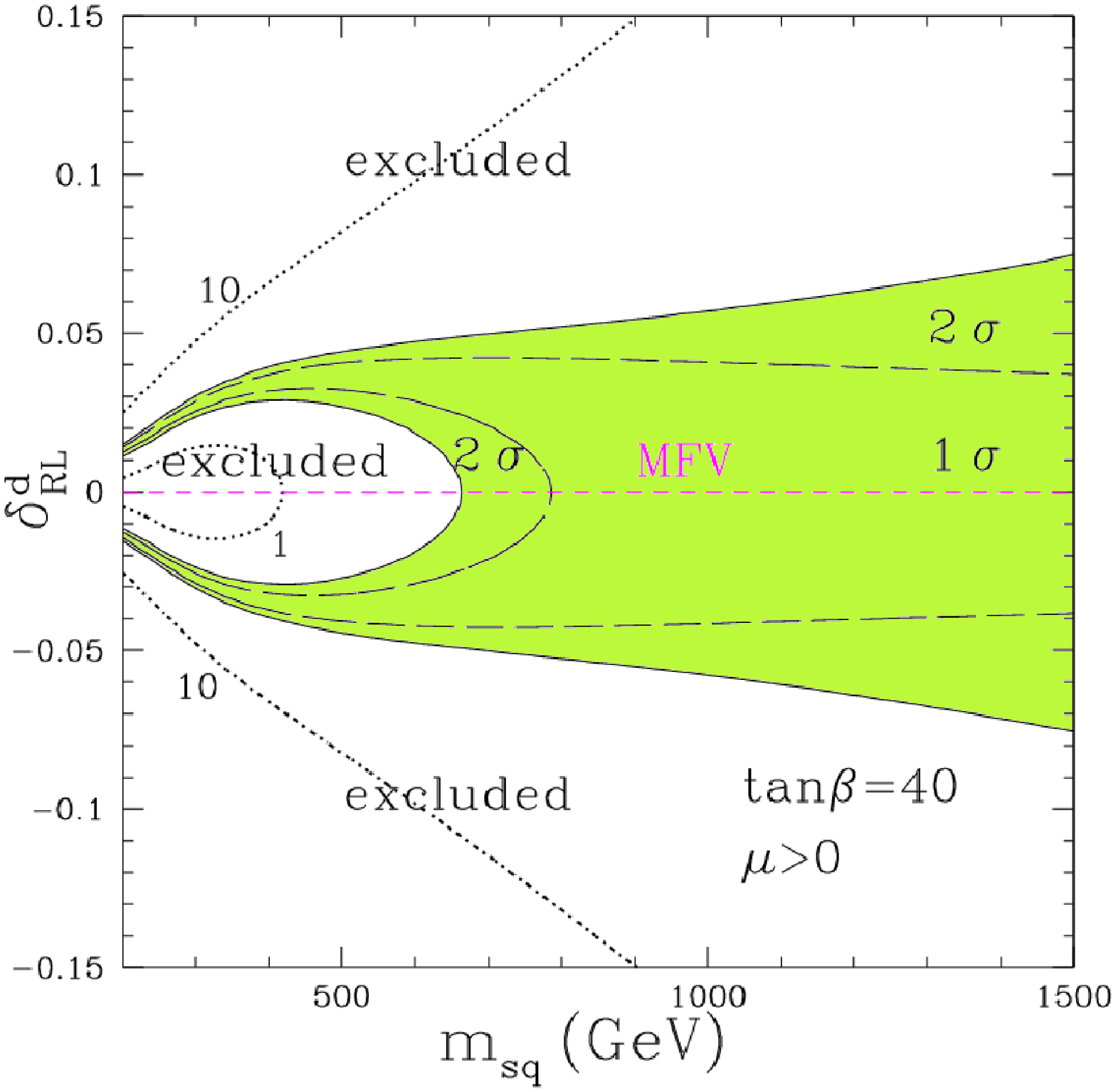, angle=0,width=6.0cm}
\hspace*{-.2cm}\psfig{figure=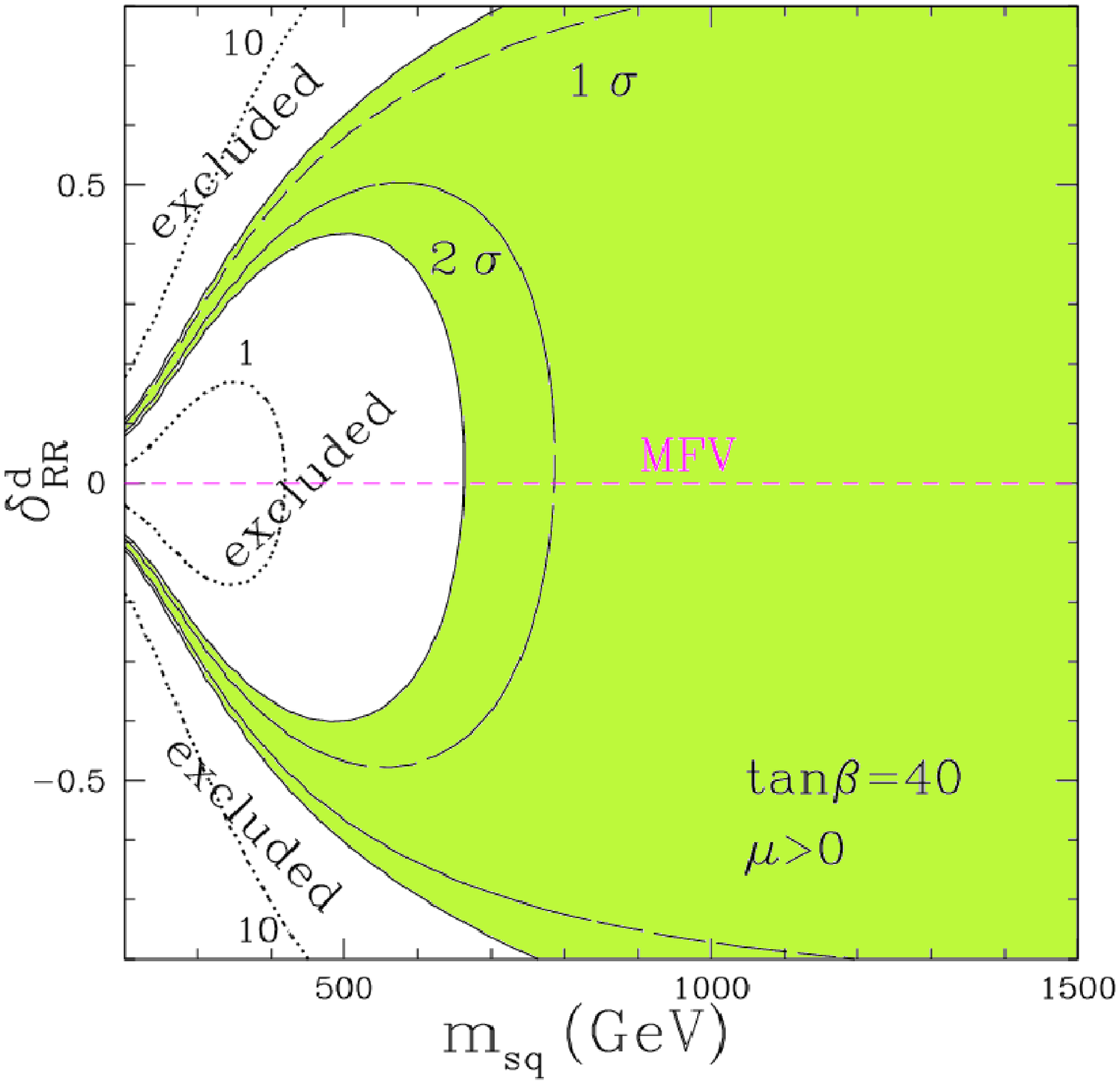, angle=0,width=6.0cm}
}
\end{minipage}
\caption{\label{fig:br-1-025-025} {\small Contours of the branching
    ratio $\brbsgamma/10^{-4}$ in the plane of $m_{\widetilde
    q}=m_{\rm sq}=\msusy$ and $\delta^d_{LL}$ (upper left),
    $\delta^d_{LR}$ (upper right), $\delta^d_{RL}$ (lower left) and
    $\delta^d_{RR}$ (lower right) for $\tanb=40$ and
    $\mgluino=m_{\widetilde q}$, $A_d=0$. All the other parameters are
    kept as in fig.~\protect\ref{fig:br-1-025-025}.  The long--dash
    (solid) curves delineate the regions which are consistent with
    experiment, eq.~(\protect\ref{bsgexptvalue:ref}), at $1\sigma$
    ($2\sigma$).
    } }
\end{center}
\end{figure}

\begin{figure}[t!]
\begin{center}
\begin{minipage}{12.0cm}  
{
\hspace*{-.2cm}\psfig{figure=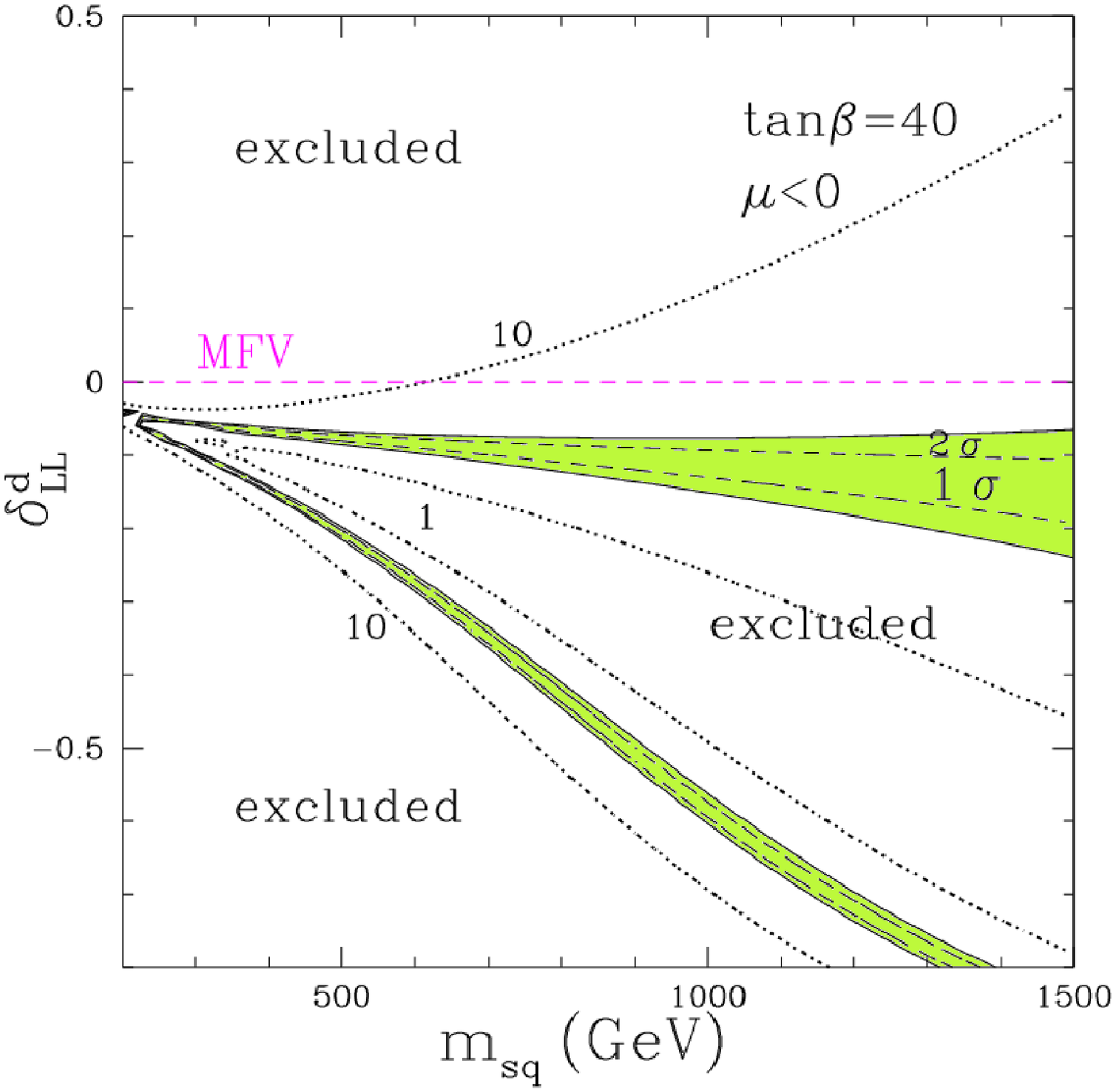, angle=0,width=6.0cm}
\hspace*{-.2cm}\psfig{figure=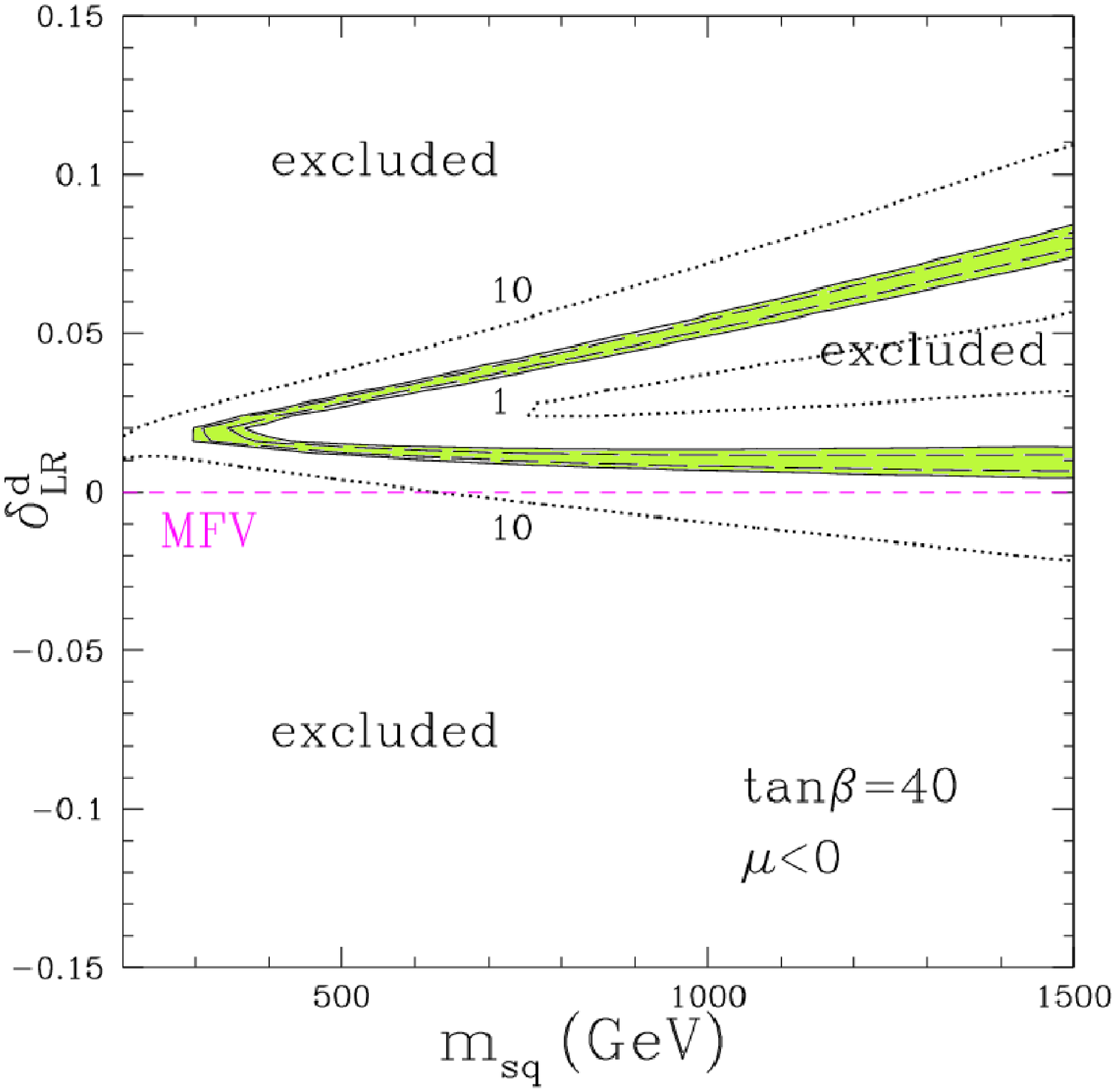, angle=0,width=6.0cm}
}
\end{minipage}
\caption{\label{fig:br-1-025-025-mun} {\small The same as in the
    upper two windows of fig.~\protect\ref{fig:br-1-025-025} but
    for $\mu<0$. In contrast, no ranges of $\delta^d_{RL}$ and
    $\delta^d_{RR}$ (lower two windows of
    fig.~\protect\ref{fig:br-1-025-025}) are consistent with
    experiment.  } }
\end{center}
\end{figure}


\vspace{0.15cm}
\noindent
\section{Results}\label{ref:results}

We will now demonstrate the effect of beyond--LO corrections to the Wilson
coefficients derived above and to $\brbsgamma$ with some representative examples.
We will work in the super--CKM basis and in deriving our numerical examples
will use the following parametrization: ${( m^{2}_{d,LL} )}_{ii}=
m^2_{\widetilde q}\,{\delta}_{ii} $, ${( m^{2}_{d,LL} )}_{ij}=
m^2_{\widetilde q}\, { (\delta^d_{LL} )}_{ij}$ (and similarly for the
$RR$ sector), ${( m^{2}_{d,LR} )}_{ii}= A_{d}\, diag(m_{d}, m_{s},
m_{b})$ and ${( m^{2}_{d,LR} )}_{ij}= m^2_{\widetilde q}\,
{(\delta^d_{LR})}_{ij}= m^2_{\widetilde q}\,
{(\delta^d_{RL})}^{\ast}_{ji}$, 
where $i,j=1,2,3$ and $i\neq j$. (Analogous parametrization is used
also in the up--sector.) In the case of $\bsgamma$, the relevant
mixings are those between the 2nd and 3rd generation squarks. For
simplicity, in this section we will use the notation
${\delta^d_{{LL}}}= \left({\delta^d_{{LL}}}\right)_{23}$, \etc\ In all
the cases presented in this section, we keep only one
$\delta^{d,u}_{.~.}$ non--zero at a time and set all other
$\delta^{d,u}_{.~.}$'s to zero.

First, we present the effect of including beyond--LO corrections
derived in the previous sections.  In the three panels of
fig.~\ref{fig:cl7gll02} we plot the Wilson coefficient $C_7(\muw)$
\vs\ $\tanb$ for $\delta^{d}_{LL}=0.2$, $\mu=500\gev$ and for other
relevant parameters as specified in the figure caption.  In
particular, $A_u$ is chosen such that an approximate cancelation
between the charged Higgs and the chargino contributions takes place
in the case of $\mu>0$ for $\delta^{d}_{LL}=0$, similarly to the case
of the Constrained MSSM (CMSSM).  The dashed (solid) line corresponds
to the LO (beyond--LO) value.  (Here the LO matching is defined so
that the SUSY contributions $\deltas C_{7,8}$ are evaluated at $\muw$,
and the NLO corrections $\deltas_{g,\widetilde{g}}C^{(1)}_{7,8}$ and
$\delta^{W,H}_{\widetilde{g}} C^{(1)}_{7,8}$ are neglected.) The
dash--dot line indicates the effect due to the NLO QCD correction
only.  Relative to the LO approximation, the chargino contribution,
which in this case dominates $C_7(\muw)$, is significantly reduced
towards zero. As a result, the overall value of $C_7(\muw)$ is also
reduced, although, like in the LO case, it does grow with
$\tanb$. (One finds a fairly similar effect for $C_8(\muw)$ as well.)

Next, in fig.~\ref{fig:cr7grl005} we plot $C_7^{\prime}(\muw)$ \vs\
$\tanb$ for $\delta^{d}_{RL}=0.05$, $\mu=500\gev$ and other parameters
as in the previous figure.  In this case, at LO the chargino contribution is zero
while beyond--LO effects generate it on the negative side, with the
magnitude growing with $\tanb$. In this sense, like before, the beyond--LO
chargino contribution is again significantly reduced relative to the
LO value.
However, at some point it out--balances the gluino
contribution which is also reduced but remains on the positive side
and is roughly independent of $\tanb$. The total $C_7^{\prime}(\muw)$
in this case decreases and at some point even changes sign. On the
other hand, the total $C_8^{\prime}(\muw)$ is reduced by a factor of
about two but remains positive and roughly independent of $\tanb$. The
main effect in $C_8^{\prime}(\muw)$ comes from reducing the gluino
part, while the chargino one is only slightly reduced down from zero.

Substantial beyond--LO effects also appear in the $LR$ and $RR$ sectors.
In order to illustrate this, in the four panels of
fig.~\ref{fig:clr78mup} we plot $C_{7,8}(\muw)$ and
$C_{7,8}^{\prime}(\muw)$ \vs\ $\tanb$ for $\delta^{d}_{LR}=0.05$ and
$\delta^{d}_{RR}=0.5$, respectively. As a rule, $C_{8}(\muw)$ and
$C_{8}^{\prime}(\muw)$, like in the previous cases, are reduced mostly because of a
strong suppression of the gluino contribution. In all the cases of
$C_{7,8}(\muw)$ and $C_{7,8}^{\prime}(\muw)$ presented in
fig.~\ref{fig:clr78mup}, one finds that, in addition, the chargino
contribution shifts from a positive, or zero, value down to an
increasingly negative one, although much less so for $C_{8}(\muw)$
than for $C_{7}(\muw)$.
 
In the case of $\mu<0$, the two competing effects from the RG running
and from the gluino--squark corrections 
lead to the
NLO--corrected Wilson coefficients either decreasing or increasing
with $\tanb$. This can be seen in the four panels of
fig.~\ref{fig:clr78mun} where all the other parameters are kept as
before. In general, one can see that beyond--LO corrections to the
Wilson coefficients are substantial. However, in contrast to the case
of $\mu>0$, the Wilson coefficients can now either be larger or
smaller relative to the LO approximation.

For comparison of the general GFM scenario with the limit of MFV, in
fig.~\ref{fig:cl7mfv} we plot $C_{7}(\muw)$ \vs\ $\tanb$ for the case
of MFV and for $\mu=\pm500\gev$ and the other parameters as
before. The coefficient $C_{8}(\muw)$ shows a similar behavior. It is
clear that the effect of including beyond--LO corrections is not as
striking as in GFM.

Large corrections to the Wilson coefficients in the case of GFM (but
not MFV) often leads to substantial changes in the predictions for
$\brbsgamma$. In fact, the overall tendency is to reduce the magnitude of
supersymmetric contributions relative to the LO, especially for large
$\tan\beta$ and $\mu>0$. This leads to the ``focusing effect'' of
concentrating on the SM value which
we have identified in the previous paper~\cite{or1}.
We illustrated the effect in fig.~\ref{fig:focus} where we plot
$\brbsgamma$ \vs\ $\msquark$ for some typical choices of parameters.
The focusing is indeed rather strong. In contrast, in the case of MFV
the effect is much less pronounced~\cite{or1}.

As explained in ref.~\cite{or1} and in sec.~\ref{sec:intro}, focusing
originates from two sources.  Firstly, renormalization group evolution
of $C_{7,8}$ from $\mususy$ to $\muw$ reduces the overall amplitude of
these coefficients.  In particular, the reduction of the strong
coupling constant at $\mususy$ has substantial effect on the gluino
contribution.  The NLO QCD matching at $\mususy$ brings further
suppression.  Secondly, the gluino loop contribution to the mass
matrix of down-type quark and the leading NLO SUSY--QCD corrections to
the Wilson coefficients at $\mususy$,~$\delta^{\chi,\wt{g}}_{\wt{g}}
C^{(\prime)(1)}_{7,8}$ have a common origin.  This considerably
reduces the LO gluino contribution to $C_{7,8}^{(\prime)}$, depending
on the sign of $\mu$.  At $\mu>0$ this correlation works as alignment,
which is enhanced by $\tan\beta$ and explains the bulk of the focusing
effect.  For $\mu<0$, SUSY--QCD corrections cause anti--alignment
instead, which competes with the overall suppression by
renormalization group evolution and results in small focusing (or even
de--focusing).  Some NLO suppression of SUSY contribution already
exits in MFV. However, flavor mixing is essential for the focusing
effect with GFM as described above.

One important consequence of the focusing effect is a significant
relaxation of bounds on the allowed amount of mixing in the squark
sector relative to the LO approximation.
We illustrate this in fig.~\ref{fig:br-2-025-025-nlovslo}
where we delineate the regions of the plane spanned by
$\delta^{d}_{LR}$ and $\msquark$ where the predicted values of
$\brbsgamma$ are consistent with experiment,
eq.~(\ref{bsgexptvalue:ref}), at $1\sigma$ (long--dashed) and
$2\sigma$ (solid). One can see a significant enlargement of the
allowed range of $\delta^{d}_{LR}$ relative to the LO approximation.
Similar strong relaxation due to NLO--level corrections is also present for $\deltadrl$
and $\deltadrr$, and somewhat less so in the case of $\deltadll$.

One should also note that, with even a small departure from the MFV
case one can significantly weaken the lower bounds on $m_{\widetilde
q}\sim\msusy$ derived in the limit of MFV. The effect is already
present in the LO approximation but it generally becomes further
strengthened by including the beyond--LO corrections considered in
this paper.  It results mostly from a competition between the chargino
and gluino contributions, the latter of which is in MFV absent.

The fact that lower limits on $\msusy$ obtained in the case of MFV are
highly unstable with respect to even small perturbations in the
defining assumptions~(\ref{eq:mfvdef}) of MFV, appears to be rather
generic. To show this, in the four panels of
fig.~\ref{fig:br-1-025-025} we plot contours of $\brbsgamma$ in the
planes of $\delta^{d}_{.~.}$ and $\msquark$ for $\tanb=40$, $\mu>0$
and all the other parameters as specified in the figure caption. The
bands delineated by dashed (solid) lines mark the $1\sigma$
($2\sigma$) regions consistent with
experiment~(\ref{bsgexptvalue:ref}). As before, only one
$\delta^{d,u}_{.~.}$ is kept different from zero in each case. Note
that in the regions in between the two allowed bands in the case of
$\delta^{d}_{LL}$ and $\delta^{d}_{LR}$ the branching ratio reaches a
shallow minimum whose position depends on the relative cancellation of
the chargino and gluino contributions, and which also depends on
$\tanb$ and other parameters. A similar shallow minimum appears in the
case of small $\delta^{d}_{RL}$ or $\delta^{d}_{RR}$ and smaller
$m_{\widetilde q}$.
(It is worth noting that the values of $\brbsgamma$ in the excluded
region between the two bands is are such that a fair decrease of the
experimental value would considerably reduce it.)  Note that the cases
of $\delta^{d}_{RL}$ and $\delta^{d}_{RR}$ the plots are almost
symmetric. This is because of the new gluino contributions that appear
in $C_{7,8}^{\prime}$ and which contribute to the branching ratio
quadratically without interfering with the MFV contributions in
$C_{7,8}$.  In these cases, new contributions always come
constructively. For $\mu<0$ (not shown in the figure) almost all of
the region is excluded.

In the plots, $\brbsgamma$ appears to be more sensitive to $\deltadlr$
and $\deltadrl$ than to $\deltadll$ and $\deltadrr$ in our
phenomenological approach.  However, note that, once we assume a
correlation between the A--terms and the Yukawa couplings, like in the
MFV, the natural scale of $\deltadlr$ ($\deltadrl$) reduces from one to
$m_b/\msusy \lsim {\cal O}(10^{-2})$.
 
Even though the focusing effect considerably reduces the gluino
contribution in the case of GFM, the branching ratio still shows
strong dependence on $\deltadll$ and $\deltadlr$ because of the
interference between the gluino contributions and the MFV
contributions to $C_{7,8}$.  In addition, a new LO chargino
contribution, other than the ones from the CKM mixing, is present in
the $\deltadll$ case because of the $SU(2)_L$ relation.

It is often claimed that the case of $\mu<0$ is inconsistent with
${\brbsgamma}_{expt}$~(\ref{bsgexptvalue:ref}). This is so because in the
case of MFV the chargino contribution adds to the SM/2HDM contribution
constructively for $\mu A_u>0$ and in unified models, like the CMSSM,
with the boundary conditions $A_0=0$ (or similar), the RG running generates
$A_u<0$ at $\mw$. As a result, theoretical predictions are
inconsistent with experiment, unless $\msusy\gsim2\tev$, or so.

However, the claim is only true in the case of the strict MFV but not
necessarily for even small departures from it, as can be seen in
fig.~\ref{fig:br-1-025-025-mun}. Indeed, for $\delta^{d}_{LL}\simeq
-0.1$ and/or $\delta^{d}_{LR}\simeq -0.02$ one can even evade the
bounds basically altogether. The effect is due to a partial
cancellation of the chargino and gluino contributions which in the
case of $\delta^{d}_{LL}$ can be efficient enough already in the LO
approximation but in the case of $\delta^{d}_{LR}$ would only allow
$m_{\widetilde q}$ down to some $500\gev$ but not lower. In contrast,
beyond LO one can have $m_{\widetilde q}$ as small as some $200\gev$,
as can be seen in fig.~\ref{fig:br-1-025-025-mun}.  On the other hand,
for no ranges of $-0.15\leq\delta^{d}_{RL}\leq0.15$ or
$-0.8\leq\delta^{d}_{RR}\leq 0.8$ and $m_{\widetilde q}\leq1.5\tev$
could we find any region consistent with experiment at either LO or
beyond--LO level.

In tables~\ref{tablell}--\ref{tablerl} we present allowed ranges of
the off--diagonal entries for a number of cases.  As can be seen in
figs.~\ref{fig:br-2-025-025-nlovslo}--\ref{fig:br-1-025-025-mun},
typically there are two bands. In the cases of $\delta^d_{LL}$ and
$\delta^d_{LR}$, one is around zero and one increasingly deviating
from zero as $\msusy$ increases. On the other hand, the allowed bands
of $\delta^d_{RL}$ and $\delta^d_{RR}$ tend to be roughly symmetric,
as mentioned above. As one can see from the tables, the allowed bands
do not necessarily increase with $\tanb$. This is because the
contributions from the chargino and the gluino show a different
dependence on $\tanb$, and their approximate cancellation, can happen
at different values of $\tanb$, as for example
fig.~\ref{fig:cr7grl005} illustrates.

\TABLE[h!] {  
 \caption{Allowed range of $\delta^d_{LL}$ beyond LO,
 which resides within the $2\sigma$ experimental error. 
 Parameters other than
 specified in the table are fixed to $M_1 = M_2 = m_{\wt{q}}$, $m_h = 0.5
 m_{\wt{q}}$, $A_u = -m_{\wt{q}}$ and $A_d = 0$.\label{tablell}}
 \begin{tabular}{ll@{}ll@{}ll@{}ll@{}l}
 \hline
 \hline
 \multicolumn{9}{l}
 {~$\delta^d_{LL}$ $( 10^{-2} )$}\\
 \hline
 \hline
$\tan\beta$
 &\multicolumn{4}{@{~}c@{~}}{ $\mu>0$}&\multicolumn{4}{c}{$\mu<0$}\\
 \cline{2-9}
$\,=40$
      &\multicolumn{2}{c@{~}}{$m_{\wt{q}}=500 {\rm GeV}$} 
       &\multicolumn{2}{c@{~}}{$m_{\wt{q}}=1000 {\rm GeV}$}
      &\multicolumn{2}{c@{~}}{$m_{\wt{q}}=500 {\rm GeV}$} 
       &\multicolumn{2}{c}{$m_{\wt{q}}=1000 {\rm GeV}$}


 \\
 \hline
 A    
 &$(-6, -3)$, &~$(21, 23)$ 
 &$(-7, 6)$, &~$(74, 77)$
 &$(-22, -20)$, &~$(-9, -7)$  
 &$(-50, -56)$, &~$(-15, -8)$\\
 B    
 &$(-7, -4)$, &~$(24, 27)$ 
 &$(-8, 7)$,  &~$(78, 82)$
 &$(-25, -23)$, &~$(-10, -8)$ 
 &$(-65, -61)$, &~$(-17, -10)$\\
 C    
 &$(-4, \phantom{-}1)$, &~$(26, 29)$ 
 &$(-6, 9)$, &~$(82, 86)$ 
 &$(-11, -10)$, &~$(-5, -4)$  
 &$(-38, -35)$, &~$(-10, -6)$\\
 D
 &$(-4, -2)$, &~$(22, 24)$
 &$(-2, 10)$, &~$(71, 75)$
 &$(-23, -22)$, &~$(-10, -9)$
 &$(-60, -56)$, &~$(-20, -14)$\\
 \hline
 \hline
$\tan\beta$ 
&\multicolumn{4}{@{~}c@{~}}{ $\mu>0$}&\multicolumn{4}{c}{$\mu<0$}\\
 \cline{2-9}
$\,=60$
      &\multicolumn{2}{c@{~}}{$m_{\wt{q}}=500 {\rm GeV}$} 
       &\multicolumn{2}{c@{~}}{$m_{\wt{q}}=1000 {\rm GeV}$}
      &\multicolumn{2}{c@{~}}{$m_{\wt{q}}=500 {\rm GeV}$} 
       &\multicolumn{2}{c}{$m_{\wt{q}}=1000 {\rm GeV}$}

 \\
 \hline
 A    
 &$(-7, -5)$, &~$(13, 15)$
 &$(-8, 1)$,  &~$(59, 61)$    
 &$\simeq -13$, &~$(-7, -6)$
 &$(-38, -36)$, &~$(-11, -8)$\\
 B    
 &$(-8, -6)$, &~$(15, 18)$
 &$(-9, 2)$, &~$(64, 69)$
 &$(-15, -14)$, &~$(-8, -7)$
 &$(-42, -39)$, &~$(-12, -9)$\\  
 C   
 &$(-5, -3)$, &~$(18, 21)$
 &$(-7,4)$, &~$(65, 69)$
 &\multicolumn{2}{c@{~}}{Not allowed} 
 &$(-14, -13)$, &~$(-5, -4)$\\
 D
 &$(-6, -4)$, &~$(13, 15)$
 &$(-5, 3)$, &~$(55, 60)$
 &$(-14, -13)$, &~$(-8, -7)$
 &$(-39, -36)$, &~$(-14, -11)$\\
 \hline
 \hline

 \multicolumn{9}{l}{}\\
 \end{tabular}

 \begin{tabular}{cccc}
 \hline
 \hline
   & $x_{\wt{g}}=(m_{\wt{g}}/m_{\wt{q}})^2$   
   & $x_{\mu}=(|\mu|/m_{\wt{q}})^2$
   & $x_{H}=(m_H/m_{\wt{q}})^2$ \\
 \hline
 Case A & 1 & 0.25 & 0.25\\ 
 Case B & 2 & 0.25 & 0.25\\
 Case C & 1 & 1 &0.25 \\
 Case D & 1 & 0.25 & 0.04\\
 \hline
 \hline

\end{tabular}

}


\TABLE[h!] {  
\label{tablerr}
\caption{Allowed range of $\delta^d_{RR}$ with the same parameter set
 as in Table \ref{tablell}.}
 \begin{tabular}{ll@{}ll@{}ll@{}ll@{}l}

 \hline
 \hline
 \multicolumn{9}{l}
 {~$\delta^d_{RR}$ $( 10^{-2} )$}\\
 \hline
 \hline
 &\multicolumn{4}{@{~}c@{~}}{ $\mu>0$}&\multicolumn{4}{c}{$\mu<0$}\\
 \cline{2-9}
\raisebox{1.5ex}[0pt]{$\tan\beta=40$}
      &\multicolumn{2}{c@{~}}{$m_{\wt{q}}=500 {\rm GeV}$} 
       &\multicolumn{2}{c@{~}}{$m_{\wt{q}}=1000 {\rm GeV}$}
      &\multicolumn{2}{c@{~}}{$m_{\wt{q}}=500 {\rm GeV}$} 
       &\multicolumn{2}{c}{$m_{\wt{q}}=1000 {\rm GeV}$}

 \\
 \hline
 A    
 &$(-59 , -41)$,&~$(43 , 61)$  
 &\multicolumn{2}{c@{~}}{$(-90 , 93)$}
 &\multicolumn{2}{c@{~}}{Excluded}
 &\multicolumn{2}{c}{Excluded}\\
 B    
 &\multicolumn{2}{c@{~}}{$(-97 , -93)$}               
 &\multicolumn{2}{c@{~}}{No constraint}
 &\multicolumn{2}{c@{~}}{Excluded}
 &\multicolumn{2}{c}{Excluded}\\
 C    
 &$(-30 , -9)$, &~$(11 , 31)$   
 &\multicolumn{2}{c@{~}}{$(-63 , 67)$}
 &\multicolumn{2}{c@{~}}{Excluded}
 &\multicolumn{2}{c}{Excluded}\\
 D
 &$(-50,-27)$, &~$(28, 51)$
 &\multicolumn{2}{c@{~}}{$(-50, 55)$}
 &\multicolumn{2}{c@{~}}{Excluded}
 &\multicolumn{2}{c@{~}}{Excluded}\\  
 \hline
 \hline
&\multicolumn{4}{@{~}c@{~}}{ $\mu>0$}&\multicolumn{4}{c}{$\mu<0$}\\
 \cline{2-9}
\raisebox{1.5ex}[0pt]{$\tan\beta=60$}
      &\multicolumn{2}{c@{~}}{$m_{\wt{q}}=500 {\rm GeV}$} 
       &\multicolumn{2}{c@{~}}{$m_{\wt{q}}=1000 {\rm GeV}$}
      &\multicolumn{2}{c@{~}}{$m_{\wt{q}}=500 {\rm GeV}$} 
       &\multicolumn{2}{c}{$m_{\wt{q}}=1000 {\rm GeV}$}

 \\
 \hline
 A    
 &$(-71 , -58)$, &~$(59 , 72)$
 &\multicolumn{2}{c@{~}}{No constraint}
 &\multicolumn{2}{c@{~}}{Excluded}
 &\multicolumn{2}{c}{Excluded}\\
 B    
 &\multicolumn{2}{c@{~}}{Excluded}
 &\multicolumn{2}{c@{~}}{No constraint}
 &\multicolumn{2}{c@{~}}{Excluded}
 &\multicolumn{2}{c}{Excluded}\\
 C    
 &$(-28 , -18)$, &~$(19 , 29)$
 &\multicolumn{2}{c@{~}}{$(-61 , 63)$}
 &\multicolumn{2}{c@{~}}{Not allowed}
 &\multicolumn{2}{c@{~}}{Excluded}\\
 D
 &$(-63,-50)$, &~$(50, 64)$
 &\multicolumn{2}{c@{~}}{$(-78, 81)$}
 &\multicolumn{2}{c@{~}}{Excluded}
 &\multicolumn{2}{c@{~}}{Excluded}\\
 \hline
 \hline
 \end{tabular}
}


 
\TABLE[h!] {  
 \label{tablelr}
\caption{Allowed range of $\delta^d_{LR}$ with the same paramter set
as in Table \ref{tablell}.}
 \begin{tabular}{ll@{}ll@{}ll@{}ll@{}l}
 \hline
 \hline
 \multicolumn{9}{l}
 {~$\delta^d_{LR}$ $( 10^{-2} )$}\\
 \hline
 \hline
 &\multicolumn{4}{@{~}c@{~}}{ $\mu>0$}&\multicolumn{4}{c}{$\mu<0$}\\
 \cline{2-9}
\raisebox{1.5ex}[0pt]{$\tan\beta=40$}
      &\multicolumn{2}{c@{~}}{$m_{\wt{q}}=500 {\rm GeV}$} 
       &\multicolumn{2}{c@{~}}{$m_{\wt{q}}=1000 {\rm GeV}$}
      &\multicolumn{2}{c@{~}}{$m_{\wt{q}}=500 {\rm GeV}$} 
       &\multicolumn{2}{c}{$m_{\wt{q}}=1000 {\rm GeV}$}
%

 \\
 \hline
 A    
 &$(-2 , -1)$, &~$(4 , 5)$      
 &$(-1 , 1)$, &~$\simeq 9$
 &$\simeq 1$, &~$\simeq 3$        
 &$\simeq 1$, &~$(5 , 6)$\\
 B    
 &$(-4 , -2 )$, &~$(8 , 9)$      
 &$(-3 , 3)$, &~$(17 , 20)$
 &$\simeq 2$, &~$\simeq 4$        
 &$(1 , 2)$, &~$\simeq 7$\\
 C    
 &$(-1 , \phantom{-}0 )$, &~$\simeq 5$      
 &$(-1 , 1)$, &~$(10 , 12)$
 &$\simeq 1$, &~$\simeq 2$        
 &$\simeq 1$, &~$(4 , 5)$\\
 D
 &$\simeq -1$, &~$(4, 5)$
 &$(0, 2)$, &~$(9, 11)$
 &$\simeq 2$, &~$\simeq 3$
 &$(1, 2)$, &~$(5, 6)$\\
 \hline
 \hline
&\multicolumn{4}{@{~}c@{~}}{ $\mu>0$}&\multicolumn{4}{c}{$\mu<0$}\\
 \cline{2-9}
\raisebox{1.5ex}[0pt]{$\tan\beta=60$}
      &\multicolumn{2}{c@{~}}{$m_{\wt{q}}=500 {\rm GeV}$} 
       &\multicolumn{2}{c@{~}}{$m_{\wt{q}}=1000 {\rm GeV}$}
      &\multicolumn{2}{c@{~}}{$m_{\wt{q}}=500 {\rm GeV}$} 
       &\multicolumn{2}{c}{$m_{\wt{q}}=1000 {\rm GeV}$}

 \\
 \hline
 A    
 &$(-3 , -2)$, &~$(4 , 5)$   
 &$(-2 ,  1)$, &~$(10, 13)$
 &$\simeq 2$, &~$\simeq 3$        
 &$\simeq 1$, &~$(4 , 5)$\\
 B    
 &$(-10 , -7)$, &~$(8 , 10)$    
 &\multicolumn{2}{c@{~}}{$(-11 , 17)$}
 &$\simeq 2$, &~$\simeq 4$        
 &$(1 , 2)$, &~$\simeq 6$\\
 C    
 &$(-2 , -1)$,  &~$(5 , 6)$    
 &$(-2 , 1)$, &~$(12 , 15)$
 &\multicolumn{2}{c@{~}}{Not allowed}       
 &$\simeq 1$, &~$\simeq 3$\\
 D
 &$(-3, -2)$, &~$(4, 5)$
 &$(-2, 1)$, &~$(10, 13)$
 &$\simeq 2$, &~$\simeq 3$
 &$\simeq 2$, &~$\simeq 5$\\
 \hline
 \hline
 \end{tabular}
}

 

\TABLE[h!] {  
\label{tablerl}
\caption{Allowed range of $\delta^d_{RL}$ with the same parameter set
 as in Table \ref{tablell}.}
\begin{tabular}{ll@{}ll@{}ll@{}ll@{}l}

 \hline
 \hline
 \multicolumn{9}{l}
 {~$\delta^d_{RL}$ $( 10^{-2} )$}\\
 \hline
 \hline
 &\multicolumn{4}{@{~}c@{~}}{ $\mu>0$}&\multicolumn{4}{c}{$\mu<0$}\\
 \cline{2-9}
\raisebox{1.5ex}[0pt]{$\tan\beta=40$}
      &\multicolumn{2}{c@{~}}{$m_{\wt{q}}=500 {\rm GeV}$} 
       &\multicolumn{2}{c@{~}}{$m_{\wt{q}}=1000 {\rm GeV}$}
      &\multicolumn{2}{c@{~}}{$m_{\wt{q}}=500 {\rm GeV}$} 
       &\multicolumn{2}{c}{$m_{\wt{q}}=1000 {\rm GeV}$}


 \\
 \hline
 A    
 &$(-5 , -3)$, &~$(3 , 4)$      
 &\multicolumn{2}{c@{~}}{$(-6 , 6)$}
 &\multicolumn{2}{c@{~}}{Excluded}
 &\multicolumn{2}{c@{~}}{Excluded}\\
 B    
 &$(-5 , -3)$, &~$(3 , 5)$      
 &\multicolumn{2}{c@{~}}{$(-6 , 5)$}
 &\multicolumn{2}{c@{~}}{Excluded}
 &\multicolumn{2}{c@{~}}{Excluded}\\
 C    
 &$(-4 , -1)$, &~$(1 , 4)$      
 &\multicolumn{2}{c@{~}}{$(-6 , 6)$}
 &\multicolumn{2}{c@{~}}{Excluded}
 &\multicolumn{2}{c@{~}}{Excluded}\\
 D
 &$(-4, -2)$, &~$(2, 4)$
 &\multicolumn{2}{c@{~}}{$(-4, 3)$}
 &\multicolumn{2}{c@{~}}{Excluded}
 &\multicolumn{2}{c@{~}}{Excluded}\\
 \hline
 \hline
&\multicolumn{4}{@{~}c@{~}}{ $\mu>0$}&\multicolumn{4}{c}{$\mu<0$}\\
 \cline{2-9}
\raisebox{1.5ex}[0pt]{$\tan\beta=60$}
      &\multicolumn{2}{c@{~}}{$m_{\wt{q}}=500 {\rm GeV}$} 
       &\multicolumn{2}{c@{~}}{$m_{\wt{q}}=1000 {\rm GeV}$}
      &\multicolumn{2}{c@{~}}{$m_{\wt{q}}=500 {\rm GeV}$} 
       &\multicolumn{2}{c}{$m_{\wt{q}}=1000 {\rm GeV}$}

 \\
 \hline
 A    
 &$(-4 , -3)$, &~$(3 , 4)$      
 &\multicolumn{2}{c@{~}}{$(-5 , 5)$}
 &\multicolumn{2}{c@{~}}{Excluded}
 &\multicolumn{2}{c@{~}}{Excluded}\\
 B    
 &$(-3 , -2)$, &~$(2 , 3)$      
 &\multicolumn{2}{c@{~}}{$(-4 , 3)$}
 &\multicolumn{2}{c@{~}}{Excluded}
 &\multicolumn{2}{c@{~}}{Excluded}\\
 C    
 &$(-7 , \phantom{-}4)$, &~$(4 , 7)$       
 &\multicolumn{2}{c@{~}}{$(-10 , 10)$}
 &\multicolumn{2}{c@{~}}{Not allowed}
 &\multicolumn{2}{c@{~}}{Excluded}\\
 D
 &$(-4, -3)$, &~$(3, 4)$
 &\multicolumn{2}{c@{~}}{$(-4, 4)$}
 &\multicolumn{2}{c@{~}}{Excluded}
 &\multicolumn{2}{c@{~}}{Excluded}\\
 \hline
 \hline

 \end{tabular}
 }
 \vspace{2em}


\begin{figure}[t!]
\begin{center}
\begin{minipage}{12.0cm}  
{
\hspace*{-.2cm}\psfig{figure=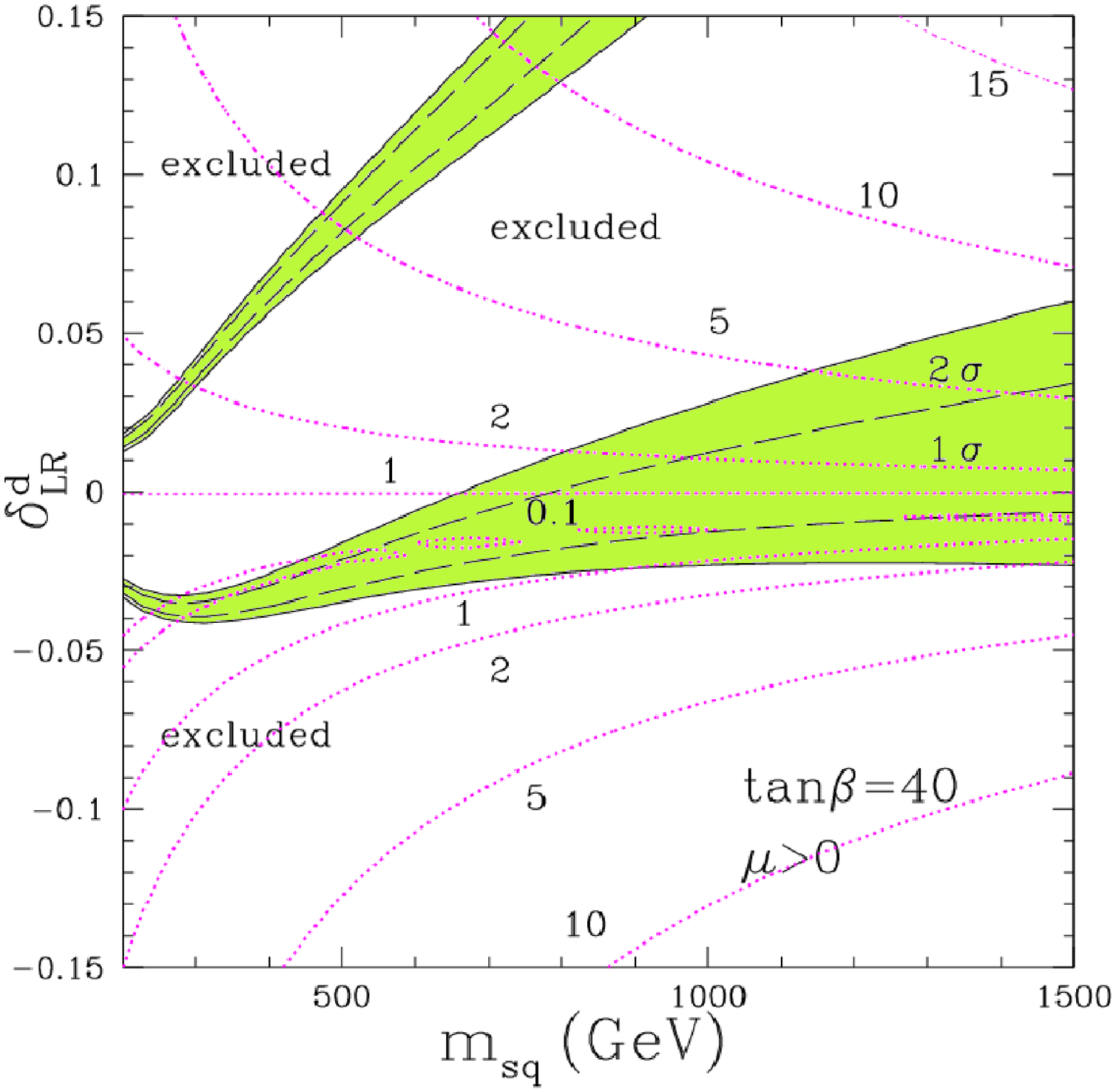, angle=0,width=6.0cm}
\hspace*{-.2cm}\psfig{figure=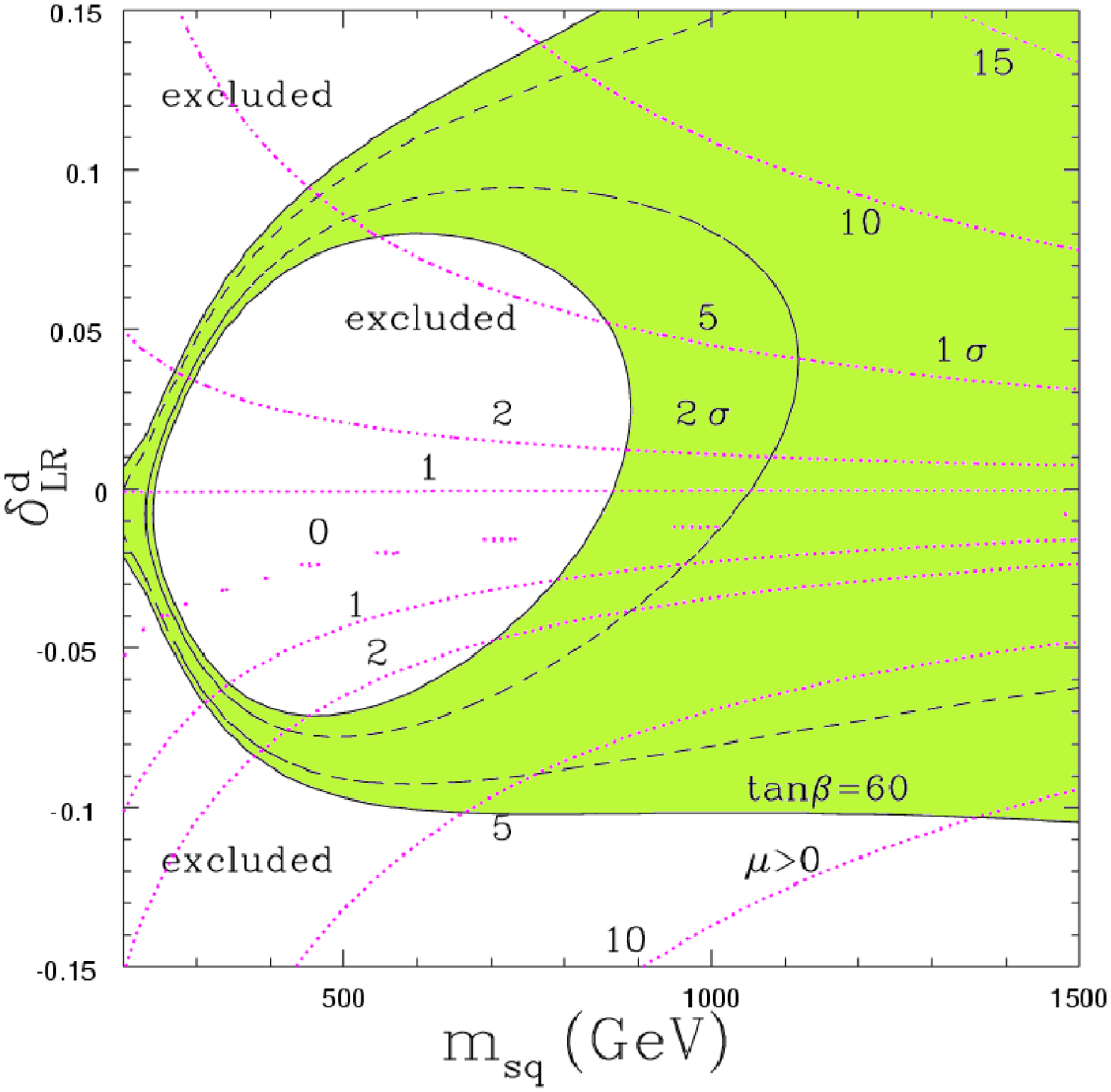, angle=0,width=6.0cm}
}
\end{minipage}
\caption{\label{fig:vcb-40-60-2-025-025-mun} {\small Contours of the
    ratio $K^{(0)}_{cb}/K_{cb}$ in the plane spanned by $m_{\widetilde
    q}=m_{\rm sq}$ and $\delta^d_{LR}$. We fix
    $\mgluino=\sqrt{2}\,m_{\widetilde q}$, $A_d=0$,
    $A_u=-m_{\widetilde q}$, $\mu=0.5\, m_{\widetilde q}$, and
    $m_{H^+}=0.5\, m_{\widetilde q}$ and show two cases $\tanb=40$
    (left) and $\tanb=60$ (right).
} }
\end{center}
\end{figure}

The significant enlargement of the allowed parameter space in the
framework of GFM, relative to MFV, leads to a strong relaxation of
constraints on the allowed ranges of the ``bare'' CKM matrix element
$K^{(0)}_{cb}$ relative to $K_{cb}\simeq 0.04 $ which is determined by
experiment. In the two windows of
fig.~\ref{fig:vcb-40-60-2-025-025-mun} we present contours of the
ratio $K^{(0)}_{cb}/K_{cb}$ in the plane spanned by $m_{\widetilde
q}=m_{\rm sq}=\msusy$ and $\delta^d_{LR}$. We fix
$\mgluino=\sqrt{2}\,m_{\widetilde q}$, $A_d=0$, $A_u=-m_{\widetilde
q}$, $\mu=0.5\, m_{\widetilde q}$, and $m_{H^+}=0.5\, m_{\widetilde
q}$ and show two cases $\tanb=40$ (left) and $\tanb=60$ (right). One
can see that the ratio $K^{(0)}_{cb}/K_{cb}$ can exceed a factor of
ten, or more, for large enough, but still allowed, ranges of $\delta^d_{LR}$.
It is also interesting that the ratio can be much smaller than one and
that it can even ``cross'' zero\footnote{By definition, $K^{(0)}_{cb}$
is positive in the standard notation~\cite{pdg}. Therefore the sign
change can always be rotated away by redefinition of (s)quark fields
and does not appear in the figure.}  in the allowed region between the
two contours of 1. In this case the measured value of $K^{(0)}_{cb}$
is of purely radiative origin and is generated by GFM in
supersymmetry~\cite{banks87}. 
On the other hand, for large $\delta^d_{LL}\lsim0.8$ the ratio does
not exceed a factor of three.

\noindent
\section{Conclusions}\label{ref:conclusions}

The squark--gluino loop correction that appears beyond the LO in the
inclusive process $\bsgamma$ shows a large effect in the MSSM with
general flavor mixing. Its main feature is that of reducing the
magnitude of supersymmetric (mostly gluino) contribution to
$\brbsgamma$ relative to the LO approximation. This focusing effect
leads to a considerable relaxation of experimental constraints on
flavor mixing terms and on the lower bounds on $\msusy$ obtained in
the limit of MFV, to the extent of even allowing small $\msusy$ at
$\mu<0$.

Given such a large effect appearing beyond LO, one may question the
validity of perturbation theory in the analysis presented
here. However, in this case the mechanism that generates it, namely the
large gluino--squark corrections, is absent at LO. We believe that the
mechanism is a dominant one at NLO, although we have examined it
within the limitations of our assumptions. It would therefore be
highly desirable to have available a complete NLO calculation of
$\brbsgamma$ in supersymmetry with GFM which would be applicable in more
general circumstances than those considered here.
It is also clear that, with improving experimental precision, 
an NLO-level analysis appears to be essential in testing SUSY models
of flavor at large $\tan\beta$, like, for example, those based on a SO(10) SUSY
GUT~\cite{bdr}. 
\section*{Acknowledgments}
We would like to thank T.~Blazek, P.~Gambino, G.F.~Giudice, T.~Hurth, 
O.~Lebedev and A.~Masiero for helpful comments.

\appendix 

\newpage

\section{Effective {\boldmath $W^{\pm}$} and {\boldmath $H^{\pm}$} Vertices}
\label{sec:muw:effective-vertices}

In this Appendix we summarize the effective vertices of
$W^{\pm}$--boson (unphysical scalar $G^{\pm}$) and charged Higgs boson
$H^{\pm}$ which arise from integrating out gluino--squark loops at the
matching scale $\mu_{SUSY}$.  All the coupling constants and masses
appearing in the following are evaluated in the effective SM at some
arbitrary scale $\mu$ although in actual expressions for Wilson
coefficients they will be evaluated at the scale $\mu_{SUSY}$.  We
work in the $\overline{MS}$--scheme except for the case explicitly
mentioned.  The renormalization group equations for evolving the
couplings in the effective SM to $\mu_{SUSY}$ from their values at
$\mu_W$ are given in~\cite{bbo93}.  We work in the physical super--CKM
basis where the loop--corrected mass matrices $\mbox{\boldmath $m_{
{d,u}}$}$ ~(\ref{eq:dquarkmassmatrix})--(\ref{eq:uquarkmassmatrix}) of
both the down-- and up--type quarks are diagonal.

\subsection{ {\boldmath $W^{\pm}$}--boson}

The effective $u\,d\,W$ vertex, evaluated at some scale $\mu$, is given by
\begin{eqnarray}
{\cal L} &=& -\frac{g_2}{\sqrt{2}} \sum^3_{i,j=1} 
\ovl{u}_i(\mu) \gamma^{\mu} W_{\mu} P_L
K^{\rm eff}_{ij} d_j + h.c. 
\label{eq:udweffcoup}\\
\end{eqnarray}
Note the effective CKM matrix $K^{\rm eff}_{ij}$ which appears
in~(\ref{eq:udweffcoup}), instead of $K_{ij}$ ($i,j=1,2,3$)
\begin{eqnarray}
\label{eq:keffdef}
K^{\rm eff}_{ij} &=& K_{ij} + \left(\frac{\alpha_s}{4\pi}\right) \Delta K_{ij}
,
\end{eqnarray}
where
\begin{eqnarray}
\Delta K_{ij} &=& 4 C_2(3) \sum^3_{k,l=1}\sum^6_{I,J=1}
{ (\Gamma_{u\,L}^{\ast}) }_{Ii} { (\Gamma_{u\,L}) }_{Ik} K_{kl} 
{ (\Gamma_{d\,L}^{\ast}) }_{Jl} { (\Gamma_{d\,L}) }_{Jj} 
 C_{24}(\mg^2,{m_{\wt{u}_I}^2},{m_{\wt{d}_J}^2};\mu) 
\nonumber\\
&& +\frac{1}{2}\sum^3_{k=1} 
     \left[ { (\Delta Z_{u\,L}^{\ast}) }_{ki} K_{kj}
         +K_{ik} { (\Delta Z_{d\,L}) }_{kj} \right].
\end{eqnarray}
The first term above is due to the $W$--boson coupling to the
intermediate squark fields, while in the remaining two the $W$--boson
couples to the external quark lines, and
\begin{eqnarray}
\label{eq:deltazuldef}
 (\Delta Z_{u\,L})_{ij}(\mu) &=& 2 C_2(3) \sum^6_{I=1}
   { (\Gamma_{u\,L}^{\ast}) }_{Ii}(\Gamma_{u\,L})_{Ij}
 B_1(\mg^2,m_{\wt{u}_I}^2, 0;\mu)
,
\\
\label{eq:deltazdldef}
(\Delta Z_{d\,L})_{ij}(\mu) &=& 2 C_2(3) \sum^6_{I=1}
   { (\Gamma_{d\,L}^{\ast}) }_{Ii} { (\Gamma_{d\,L}) }_{Ij}
 B_1(\mg^2,m_{\wt{d}_J}^2, 0;\mu)
.
\end{eqnarray}

It is the elements $K^{\rm eff}_{ij}$ that
are equal to the experimentally determined elements of the CKM
matrix. The elements of $K_{ij}$ are then determined
by~(\ref{eq:keffdef}) but the difference with $K^{\rm eff}_{ij}$ is
rather tiny.

Note that the effective vertex defined by~(\ref{eq:udweffcoup}) gives
finite contributions since $\ln\mu$ and infinite terms which appear in
the Passarino--Veltman functions, which are collected in
appendix~\ref{sec:PV-functions}, disappear in actual calculations.

\subsection{Unphysical scalar {\boldmath $G^{\pm}$}}

The effective vertices of the unphysical scalar $G^{\pm}$ with $d$ and
$u$, evaluated at the scale $\mu_W$ in the 't Hooft--Feynman gauge, are
given by

\begin{eqnarray}
{\cal L} &=& \sum^3_{i,j=1}\ovl{u}_i \left( 
C^{Gu_id_j}_L P_L + C^{Gu_id_j}_R P_R\right) G^{+} d_j + h.c. 
,
\end{eqnarray}
where the couplings $C^{G u_i d_j}_{ L,R}$ receive contributions from
the tree--level $C^{G u_i d_j (0)}_{ L,R}$ and from
squark--gluino loops $C^{G u_i d_j (1)}_{ L,R}$

\begin{eqnarray}
C^{Gu_id_j}_{L,R} &=& C^{Gu_id_j (0)}_{L,R} + \frac{\alpha_s(\mu_W)}{4\pi}
C^{Gu_id_j (1)}_{L,R}.
\end{eqnarray}
The former are given by
  
\beq
\label{eq:cduddef}
C^{Gu_id_j (0)}_L = \frac{g_2 {\ovl{m}_u}_i(\muw) K_{ij}}{\sqrt{2} m_W}
,~~~~~~~~~~~~~~
C^{Gu_id_j (0)}_R = -\frac{g_2 K_{ij} {\ovl{m}_d}_j(\muw)}{\sqrt{2} m_W}
,
\eeq
where ${\ovl{m}_u}_j(\muw)$ and ${\ovl{m}_d}_j(\muw)$ denote the
running quark masses at $\mu_W$.  For simplicity we have neglected
here a small rotation of the quark mass basis due to the
renormalization group evolution between $\muw$ and $\mususy$.

The corrections are given by

\begin{eqnarray}
C^{Gu_id_j (1)}_L &=& \eta^{\frac{3}{7}} \left[- 2 C_2(3) \sum^6_{I,J=1}
{ (\Gamma_{u\,R}^{\ast}) }_{Ii} { (\Gamma_{d\,L}) }_{Jj}
\left(C^{G\wt{d}_J\wt{u}_I}\right)^{\ast} 
\mg C_0(m_{\wt{u}_I}^2,m_{\wt{d}_J}^2,\mg^2) 
\right.
\nonumber\\
&&
\phantom{eta^{\frac{3}{7}}}
\left.
  -\frac{g_2}{\sqrt{2}m_W} \sum_{k=1}^3 {(\delta {\muq})}_{ik} K_{kj}
\right.
\nonumber\\
&&                    
\phantom{eta^{\frac{3}{7}}}
\left.
+\frac{1}{2}\sum^3_{k,l,m=1} C^{Gu_id_k (0)}_L
  \left[ (\Delta Z_{d\,L})_{kj}-K^{\ast}_{lk}{(\Delta Z_{u\,L})}_{lm}
    K_{mj} \right] 
\right],
\\
C^{Gu_id_j (1)}_R &=& \eta^{\frac{3}{7}}\left[ 2 C_2(3) \sum^6_{I,J=1}
  { (\Gamma_{u\,L}^{\ast}) }_{Ii} { (\Gamma_{d\,R}) }_{Jj}
  \left(C^{G\wt{d}_J\wt{u}_I}\right)^{\ast} 
                   \mg C_0(m_{\wt{u}_I}^2,m_{\wt{d}_J}^2,\mg^2) 
\right.
\nonumber\\
&&
\phantom{eta^{\frac{3}{7}}}
\left.
  +\frac{g_2}{\sqrt{2}m_W}\sum_{k=1}^3 K_{ik} {(\delta {\mdq^{\ast}})}_{jk}
\right.
\nonumber\\
&&
\phantom{eta^{\frac{3}{7}}}
\left.
                    -\frac{1}{2}\sum^3_{k,l,m=1}
                       \left[ (\Delta Z_{u\,L}^{\ast})_{ki}
                        -K_{il}(\Delta Z_{d\,L}^{\ast})_{ml}K^{\ast}_{km} \right]
                       C^{Gu_kd_j (0)}_R
\right]
,
\end{eqnarray}
where $C_0$ is a Passarino--Veltman function and the
expressions for $(\delta m_{d,u})_{ik}$ are given in and below
eq.~(\ref{eq:deltamq}). 

The factor $\eta= \alpha_s(\mususy)/\alpha_s(\muw)$ appears as the
result of running the couplings $C^{Gu_id_j (1)}_{L,R}$, which are
initially evaluated at $\mususy$, down to $\muw$, with the SM QCD RGE's
including six quark flavors.

The functions $\Delta Z_{u\,L}$ and $\Delta
Z_{d\,L}$ are defined in~(\ref{eq:deltazuldef})
and~(\ref{eq:deltazdldef}), and
\begin{eqnarray}
C^{G\wt{d}_I\wt{u}_J} &=& 
                           -\frac{g_2}
                             {\sqrt{2}m_W}
                            \sum^3_{k,l,m=1}
                            { (\Gamma_{d\,R}) }_{Ik}
                           \left[ { (m^{2\ast}_{d,LR}) }_{lk}
                                  -m_{d_k} \mu^{\ast}\tan\beta\, \delta_{kl}  
                           \right]K^{\ast}_{ml} { (\Gamma_{u\,L}^{\ast}) }_{Jm}
\nonumber\\
                       && +
                           \frac{g_2}
                           {\sqrt{2} m_W}
                           \sum^3_{k,l,m=1}
                            { (\Gamma_{d\,L}) }_{Ik} K^{\ast}_{lk}
                           \left[{ (m^2_{u,LR}) }_{lm}
                                  -m_{u_l} \mu\cot\beta\, \delta_{lm}
                           \right]
                            { (\Gamma_{u\,R}^{\ast}) }_{Jm}
.
\end{eqnarray}

Like in the case of the effective coupling of the $W$--boson above,
$\ln\mu$ and infinite terms in $B_0$ effectively do not contribute 
because of the  unitarity of the diagonalization matrices $\mbox{\boldmath $\Gamma$}$.

\subsection{Charged Higgs boson {\boldmath $H^{\pm}$}}

The effective $H^{\pm}$ vertices with $d$ and $u$, evaluated at the $\muw$, are given by
\begin{eqnarray}
{\cal L} &=& \sum^3_{i,j=1}\ovl{u}_i \left( C^{Hu_id_j}_L P_L
                   +C^{Hu_id_j}_R P_R \right) H^{+} d_j  + h.c.
,
\end{eqnarray}
where, similarly to the case of the unphysical scalar before, the
couplings $C^{H u_i d_j L,R}$ receive contributions from the
tree--level $C^{H u_i d_j (0)}_{L,R}$ and from
squark--gluino loops $C^{H u_i d_j (1)}_{L,R}$
\begin{eqnarray}
C^{Hu_id_j}_{L,R} &=& C^{Hu_id_j (0)}_{L,R} +
\frac{\alpha_s(\mu_W)}{4\pi}C^{Hu_id_j (1)}_{L,R} 
. 
\end{eqnarray}
The former are given by 
\beq
C^{Hu_id_j (0)}_L = \frac{g_2 {\muqw}_i K_{ij}}{\sqrt{2}
m_W}\cot\beta
,~~~~~~~~~~~~~~
C^{Hu_id_j (0)}_R = \frac{g_2 K_{ij} {\mdqw}_j}{\sqrt{2} m_W}\tan\beta
,
\eeq
where ${\muqw}_i$ and ${\mdqw}_i$ have been defined below
eq.~(\ref{eq:cduddef}).
The corrections are given by 
\begin{eqnarray}
C^{Hu_id_j (1)}_L &=& \eta^{\frac{3}{7}} \left[
                   2 C_2(3) \sum^6_{I,J=1}
  { (\Gamma_{u\,R}^{\ast}) }_{Ii} { (\Gamma_{d\,L}) }_{Jj}
                   \left(C^{H\wt{d}_J\wt{u}_I}\right)^{\ast} 
                   \mg C_0(m_{\wt{u}_I}^2,m_{\wt{d}_J}^2,\mg^2) 
\right.
\nonumber\\
&&
\phantom{\eta^{\frac{3}{7}}}
\left.
   +\frac{g_2\cot\beta}{\sqrt{2}m_W}
                        \sum_{k=1}^3 {(\Delta {\muq})}_{ik}  K_{kj}
\right.
\nonumber\\
&&
\phantom{\eta^{\frac{3}{7}}}
\left.
                    +\frac{1}{2}\sum^3_{k,l,m=1} C^{Hu_id_k (0)}_L
                       \{ (\Delta Z_{d\,L})_{kj}-K^{\ast}_{lk}
                                               (\Delta Z_{u\,L})_{lm}
                                               K_{mj} \}
\right]
,
\\
C^{Hu_id_j (1)}_R &=& \eta^{\frac{3}{7}}
\left[ -2 C_2(3) \sum^6_{I,J=1}
                   { (\Gamma_{u\,L}^{\ast}) }_{Ii} { (\Gamma_{d\,R}) }_{Jj}
                   \left(C^{H\wt{d}_J\wt{u}_I}\right)^{\ast}
                   \mg C_0(m_{\wt{u}_I}^2,m_{\wt{d}_J}^2,\mg^2) 
\right.
\nonumber\\
&&
\phantom{\eta^{\frac{3}{7}}}
\left.
                  +\frac{g_2\tan\beta}{\sqrt{2}m_W}
                    \sum_{k=1}^3K_{ik}{(\Delta {{\mdq}^{\ast}})}_{jk}
\right.
\nonumber\\
&&
\phantom{\eta^{\frac{3}{7}}}
\left.
                    -\frac{1}{2}\sum^3_{k,l,m=1}
                       \{ (\Delta Z_{u\,L}^{\ast})_{ki}-K_{il}
                                     (\Delta Z_{d\,L}^{\ast})_{ml}
		       K^{\ast}_{km} \} 
                       C^{Hu_kd_j (0)}_R 
\right]
,
\end{eqnarray}
As in the previous subsection, the factor $\eta$ appears as the
result of running the couplings $C^{Hu_id_j (1)}_{L,R}$, which are
initially evaluated at $\mususy$, down to $\muw$, with SM QCD RGE's
assuming six flavors.

\begin{eqnarray}
C^{H\wt{d}_I\wt{u}_J} &=& \frac{g_2}
                          {\sqrt{2}m_W}
                          \sum^3_{k,l,m=1}
                          { (\Gamma_{d\,R}) }_{Ik}
                          \left[ { (m^{2\ast}_{d,LR}) }_{lk}\tan\beta
                                   -\mu^{\ast} {m_{d_k}} \delta_{kl} 
                           \right] K^{\ast}_{ml}{ (\Gamma_{u\,L}^{\ast})} _{Jm}
\nonumber\\
                       &&  
                           +\frac{g_2}
                           {\sqrt{2} m_W}
                           \sum^3_{k,l,m=1}
                           { (\Gamma_{d\,L}) }_{Ik} K^{\ast}_{lk}
                           \left[ { (m^2_{u,LR}) }_{lm}\cot\beta  
                                  - \mu {m_{u_l}} \delta_{lm} 
                           \right]
                           { (\Gamma_{u\,R}^{\ast}) }_{Jl}
.
\end{eqnarray}

\section{Effective {\boldmath $\chi^{\pm}$}, {\boldmath $\chi^0$}
                          and {\boldmath $\wt{g}$} Vertices}
\label{sec:Yukawa-correction}

We summarize here the effective vertices of the chargino $\chi^{\pm}$,
neutralino $\chi^0$ and gluino $\wt{g}$ fields, evaluated at
$\mususy$. Notice that the corrections to the tree--level vertices
come as a result of using gluino--squark loop--corrected values for the
quark and squark quantities appearing in the expressions below.

\subsection{Chargino {\boldmath $\chi^{\pm}$}}

The effective quark--squark--chargino vertices are given by
\begin{eqnarray}
{\cal L} &=&\sum^3_{i=1}\sum^2_{a=1}\sum^6_{I=1}
 \ovl{d}_i\left[ { (C_{d\,L}) }_{iaI} P_L
+ { (C_{d\,R}) }_{iaI} P_R \right] \chi^{-}_a
\wt{u}_I + h.c.
,
\end{eqnarray}
where 
\begin{eqnarray}
{ (C_{d\,L}) }_{iaI} &=& \frac{g_2}{\sqrt{2}m_W} 
  \sum^3_{k=1} \frac{ {(m_d^{(0)})}_{ik}}{ \cos\beta}
                     U_{a2} { (\Gamma_{u\,L}^{\ast}) }_{Ik}                    
,
\\
{ (C_{d\,R}) }_{iaI} &=& -g_2 V_{a1} { (\Gamma_{u\,L}^{\ast}) }_{Ii}
 + \frac{g_2}{\sqrt{2}m_W} \sum^3_{k=1}\frac{ {(m_u^{(0)\ast})}_{ki}}
                                      {\sin\beta}
                       V_{a2} { (\Gamma_{u\,R}^{\ast}) }_{Ik}
,
\end{eqnarray}
and $V$ and $U$ are the $2\times2$ two matrices with which one
diagonalizes the chargino mass matrix.

\subsection{Neutralino {\boldmath $\chi^0$}}

The corrected quark--squark--neutralino vertices are given by

\begin{eqnarray}
{\cal L} &=&\sum^3_{i=1}\sum^4_{r=1}\sum^6_{I=1}
 \ovl{d}_i\left[(N_{d\,L})_{irI} P_L
+(N_{d\,R})_{irI} P_R \right] \chi^{0}_r \wt{d}_I  + h.c.
,
\end{eqnarray}
where
\begin{eqnarray}
(N_{d\,L})_{irI} &=& -\sqrt{2}g_2\left[\frac{1}{3}\tan\theta_W
                           N_{r1}^{\ast}{ (\Gamma_{d\,R}^{\ast}) }_{Ii}
                           +\frac{(m^{(0)}_d)_{ik}}{2m_W\cos\beta}
                           N_{r3}^{\ast}{ (\Gamma_{d\,L}^{\ast}) }_{Ik}
                                \right]
,
\\
(N_{d\,R})_{irI} &=& -\sqrt{2}g_2\left[ \left(-\frac{1}{2}N_{r2}
                                    +\frac{1}{6}\tan\theta_W N_{r11}\right)
                                  { (\Gamma_{d\,L}^{\ast}) }_{Ii}
\right.
\nonumber\\
&&\left.
\phantom{-\sqrt{2}g_2[}
                                 +\frac{(m^{(0)\ast}_d)_{ki}}{2m_W \cos\beta}
                                  N_{r3}{ (\Gamma_{d\,R}^{\ast}) }_{Ik} 
                                \right],
\end{eqnarray}
and $N$ is the  $4\times4$ matrix with which one
diagonalizes the neutralino mass matrix~\cite{gh84}.

\subsection{Gluino {\boldmath $\wt{g}$}}
The effective quark--squark--gluino vertices are given by
\begin{eqnarray}
{\cal L} &=& \sum^3_{i=1}\sum^6_{I=1}
\ovl{d}_i
\left[
 { (G_{d\,L}) }_{iI} P_L + { (G_{d\,R}) }_{iI} P_R
\right] \wt{d}_I \wt{g}
\nonumber\\
&+&\sum^3_{i=1}\sum^6_{I=1}
\ovl{u}_i
\left[
 { (G_{u\,L}) }_{iI} P_L + { (G_{u\,R}) }_{iI} P_R
\right] \wt{u}_I \wt{g}
+ h.c.
,
\end{eqnarray}
where
\begin{equation}
{ (G_{dL}) }_{iI} = -\sqrt{2} g_s { (\Gamma_{dR}^{\ast}) }_{Ii}
,~~~~~~~
{ (G_{dR}) }_{iI} =  \sqrt{2} g_s { (\Gamma_{dL}^{\ast}) }_{Ii}
,
\end{equation}
and analogously for the up--type sector.

The above effective couplings for the chargino,
neutralino and gluino reduce to their tree--level values by making the
following replacements: $\left(\Gamma_{du\,LR}\right)_{Ii}\ra
\left(\Gamma_{du\,LR}^{(0)}\right)_{Ii}$ and
$\left(m^{(0)}_{d,u}\right)_{ij}\ra
\left(m_{d,u}\right)_{ij}\delta_{ij}$.



\section{Passarino-Veltman functions}
\label{sec:PV-functions}

We present here explicit expressions for the Passarino-Veltman
functions~\cite{vpfunction} which appear in our calculation.
We take all the external momenta equal to zero.

\begin{eqnarray}
B_0(x,y;\mu) &=&
 \frac{1}{\epsilon}+2\ln{\mu}
 -1-\frac{x\ln{x}-y\ln{y}}{x-y}
,
\\
B_1(x,y;\mu) &=& 
 -\frac{1}{2}\frac{1}{\epsilon}-\ln{\mu} -\frac{1}{4}\frac{3x-y}{x-y}
\nonumber\\
&&+\frac{x^2}{(x-y)^2}\ln{x}-\frac{1}{2}\frac{(2x-y)y}{(x-y)^2}\ln{y}
,
\\
C_0(x,y,z) &=& -\left\{ \frac{x\ln{x}}{(y-x)(z-x)} \right.
\nonumber\\
                 && \left.            +\frac{y\ln{y}}{(z-y)(x-y)}
                             +\frac{z\ln{z}}{(x-z)(y-z)} \right\}
,
\\
C_{24}(x,y,z;\mu) &=& \frac{1}{4}\left\{
           \frac{1}{\epsilon}+2\ln{\mu}
           +\frac{3}{2} \right.
           -\frac{x^2\ln{x}}{(x-y)(x-z)}
\nonumber\\
        &&   \left. -\frac{y^2\ln{y}}{(y-x)(y-z)}
           -\frac{z^2\ln{z}}{(z-x)(z-y)}
           \right\}
.
\end{eqnarray}

\section{Mass functions for the Wilson coefficients}
\label{sec:mass-functions}

In this section we collect several auxiliary functions which appear in
sec.~\ref{sec:wilson}.  The mass functions for the SM and 2HDM
contributions to $C_{7,8}$ are given in~\cite{cdgg97},
\begin{align}
F_7^{(1)}(x) =&
\frac{x(7-5x-8x^2)}{24(x-1)^3}+\frac{x^2(3x-2)}{4(x-1)^4}\ln x ,\\
F_8^{(1)}(x) =&
\frac{x(2+5x-x^2)}{8(x-1)^3}-\frac{3x^2}{4(x-1)^4}\ln x ,\\ 
%
%
F_7^{(2)}(x) =& \frac{x(3-5x)}{12(x-1)^2}+\frac{x(3x-2)}{6(x-1)^3}\ln
x ,\\
F_8^{(2)}(x) =& \frac{x(3-x)}{4(x-1)^2}-\frac{x}{2(x-1)^3}\ln x .
\end{align}

The mass functions and related constants for the $\chi^-$, $\chi^0$ and
 $\wt{g}$ contributions to $C_{7,8}$ are given in~\cite{bmu00},

\begin{align}
H^{[7]}_1(x) =& \frac{-3x^2+2x}{6(1-x)^4}\ln x 
+ \frac{-8x^2-5x+7}{36(1-x)^3} ,\\
H^{[8]}_1(x) =& \frac{x}{2(1-x)^4}\ln x
+\frac{-x^2+5x+2}{12(1-x)^3} ,\\
%
H^{[7]}_2(x) =& \frac{-3x^2+2x}{3(1-x)^3}\ln x
+\frac{-5x^2+3x}{6(1-x)^2},\\
H^{[8]}_2(x) =& \frac{x}{(1-x)^3}\ln
x+\frac{-x^2+3x}{2(1-x)^2},\\
%
H^{[7]}_3(x) =& -\frac{1}{3}H^{[8]}_1(x), \\
H^{[8]}_3(x) =& H^{[8]}_1(x),\\
%
H^{[7]}_4(x) =& -\frac{1}{3}\left( H^{[8]}_2(x)+\frac{1}{2}\right), \\
H^{[8]}_4(x) =& H^{[8]}_2(x)+\frac{1}{2},\\
%
H^{[7]}_5(x) =& -\frac{1}{3}H^{[8]}_1(x),\\
H^{[8]}_5(x) =& \frac{9x^2-x}{16(1-x)^4}\ln
x+\frac{19x^2+40x-11}{96(1-x)^3},\\
%
H^{[7]}_6(x) =& -\frac{1}{3}\left(H^{[8]}_2(x)+\frac{1}{2}\right),\\
H^{[8]}_6(x) =& \frac{9x^2-x}{8(1-x)^3}\ln x+\frac{13x-5}{8(1-x)^2},\\
\lambda^{[7]} =& \frac{5}{6},~~~~~\lambda^{[8]} = \frac{1}{2}.
\end{align}

\section{NLO QCD corrections to the Wilson coefficients}
\label{sec:nlo-qcd-corrections}

For completeness, in this section we collect the NLO QCD corrections
to the Wilson coefficients given in~\cite{cdgg97,bmu00}.

The NLO QCD corrections to the Wilson
coefficients in the SM are summarized as~\cite{cdgg97},
\begin{equation}
\delta^{W}_{g} C_{i}^{(1)}(\muw) =
\begin{cases}
 15+6\ln \frac{\muw^2}{m_W^2}, & \text{for}~~~i=1, \\
 0,                            & \text{for}~~~i=2,3,5,6, \\
 E\left(\frac{\ovl{m}_t^2(\muw)}{m_W^2}\right)-\frac{2}{3}+\frac{2}{3}\ln
 \frac{\muw^2}{m_W^2}, &  \text{for}
 ~~~i=4, \\
 G_{i}\left(\frac{\ovl{m}_t^2(\muw)}{m_W^2}\right) 
+ \Delta_{i}\left(\frac{\ovl{m}_t^2(\muw)}{m_W^2}\right)
\ln \frac{\muw^2}{m_W^2}, & \text{for}~~~i=7,8,
\end{cases}
\end{equation}
 where
\begin{align}
\begin{split}
E(x) =& \frac{x(-18+11 x+x^2)}{12(x-1)^3}
+\frac{x^2(15-16 x+4 x^2)}{6(x-1)^4}\ln x
-\frac{2}{3}\ln x,
\end{split} \\
%
\begin{split}
G_7(x) =& \frac{-16 x^4-122 x^3+80 x^2-8 x}{9(x-1)^4} {\rm Li_2}
\left(1-\frac{1}{x}\right) \\
&
+\frac{6 x^4+46 x^3 -28 x^2}{3(x-1)^5}\ln^2 x \\
&+\frac{-102 x^5 -588 x^4 -2262 x^3 +3244 x^2 -1364 x
+208}{81(x-1)^5}\ln x \\
&+\frac{1646 x^4+12205 x^3-10740 x^2+2509x-436}{486(x-1)^4}, 
\end{split}\\
%
\begin{split}
\Delta_7(x) =& \frac{208-1111 x+1086 x^2+383 x^3+82 x^4}{81(x-1)^4}\\
&
+\frac{2x^2(14-23 x-3 x^2)}{3(x-1)^5}\ln x,
\end{split}\\
%
\begin{split}
G_8(x) =& \frac{-4 x^4+40 x^3+41 x^2+x}{6(x-1)^4}{\rm Li_2}\left(
1-\frac{1}{x} \right) \\
&
 +\frac{-17 x^3-31 x^2}{2(x-1)^5} \ln^2 x \\
& +\frac{-210 x^5+1086 x^4+4893 x^3+2857 x^2-1994 x+280}{216(x-1)^5}
\ln x \\
& +\frac{737 x^4-14102 x^3-28209 x^2+610 x-508}{1296(x-1)^4},
\end{split}\\
%
\begin{split}
\Delta_8(x) =& \frac{140-902 x-1509 x^2-398 x^3+77 x^4}{108(x-1)^4}\\
&
+\frac{x^2(31+17 x)}{2(x-1)^5}\ln x.
\end{split}
\end{align}
%

Similarly, the NLO QCD corrections to the charged Higgs boson contributions
 to the Wilson coefficients are given in~\cite{cdgg97}, 
\begin{gather}
\begin{split}
\delta^H_g C^{(1)}_i(\muw) =
\begin{cases}
0, &\text{for}~~~i=1,2,3,5,6, \\
E^H\left(y\right), 
 & \text{for}~~~i=4, \\
G^H_i\left(y\right)
+ \Delta^H_i\left(y\right)
\ln \frac{\muw^2}{m_H^2} 
-\frac{4}{9} E^H\left(y\right),
 &\text{for}~~~i=7, \\
G^H_i\left(y\right)
+ \Delta^H_i\left(y\right)
\ln \frac{\muw^2}{m_H^2} 
-\frac{1}{6} E^H\left(y\right), 
 &\text{for}~~~i=8,
\end{cases}
\end{split}
\end{gather}
where, $y \equiv \ovl{m}_t^2(\muw)/m_H^2$,
\begin{align}
\begin{split}
E^H(y) =& \frac{y(16-29 y+7 y^2)}{36(y-1)^3}
+\frac{y(3y-2)}{6(y-1)^4} \ln y, 
\end{split}\\
%
\begin{split}
G^H_7(y) =& -\frac{4}{3}y\left[
\frac{4(-3+7y-2y^2)}{3(y-1)^3}{\rm Li_2}\left(1-\frac{1}{y}\right)
+\frac{8-14y-3y^2}{3(y-1)^4}\ln^2 y \right. \\
&
\phantom{=-\frac{4}{3}y}\left.
+\frac{2(-3-y+12 y^2-2 y^3)}{3(y-1)^4}\ln y
+\frac{7-13 y+2 y^2}{(y-1)^3} \right] \\
&+\frac{2}{9}y\left[ \frac{y(18-37y+8y^2)}{(y-1)^4}{\rm
Li_2}\left(1-\frac{1}{y}\right)
+\frac{y(-14+23y+3y^2)}{(y-1)^5}\ln^2 y \right. \\
&
\phantom{=+\frac{2}{9}y}
\left.
+\frac{-50+251 y-174y^2-192y^3+21y^4}{9(y-1)^5}\ln y
\right.\\
&
\phantom{=+\frac{2}{9}y}
\left.
+\frac{797-5436y+7569y^2-1202y^3}{108(y-1)^4}\right],
\end{split}\\
%
\begin{split}
\Delta^H_7(y) =&
-\frac{2}{9}y\left[\frac{21-47y+8y^2}{(y-1)^3}+\frac{2(-8+14y+3y^2)}{(y-1)^4}\ln
y\right]\\
&
+\frac{2}{9}y\left[
\frac{-31-18y+135y^2-14y^3}{6(y-1)^4}+\frac{y(14-23y-3y^2)}{(y-1)^5}\ln
y \right],
\end{split}\\
%
\begin{split}
G^H_8(y) =& -\frac{1}{3}y\left[ 
\frac{-36+25 y-17 y^2}{2(y-1)^3}{\rm Li_2}\left( 1-\frac{1}{y} \right)
+\frac{y(19+17y)}{(y-1)^4}\ln^2 y \right. \\
&
\phantom{=-\frac{1}{3}y}
\left.
+\frac{-3-187y+12y^2-14y^3}{4(y-1)^4}\ln y
+\frac{3(143-44 y+29y^2)}{8(y-1)^3} \right] \\
&
+\frac{1}{6}y\left[ \frac{y(30-17y+13y^2)}{(y-1)^4}{\rm Li_2}
\left(1-\frac{1}{y}\right)
-\frac{y(31+17y)}{(y-1)^5}\ln^2 y \right. \\
&
\phantom{=+\frac{1}{6}y}
+\frac{-226+817y+1353y^2+318y^3+42y^4}{36(y-1)^5}\ln y \\
&
\phantom{=+\frac{1}{6}y}
\left.
+\frac{1130-18153 y+7650y^2-4451y^3}{216(y-1)^4} \right], 
\end{split}\\
%
\begin{split}
\Delta^H_8(y) =& -\frac{1}{3}y\left[\frac{81-16y+7y^2}{2(y-1)^3}
-\frac{19+17y}{(y-1)^4}\ln y \right] \\
&
+\frac{1}{6}y\left[\frac{-38-261y+18y^2-7y^3}{6(y-1)^4}
+\frac{y(31+17y)}{(y-1)^5}\ln y \right].
\end{split}
\end{align}
%

The NLO matching conditions for $\bsgamma$ in the MSSM have been
calculated in the limit of $\mg \to \infty$ by Bobeth, \etal,
in~\cite{bmu00}.  We apply their calculation on the 2-loop gluon
corrections to $C_{7,8}$ at $\mususy$, while distinguishing between
the two scales $\muw$ and $\mususy$.  We neglect the subdominant
contribution from $C_4$ and do not include the corrections from
quartic squark vertex in light of the different assumptions about the
mass hierarchy.

The NLO QCD correction\footnote{Strictly speaking, higher--order SQCD
 correction is also included in the formulae below through
 gluino-squark corrections to the squark vertices and masses.  } to
 the chargino contribution to $C_{7,8}$ is given in~\cite{bmu00},
\begin{align}
\begin{split}
\delta^{\chi^-}_g C^{(1)}_{7,8}(\mususy)
 =& \frac{1}{g_2^2 K^{\ast}_{ts}K_{tb}}\sum^6_{I=1}\sum^2_{a=1}
\frac{m_W^2}{\mca^2} \\
&
\times
\left[
(C_{dR})_{2aI}(C_{dR})^{\ast}_{3aI}
\left\{ H^{[7,8]\prime}_1(\xsuica)+ H^{[7,8]\prime\prime}_1(\xsuica)\ln
 \left(\frac{\mususy^2}{{m_{\wt{u}_I}}^2}\right) \right\}
\right.\\
&
\phantom{\times}
~~+\frac{\mca}{m_b}(C_{dR})_{2aI}(C_{dL})^{\ast}_{3aI} \\
&
\phantom{\times+}
~~\left.\times
\left\{ H^{[7,8]\prime}_2(\xsuica)+ H^{[7,8]\prime\prime}_2(\xsuica)\ln
 \left(\frac{\mususy^2}{{m_{\wt{u}_I}}^2}\right) \right\}
\right],
\end{split}
\end{align}
where
$\xsuica \equiv {m_{\wt{u}_I}}^2/{m_{\chi^-_a}}^2$ and
\begin{align}
\begin{split}
H^{[7]\prime}_1(x) =& \frac{24 x^3+52 x^2-32 x}{9(1-x)^4}
{\rm Li_2}\left(1-\frac{1}{x}\right) 
+\frac{-189 x^3-783 x^2+425 x+43}{81(1-x)^5}\ln x \\
&
+\frac{-1030x^3-1899x^2+1332x+85}{243(1-x)^4}, 
\end{split}\\
\begin{split}
H^{[7]\prime\prime}_1(x) =& \frac{6x^3-62x^2+32x}{9(1-x)^5}\ln x
+\frac{28x^3-129x^2-12x+41}{27(1-x)^4}, 
\end{split}\\
%
\begin{split}
H^{[7]\prime}_2(x) =& \frac{112x^2-48x}{9(1-x)^3}{\rm Li_2}
\left(1-\frac{1}{x}\right)+\frac{12x^3-176x^2+64x+16}{9(1-x)^4}\ln x
\\
&
+\frac{-170x^2+66x+20}{9(1-x)^3},
\end{split}\\
\begin{split}
H^{[7]\prime\prime}_2(x) =& \frac{12x^3-88x^2+40x}{9(1-x)^4}\ln x
+\frac{-14x^2-54x+32}{9(1-x)^3}, 
\end{split}\\
%
\begin{split}
H^{[8]\prime}_1(x) =& \frac{-9x^3-46x^2-49x}{12(1-x)^4}{\rm Li_2}
\left(1-\frac{1}{x}\right)+\frac{81x^3+594x^2+1270x+71}{108(1-x)^5}
\ln x \\
&
+\frac{923x^3+3042x^2+6921x+1210}{648(1-x)^4}, 
\end{split}\\
\begin{split}
H^{[8]\prime\prime}_1(x) =& \frac{5x^2+19x}{3(1-x)^5}\ln x
+\frac{7x^3-30x^2+141x+26}{18(1-x)^4}, 
\end{split}
\end{align}
\begin{align}
\begin{split}
H^{[8]\prime}_2(x) =& \frac{-16x^2-12x}{3(1-x)^3}{\rm
Li_2}\left(1-\frac{1}{x}\right)+\frac{52x^2+109x+7}{6(1-x)^4}\ln x \\
&
+\frac{95x^2+180x+61}{12(1-x)^3}, 
\end{split}\\
\begin{split}
H^{[8]\prime\prime}_2(x) =& \frac{10x^2+26x}{3(1-x)^4}\ln x
+\frac{-x^2+30x+7}{3(1-x)^3}.
\end{split}
\end{align}

The NLO QCD correction to the neutralino contribution to $C_{7,8}$
 is given in~\cite{bmu00},
\begin{align}
\begin{split}
\delta^{\chi^0}_g C^{(1)}_{7,8}(\mususy) =& \frac{1}{g_2^2 K_{ts}^{\ast}
K_{tb}}\sum^6_{I=1}\sum^4_{r=1}\frac{m_W^2}{\mnr^2} \\
&
 \times\left[
(N_{dR})_{2rI}(N_{dR})^{\ast}_{3rI}
\left\{ H^{[7,8]\prime}_3(\xsdinr)+H^{[7,8]\prime\prime}_3(\xsdinr)\ln
\left(\frac{\mususy^2}{{\msd}_I^2}\right) \right\} \right. \\
&
\phantom{\times}
~~+\frac{\mnr}{m_b}(N_{dR})_{2rI} (N_{dL})^{\ast}_{3rI} \\
&
\phantom{\times\times}
~~\left.
\times\left\{H^{[7,8]\prime}_4(\xsdinr)+H^{[7,8]\prime\prime}_4(\xsdinr)
\ln\left(\frac{\mususy^2}{{\msd}_I^2}\right) \right\} \right],
\end{split} 
\end{align}
where $\xsdinr \equiv {m_{\wt{d}_I}}^2/{m_{\chi^0_r}}^2$ and
\begin{align}
\begin{split}
H^{[7]\prime}_3(x) =& \frac{16x^2+28x}{9(1-x)^4}{\rm Li_2}\left(
1-\frac{1}{x} \right) + \frac{-108x^2-358x-38}{81(1-x)^5}\ln x
\\
&
+\frac{23x^3-765x^2-693x-77}{243(1-x)^4}, 
\end{split}\\
\begin{split}
H^{[7]\prime\prime}_3(x) =& \frac{4x^2-28x}{9(1-x)^5}\ln x
+\frac{-8x^3+42x^2-84x-22}{27(1-x)^4}, 
\end{split}\\
%
\begin{split}
H^{[7]\prime}_4(x) =& \frac{16x^2+48x}{9(1-x)^3}{\rm Li_2}
\left(1-\frac{1}{x}\right)+\frac{-8x^2-68x-8}{9(1-x)^4}\ln x \\
&
+\frac{-26x^2-54x-4}{9(1-x)^3}, 
\end{split}\\
\begin{split}
H^{[7]\prime\prime}_4(x) =& \frac{8x^2-44x}{9(1-x)^4}\ln x 
+\frac{10x^2-30x-16}{9(1-x)^3}, 
\end{split}\\
%
\begin{split}
H^{[8]\prime}_{3,4}(x) =& H^{[8]\prime}_{1,2}(x), 
\end{split}\\
\begin{split}
H^{[8]\prime\prime}_{3,4}(x) =& H^{[8]\prime\prime}_{1,2}(x), 
\end{split}
\end{align}

The NLO QCD corrections to the gluino contribution to $C_{7,8}$ is
given by~\cite{bmu00},
\begin{gather}
\begin{split}
\delta^{\wt{g}}_g C^{(1)}_{7,8}(\mususy) =&
\frac{4}{3g_2^2K_{ts}^{\ast} K_{tb}} \frac{m_W^2}{\mg^2}\sum^6_I \\
&
\times\left[
(G_{dR})_{2I}(G_{dR})^{\ast}_{3I}
\left\{H^{[7,8]\prime}_5(\xsdig)+H^{[7,8]\prime\prime}_5(\xsdig)\ln\left(\frac{\mususy^2}{{\msd}_I^2}\right)
\right\} \right. \\
&
\phantom{\times}
~~+\frac{\mg}{m_b}(G_dR)_{2I}(G_{dL})^{\ast}_{3I} \\
&
\phantom{\times\times}
~~\left.\times\left\{
H^{[7,8]\prime}_6(\xsdig)+H^{[7,8]\prime\prime}_6(\xsdig)\ln\left(\frac{\mususy^2}{{\msd}_I^2}\right)\right\}
\right]
\end{split}
\end{gather}
where
$\xsdig \equiv {m_{\wt{d}_I}}^2/\mg^2$ and
\begin{align}
\begin{split}
H^{[7]\prime}_5(x) =& \frac{17x^2+86x-15}{18(1-x)^4}{\rm
Li_2}\left(1-\frac{1}{x}\right)+\frac{6x^3+45x^2+66x-5}{12(1-x)^5}\ln^2
x \\
&
+\frac{-36x^4-315x^3+1161x^2+751x+23}{162(1-x)^5}\ln x \\
&
+\frac{-799x^3+1719x^2+10431x-1847}{972(1-x)^4}, 
\end{split}\\
\begin{split}
H^{[7]\prime\prime}_5(x) =& \frac{18x^3+107x^2+43x}{18(1-x)^5}\ln x
+\frac{-5x^3+384x^2+699x+20}{108(1-x)^4}, 
\end{split}\\
%
\begin{split}
H^{[7]\prime}_6(x) =& \frac{19x^2+60x-15}{9(1-x)^3}{\rm Li_2}\left(
1-\frac{1}{x} \right)+\frac{6x^3+36x^2+48x-5}{6(1-x)^4}\ln^2 x \\ 
&
+\frac{-27x^3+106x^2+52x+1}{9(1-x)^4}\ln
x+\frac{14x^2+333x-83}{18(1-x)^3} x, 
\end{split}\\
\begin{split}
H^{[7]\prime\prime}_6(x) =& \frac{18x^3+80x^2+28x}{9(1-x)^4}\ln
x+\frac{55x^2+69x+2}{9(1-x)^3}, 
\end{split}\\
%
\begin{split}
H^{[8]\prime}_5(x) =& \frac{45x^3-1208x^2+901x-570}{96(1-x)^4}
{\rm Li_2}\left(
1-\frac{1}{x}\right) 
\\
&
+\frac{-237x^3-846x^2+282x-95}{32(1-x)^5}\ln^2 x 
\\
&
+\frac{2520x^4-10755x^3-10638x^2-6427x-44}{864(1-x)^5}\ln x
\\
&
+\frac{5359x^3-241425x^2+143253x-59251}{5184(1-x)^4}, 
\end{split}\\
\begin{split}
H^{[8]\prime\prime}_5(x) =& \frac{-747x^3-640x^2+43x}{48(1-x)^5}\ln x 
+\frac{-779x^3-7203x^2-93x+11}{288(1-x)^4},
\end{split}
\end{align}
\begin{align}
\begin{split}
H^{[8]\prime}_6(x) =& \frac{-359x^2+339x-204}{24(1-x)^3}{\rm
Li_2}\left(1-\frac{1}{x}\right)\\
&
+\frac{-78x^3-333x^2+105x-34}{8(1-x)^4}\ln^2 x \\
&
+\frac{-207x^3-1777x^2+23x-151}{48(1-x)^4} \ln x \\
&
+\frac{-1667x^2+990x-379}{24(1-x)^3}, 
\end{split}\\
\begin{split}
H^{[8]\prime\prime}_6(x) =& \frac{-126x^3-133x^2+7x}{6(1-x)^4}\ln x+\frac{-553x^2+84x-35}{12(1-x)^3}.
\end{split}
\end{align}
Similar expressions for $\delta^{\chi^-,\chi^0,\wt{g}}_{g} C^{\prime
(1)}_{7,8}$ can be obtained by interchanging the indices $L$ and $R$ in
the above formulae.



\begin{thebibliography}{99}


\bibitem{hurth02}
For an excellent recent review, see T.~Hurth, ``Present Status of
Inclusive Rare B Decays'', \hepph{0212304}, to appear in Reviews of Modern
Physics. 

\bibitem{exp}
S.~Chen {\it et al.}  [CLEO Collaboration],
Phys.\ Rev.\ Lett.\  {\bf 87}, 251807 (2001)
[arXiv:hep-ex/0108032];
%
R.~Barate {\it et al.}  [ALEPH Collaboration],
Phys.\ Lett.\ B {\bf 429}, 169 (1998);
%
K.~Abe {\it et al.}  [Belle Collaboration],
Phys.\ Lett.\ B {\bf 511}, 151 (2001)
[arXiv:hep-ex/0103042];
%
%
B.~Aubert {\it et al.}  [BABAR Collaboration],
arXiv:hep-ex/0207074;
%
B.~Aubert {\it et al.}  [BaBar Collaboration],
arXiv:hep-ex/0207076.

\bibitem{expaverage}
C.~Jessop, ``A world average for $B \to X_s\gamma$,''
SLAC-PUB-9610.


\bibitem{bm02}
A.~J.~Buras and M.~Misiak,
``Anti-B $\to$ X/s gamma after completion of the NLO QCD calculations,''
Acta Phys.\ Polon.\ B {\bf 33}, 2597 (2002)
[arXiv:hep-ph/0207131]
 and references therein.

\bibitem{ggh03}
P.~Gambino, M.~Gorbahn and U.~Haisch,
``Anomalous dimension matrix for radiative and rare semileptonic B
	decays  up to three loops,''
arXiv:hep-ph/0306079.


\bibitem{bcmu02}
A.~J.~Buras, A.~Czarnecki, M.~Misiak and J.~Urban,
``Completing the NLO QCD calculation of anti-B $\to$ X/s gamma,''
Nucl.\ Phys.\ B {\bf 631}, 219 (2002)
[arXiv:hep-ph/0203135].




\bibitem{2hdnlo}
F.~M.~Borzumati and C.~Greub,
Phys.\ Rev.\ D {\bf 58}, 074004 (1998)
[arXiv:hep-ph/9802391]
%
;
%
P.~Ciafaloni, A.~Romanino and A.~Strumia,
Nucl.\ Phys.\ B {\bf 524}, 361 (1998)
[arXiv:hep-ph/9710312].

\bibitem{cdgg97}
M.~Ciuchini, G.~Degrassi, P.~Gambino and G.~F.~Giudice,
``Next-to-leading QCD corrections to B $\to$ X/s gamma: Standard model
and two-Higgs doublet model,'' Nucl.\ Phys.\ B {\bf 527}, 21 (1998)
[arXiv:hep-ph/9710335].

\bibitem{gm01}
P.~Gambino and M.~Misiak,
``Quark mass effects in anti-B $\to$ X/s gamma,''
Nucl.\ Phys.\ B {\bf 611}, 338 (2001)
[arXiv:hep-ph/0104034].

\bibitem{bbmr90}
S.~Bertolini, F.~Borzumati, A.~Masiero and G.~Ridolfi,
``Effects Of Supergravity Induced Electroweak Breaking On Rare B Decays And Mixings,''
Nucl.\ Phys.\ B {\bf 353}, 591 (1991).

%
\bibitem{hrs}
L.~J.~Hall, R.~Rattazzi and U.~Sarid,
Phys.\ Rev.\ D {\bf 50}, 7048 (1994)
[arXiv:hep-ph/9306309]
;
R.~Rattazzi and U.~Sarid,
Nucl.\ Phys.\ B {\bf 501}, 297 (1997)
[arXiv:hep-ph/9612464].



\bibitem{br99} 
T.~Blazek and S.~Raby,
``b $\to$ s gamma with large tan(beta) in MSSM analysis constrained by a  realistic SO(10) model,''
Phys.\ Rev.\ D {\bf 59}, 095002 (1999)
[arXiv:hep-ph/9712257].

\bibitem{cgnw99}
M.~Carena, D.~Garcia, U.~Nierste and C.~E.~Wagner,
``Effective Lagrangian for the anti-t b H+ interaction in the MSSM and  charged Higgs phenomenology,''
Nucl.\ Phys.\ B {\bf 577}, 88 (2000)
[arXiv:hep-ph/9912516].

%
\bibitem{dgg00}
G.~Degrassi, P.~Gambino and G.~F.~Giudice,
JHEP {\bf 0012}, 009 (2000)
[arXiv:hep-ph/0009337].


\bibitem{cgnw00}
M.~Carena, D.~Garcia, U.~Nierste and C.~E.~Wagner,
``b $\to$ s gamma and supersymmetry with large tan(beta),''
Phys.\ Lett.\ B {\bf 499}, 141 (2001)
[arXiv:hep-ph/0010003].

\bibitem{mfv:ref}
E.~Gabrielli and G.~F.~Giudice,
``Supersymmetric corrections to epsilon'/epsilon at the leading  order in QCD and QED,''
Nucl.\ Phys.\ B {\bf 433}, 3 (1995)
[Erratum-ibid.\ B {\bf 507}, 549 (1997)]
[arXiv:hep-lat/9407029].


\bibitem{ggms96}
F.~Gabbiani, E.~Gabrielli, A.~Masiero and L.~Silvestrini,
``A complete analysis of FCNC and CP constraints in general SUSY extensions of the standard model,''
Nucl.\ Phys.\ B {\bf 477}, 321 (1996)
[arXiv:hep-ph/9604387].

\bibitem{cfms03}
M.~Ciuchini, E.~Franco, A.~Masiero and L.~Silvestrini,
Phys.\ Rev.\ D {\bf 67}, 075016 (2003)
[arXiv:hep-ph/0212397].

\bibitem{agis02}
G.~D'Ambrosio, G.~F.~Giudice, G.~Isidori and A.~Strumia,
``Minimal flavour violation: An effective field theory approach,''
Nucl.\ Phys.\ B {\bf 645}, 155 (2002)
[arXiv:hep-ph/0207036].

\bibitem{ciuchinietal98}
M.~Ciuchini {\it et al.},
``Delta M(K) and epsilon(K) in SUSY at the next-to-leading order,''
JHEP {\bf 9810}, 008 (1998)
[arXiv:hep-ph/9808328].

\bibitem{en81}
J.~R.~Ellis and D.~V.~Nanopoulos,
``Flavor Changing Neutral Interactions In Broken Supersymmetric Theories,''
Phys.\ Lett.\ B {\bf 110}, 44 (1982).

\bibitem{bbm86}
S.~Bertolini, F.~Borzumati and A.~Masiero,
Phys.\ Lett.\ B {\bf 192}, 437 (1987);
%
A.~Masiero and G.~Ridolfi,
Phys.\ Lett.\  {\bf 212B}, 171 (1988)
[Addendum-ibid.\  {\bf 213B}, 562 (1988)];
%
F.~Gabbiani and A.~Masiero,
Nucl.\ Phys.\ B {\bf 322}, 235 (1989).

\bibitem{ht92}
J.~S.~Hagelin, S.~Kelley and T.~Tanaka,
``Supersymmetric flavor changing neutral currents: Exact amplitudes and phenomenological analysis,''
Nucl.\ Phys.\ B {\bf 415}, 293 (1994).

\bibitem{bghw99}
F.~Borzumati, C.~Greub, T.~Hurth and D.~Wyler,
Phys.\ Rev.\ D {\bf 62}, 075005 (2000)
[arXiv:hep-ph/9911245]
;
T.~Besmer, C.~Greub and T.~Hurth,
Nucl.\ Phys.\ B {\bf 609}, 359 (2001)
[arXiv:hep-ph/0105292].

\bibitem{ekrww02}
L.~Everett, G.~L.~Kane, S.~Rigolin, L.~T.~Wang and T.~T.~Wang,
``Alternative approach to b $\to$ s gamma in the uMSSM,''
JHEP {\bf 0201}, 022 (2002)
[arXiv:hep-ph/0112126].

\bibitem{or1}
K.~Okumura and L.~Roszkowski,
``De-constraining supersymmetry from b $\to$ s gamma?,''
arXiv:hep-ph/0208101.

\bibitem{bhy03}
F.~Borzumati, C.~Greub and Y.~Yamada,
``Towards an exact evaluation of the supersymmetric O(alpha(s) tan(beta))  corrections to anti-B $\to$ X/s gamma,''
arXiv:hep-ph/0305063.


\bibitem{cdgg98}
M.~Ciuchini, G.~Degrassi, P.~Gambino and G.~F.~Giudice,
``Next-to-leading {QCD} corrections to B $\to$ X/s gamma in supersymmetry,''
Nucl.\ Phys.\ B {\bf 534}, 3 (1998)
[arXiv:hep-ph/9806308].

\bibitem{banks87}
T.~Banks,
``Supersymmetry And The Quark Mass Matrix,''
Nucl.\ Phys.\ B {\bf 303}, 172 (1988).

\bibitem{for03}
J.~Foster, K.~Okumura and L.~Roszkowski, in preparation.


\bibitem{babukolda99}
K.S.~Babu and C.~Kolda, hep-ph/9909476.
K.~S.~Babu and C.~F.~Kolda,
``Higgs-mediated B0 $\to$ mu+ mu- in minimal supersymmetry,''
Phys.\ Rev.\ Lett.\  {\bf 84}, 228 (2000) [arXiv:hep-ph/9909476].

\bibitem{ccb}
J.~A.~Casas and S.~Dimopoulos,
Phys.\ Lett.\ B {\bf 387}, 107 (1996)
[arXiv:hep-ph/9606237]
;
H.~Baer, M.~Brhlik and D.~Castano,
Phys.\ Rev.\ D {\bf 54}, 6944 (1996)
[arXiv:hep-ph/9607465].

\bibitem{pbmz97}
D.~M.~Pierce, J.~A.~Bagger, K.~T.~Matchev and R.~j.~Zhang,
``Precision corrections in the minimal supersymmetric standard model,''
Nucl.\ Phys.\ B {\bf 491}, 3 (1997)
[arXiv:hep-ph/9606211].

\bibitem{gh84}
J.~F.~Gunion and H.~E.~Haber,
Nucl.\ Phys.\ B {\bf 272}, 1 (1986);
%
Nucl.\ Phys.\ B {\bf 278}, 449 (1986);
%
Nucl.\ Phys.\ B {\bf 307}, 445 (1988);
[Erratum-ibid.\ B {\bf 402}, 569 (1993)].

\bibitem{br95}
T.~Blazek, S.~Raby and S.~Pokorski,
``Finite supersymmetric threshold corrections to CKM matrix elements
in the large tan beta regime,'' 
Phys.\ Rev.\ D {\bf 52}, 4151 (1995)
[arXiv:hep-ph/9504364].

\bibitem{dgh84}
M.~Dugan, B.~Grinstein and L.~J.~Hall,
``CP Violation In The Minimal N=1 Supergravity Theory,''
Nucl.\ Phys.\ B {\bf 255}, 413 (1985).

\bibitem{hkr85}
L.~J.~Hall, V.~A.~Kostelecky and S.~Raby,
``New Flavor Violations In Supergravity Models,''
Nucl.\ Phys.\ B {\bf 267}, 415 (1986).

\bibitem{cmm97}
K.~G.~Chetyrkin, M.~Misiak and M.~Munz,
``Weak radiative B-meson decay beyond leading logarithms,''
Phys.\ Lett.\ B {\bf 400}, 206 (1997)
[Erratum-ibid.\ B {\bf 425}, 414 (1998)]
[arXiv:hep-ph/9612313].


\bibitem{misiakmunz95}
M.~Misiak and M.~Munz,
``Two loop mixing of dimension five flavor changing operators,''
Phys.\ Lett.\ B {\bf 344}, 308 (1995).

\bibitem{buras92}
A.J.~Buras and M.K.~Harlander, 1992,
in Heavy Flavours, edited by A.J.~Buras and M.~Lindner
 (World Scientific, Singapore),p. 58.

\bibitem{bmu00}
C.~Bobeth, M.~Misiak and J.~Urban,
``Matching conditions for b $\to$ s gamma and b $\to$ s gluon in extensions  of the standard model,''
Nucl.\ Phys.\ B {\bf 567}, 153 (2000)
[arXiv:hep-ph/9904413].

\bibitem{inamilim80}
T.~Inami and C.~S.~Lim,
``Effects Of Superheavy Quarks And Leptons In Low-Energy Weak
Processes K(L) $\to$ Mu Anti-Mu, K+ $\to$ Pi+ Neutrino Anti-Neutrino
And K0 $\leftrightarrow$ Anti-K0,''
Prog.\ Theor.\ Phys.\  {\bf 65}, 297 (1981)
[Erratum-ibid.\  {\bf 65}, 1772 (1981)].

\bibitem{pdg}
K.~Hagiwara {\it et al.}  [Particle Data Group Collaboration],
Phys.\ Rev.\ D {\bf 66}, 010001 (2002);
L.~L.~Chau and W.~Y.~Keung,
Phys.\ Rev.\ Lett.\  {\bf 53}, 1802 (1984).

\bibitem{bbo93}
V.~D.~Barger, M.~S.~Berger and P.~Ohmann,
``The Supersymmetric particle spectrum,''
Phys.\ Rev.\ D {\bf 49}, 4908 (1994)
[arXiv:hep-ph/9311269].


\bibitem{vpfunction}
G.~'t Hooft and M.~J.~Veltman,
Nucl.\ Phys.\ B {\bf 153}, 365 (1979).
;
G.~Passarino and M.~J.~Veltman,
Nucl.\ Phys.\ B {\bf 160}, 151 (1979).
;
We follw the definition in 
K.~Hagiwara, S.~Matsumoto, D.~Haidt and C.~S.~Kim,
Z.\ Phys.\ C {\bf 64}, 559 (1994)
[Erratum-ibid.\ C {\bf 68}, 352 (1995)]
[arXiv:hep-ph/9409380].

\bibitem{bdr}
T.~Blazek, R.~Dermisek and S.~Raby,
Phys.\ Rev.\ Lett.\  {\bf 88}, 111804 (2002)
[arXiv:hep-ph/0107097]
;
Phys.\ Rev.\ D {\bf 65}, 115004 (2002)
[arXiv:hep-ph/0201081].





\end{thebibliography}
\end{document}